\documentclass[12pt]{article}

\usepackage{subfigure}
\usepackage{amssymb}
\usepackage{amsfonts}
\usepackage{bm}
\usepackage{color}
\usepackage{calc}
\usepackage[latin1]{inputenc}
\usepackage{amsfonts}
\usepackage{amsmath}
\usepackage[english]{babel}
\usepackage[T1]{fontenc}
\usepackage{verbatim}
\usepackage{subfigure}
\usepackage{enumerate}
\usepackage{graphicx}
\usepackage{natbib}
\usepackage{url}
\usepackage{algorithmic,algorithm}
\usepackage{arydshln}

\def\d{{\mathrm d}}

\def\X{{\bf X}}
\def\A{{\bf A}}
\def\B{{\bf B}}
\def\C{{\bf C}}

\def\G{{\bf G}}

\def\MM{{M}}
\def\MN{{v}}

\def\P{{\bf P}}
\def\SS{{\bf S}}

\def\K{{\bf K}}
\def\L{{\bf L}}

\def\0{{\bf 0}}

\def\N{{\mathcal N}}
\def\HH{{\mathcal H}}
\def\J{{\mathcal J}}

\newcommand{\rank}{\operatorname{rank}}
\def\bxi{\pmb{\xi}}
\def\bpsi{\boldsymbol{\psi}}

\let\Re\relax
\DeclareMathOperator{\Re}{Re}
\let\Im\relax
\DeclareMathOperator{\Im}{Im}

\def\Bmp#1{ \begin{minipage}{#1} }
\def\Emp{ \end{minipage} }
\def\Bmpc#1{ \begin{minipage}[c]{#1} }
\def\Bmpt#1{ \begin{minipage}[t]{#1} }
\def\Bmpb#1{ \begin{minipage}[b]{#1} }

\begin{document}
\title{Linear feedback stabilization of point vortex equilibria near a Kasper Wing}

\author{R.~Nelson$^{1,}$\thanks{Email address for correspondence: rnelson@ic.ac.uk}, \ B.~Protas$^{2}$ and T.~Sakajo$^{3}$
\\ \\ 
$^1$ Department of Mathematics, Imperial College London \\ 
{180 Queen's Gate, London, SW7 2AZ}, UK
\\ \\
$^2$ Department of Mathematics and Statistics, McMaster University \\
 Hamilton, Ontario L85 4K1, Canada
\\ \\ 
$^3$ Department of Mathematics, Kyoto University  \\
{Kitashirakawa Oiwake-cho, Sakyo-ku, Kyoto, 606-8502}, Japan
}

\date{\today}
\maketitle

\begin{abstract}
  This paper concerns feedback stabilization of point vortex
  equilibria above an inclined thin plate and a three-plate
  configuration known as the Kasper Wing in the presence of an
  oncoming uniform flow. The flow is assumed to be potential and is
  modeled by the two-dimensional incompressible Euler equations.
  Actuation has the form of blowing and suction localized on the main
  plate and is represented in terms of a sink-source singularity,
  whereas measurement of pressure across the plate serves as system
  output. We focus on point-vortex equilibria forming a one-parameter
  family with locus approaching the trailing edge of the main plate
  and show that these equilibria are either unstable or neutrally
  stable. Using methods of linear control theory we find that the
  system dynamics linearised around these equilibria are both
  controllable and observable for almost all actuator and sensor
  locations. The design of the feedback control is based on the
  Linear-Quadratic-Gaussian (LQG) compensator.  Computational results
  demonstrate the effectiveness of this control and the key finding of
  this study is that Kasper Wing configurations are in general not
  only more controllable than their single plate counterparts, but
  also exhibit larger basins of attraction under LQG feedback control.
  The feedback control is then applied to systems with additional
  perturbations added to the flow in the form of random fluctuations
  of the angle of attack and a vorticity shedding mechanism.  Another
  important observation is that, in the presence of these additional
  perturbations, the control remains robust, provided the system does
  not deviate too far from its original state. Furthermore, except in
  a few isolated cases, introducing a vorticity shedding mechanism
  enhanced the effectiveness of the control. Physical interpretation
  is provided for the results of the controllability and observability
  analysis as well as the response of the feedback control to
  different perturbations.
\end{abstract}

\begin{flushleft}
\end{flushleft}



\section{Introduction}\label{intro}

Flows over sharp-edged aerofoils {are} in general separated.
{Manipulation of} this separation in order to achieve beneficial
aerodynamic properties is of great practical importance and this topic
is therefore subject to a great deal of attention in the scientific
{and engineering} literature. This control of the flow separation
is generally aimed at increasing the lift and decreasing the drag
experienced by the aerofoil and the {control} strategies used can
broadly be classified as either \textit{passive} or \textit{active}.
Passive control strategies {usually} involve some form of
geometrical modification to the aerofoil, for example, {by
  deployment of} vortex generators such as Gurney flaps
\citep{storms:lenh}. Active control strategies, on the other hand,
involve the injection or extraction of energy. {For instance,}
synthetic jet actuators are commonly used to perform such flow control
over aerofoils \citep{smith98}. Another approach is to use some form
of actuation to ``simulate'' a passive control, for example, by using
plasma actuators to imitate Gurney flaps \citep{feng15}.

From a physical perspective, the goal of all such devices is to
{allow shed vortices to convect over the airfoil, past the
  trapped vortices whose role is to increase} the circulation around
the aerofoil. {For example, in the} flow past an inclined flat plate
at high Reynolds number, large-scale vortex structures are
periodically shed from both the leading and trailing edges of the
plate. If a leading edge vortex could be trapped above the plate, a
potentially large gain in lift could be achieved for little cost. It
was with this motivation in mind that Witold Kasper designed, and was
subsequently granted a patent for, his ``aircraft wing with vortex
generation'' \citep{kasper1974aircraft}.  Following test-flights
investigating the stall characteristics of his BKB-1 glider, Kasper
reported lift-to-drag ratios of $L/D=17.6$ at {the speed of}
$32\mathrm{kmh^{-1}}$ and angle of attack of 35 degrees, considered
remarkable at such low speeds \citep{kruppa}.  If such results could
be {reproduced and} verified, they {would} likely indicate a
previously undocumented phenomenon operating on Kasper's glider.
Kasper {conjectured} that a vortex tube formed above the wing
concomitantly increasing the lift and decreasing the drag, a
phenomenon coined ``vortex lift''. He further speculated that this
effect could be enhanced by positioning additional ``flaps'' or
auxiliary aerofoils close to the main aerofoil to control the feeding
and shedding of vortices in the vicinity of the wing. This
configuration, where two additional auxiliary aerofoils are placed off
the rear of the main aerofoil is what is referred to as a ``Kasper
Wing''.

Among the first mathematical investigations into the phenomenon of
vortex lift was {the study} carried out by
\citet{saffman:attachedvortex}, where solutions for steady
{potential} flows consisting of a single point vortex located
above a thin flat plate in the presence of a {uniform stream
  were} given explicitly. In that study, it {was} shown that
leading and trailing edge loci {exist} for the point vortex
position that satisfy the steady state condition and that these
equilibria are force-enhancing, that is, the presence of the point
vortex enhances the lift experienced by the aerofoil. Further, a
preliminary stability analysis indicated that linearly stable
solutions exist only when the vortex is located on the trailing-edge
locus not too close to the plate and the angle of attack is {greater} than
roughly $8^{\circ}$. A similar analysis was carried out for the
Joukowski aerofoil by \citet{huang}. {Mathematical and
  computational aspects of control problems involving vortex flows
  were surveyed by \citet{p08a}.}

Some other studies motivated by vortex trapping mechanisms in the
presence of aerofoils include {the work of
  \citet{zannetti:suction} who considered} the effect of leading-edge
wall suction {on stabilization of} vortex shedding in the flow
over a flat plate at incidence, and {the work of}
\citet{xia:flapping} where similar techniques {were} extended to
consider a flapping plate.  Investigations of some properties of
vortex cells, where an aerofoil contains a cavity specifically
designed to trap a vortex, were carried out by
\citet{bunyakin:batchelor} and more recently by
\citet{donelli:vortexcell}.  In the latter study, results from
point-vortex, Prandtl-Batchelor flow, and Reynolds-averaged
Navier-Stokes models were compared to ascertain the usefulness of
inviscid {flow} models in the design of vortex cells. It was seen
that the point-vortex model {produced} qualitatively similar
results {to {those} obtained using the Reynolds-averaged
  Navier-Stokes system}, whilst the Prandtl-Batchelor flow model gave
an acceptable representation of {these} solutions. However, the
authors also comment that the vortex
is expected to be unstable {in the configurations considered}.

{The trapped vortex is formed from the vorticity in the boundary
  layer separating from the leading edge of the airfoil which then
  undergoes the Kelvin-Helmholtz instability, an effect which is not
  explicitly accounted for by inviscid point-vortex and vortex-patch
  models. However, potential-flow models typically involve free
  parameters (related to circulations around contours) which make it
  possible to impose certain additional conditions reflecting viscous
  effects.  In the studies referenced above}, the location of the
separating streamline is subject to the so-called Kutta condition.
Inviscid flows over sharp edges are {characterized by unbounded
  velocity on the boundary} in the absence of separation and imposing
a Kutta condition at the desired point simultaneously forces the flow
to be regular.  \citet{gallizio10} derived precise conditions under
which vortex-patch solutions of the two-dimensional Euler equations
can be continued {with respect to parameters such that nearby
  solutions also satisfy the Kutta condition}.  These conditions
{were} illustrated by computing a continuous family of patch solutions
connecting the point vortex and Prandtl-Batchelor solution.

{Given that flow configurations with trapped vortices tend to be
  unstable, the purpose of the present study {is} to propose {and
    validate} practical stabilization strategies}. Flow configurations
will be based on those of \citet{saffman:attachedvortex} for the
single plate case and those of \citet{nelson:trapped-vortices} for the
Kasper Wing case.  {We will first provide a control-oriented
  characterization of the stability of these flow equilibria and will
  then design a Linear-Quadratic-Gaussian (LQG) compensator for their
  stabilization.}  Flow actuation will be carried out by placing a
sink-source singularity at a chosen fixed location on the main plate.
Use of the sink-source as an actuation mechanism is intended to mimic
the effect of {blowing and suction} commonly used in control
strategies in which viscosity is taken into account. Pressure
difference across the plate at a chosen location will be used as the
measurement. {By analyzing the performance of the closed-loop control
  systems we will demonstrate that in fact the Kasper wing
  configuration has a positive effect on the robustness of the
  trapped-vortex equilibria, which is a key finding of this
  investigation.}

The structure of the paper is as follows: in \S \ref{uc} {we
  state} the model equations in the absence of any control
{actuation, discuss the stability of the different equilibria and
  identify the equilibrium configurations which will be} subject to
further analysis. In \S \ref{ctr-char} we characterize the model from
a control-theoretic perspective and in \S \ref{sec:lqg} derive the LQG
compensator. {Stability properties of the closed-loop system are
  investigated computationally in \S \ref{nres}, whereas in \S
  \ref{stochastic} the performance of the LQG control is studied in
  the presence of different perturbations.}  Further points of
discussion and final conclusions are given in \S\ref{discussion}
and \S\ref{conclusions}.
 
\section{Flow models in the uncontrolled setting}\label{uc}

{We begin by stating the governing equations for} both the single
plate and Kasper Wing systems {and following this survey the stability of some 
of the equilibria these systems admit}. A
detailed derivation of these equations is presented in
\citet{nelson:trapped-vortices}, {whereas} the stability and
non-linear robustness of various configurations is examined in
\citet{nelson:robustness}. Here, attention is restricted to
configurations which {will be studied in the controlled setting.}

\subsection{Governing equations}
\label{sec:eqns}

\begin{figure}
 \centering
\mbox{
  \subfigure[]{\label{configz}\includegraphics[width=0.55\textwidth]{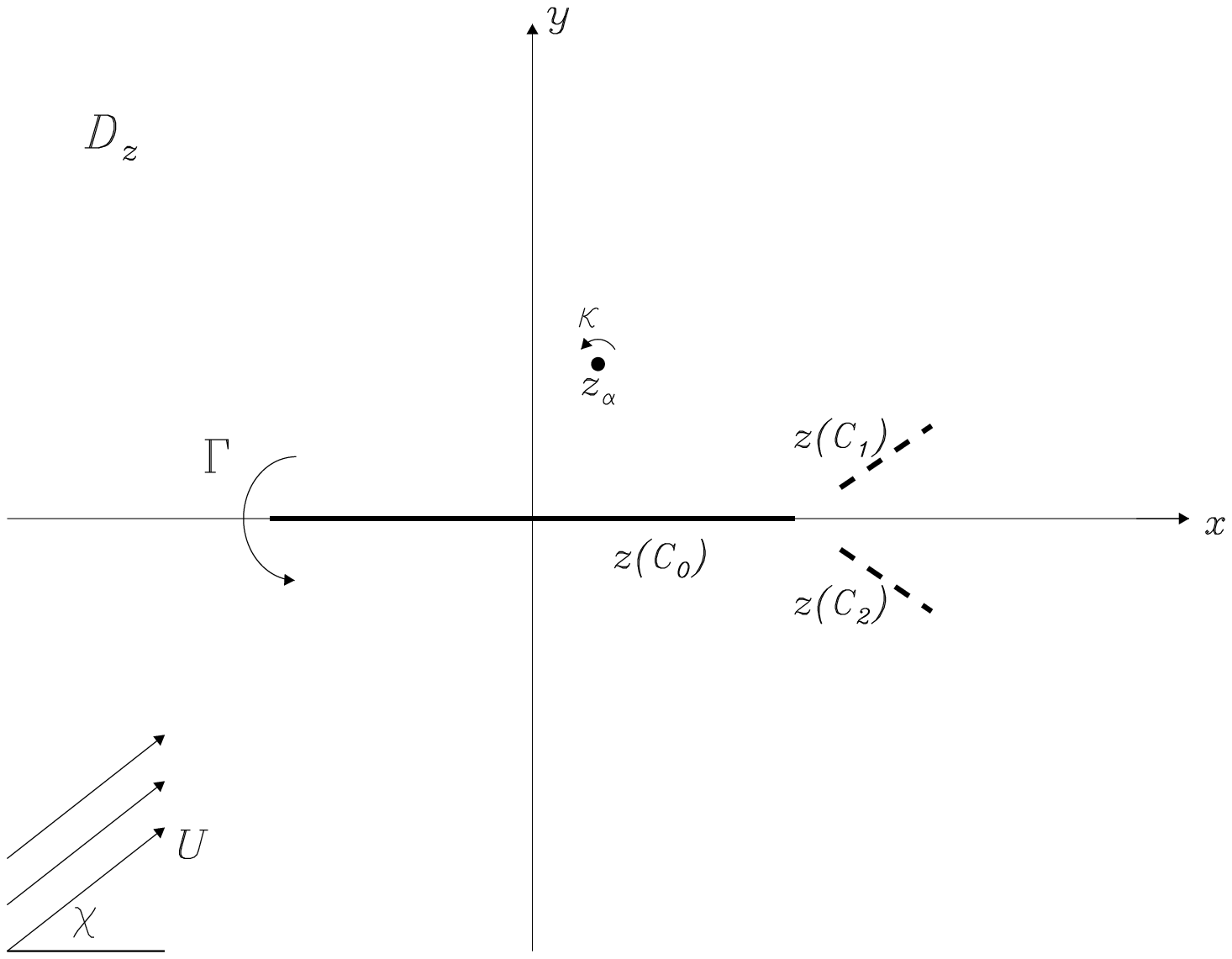}}\qquad
  \subfigure[]{\label{configzeta}\includegraphics[width=0.4\textwidth]{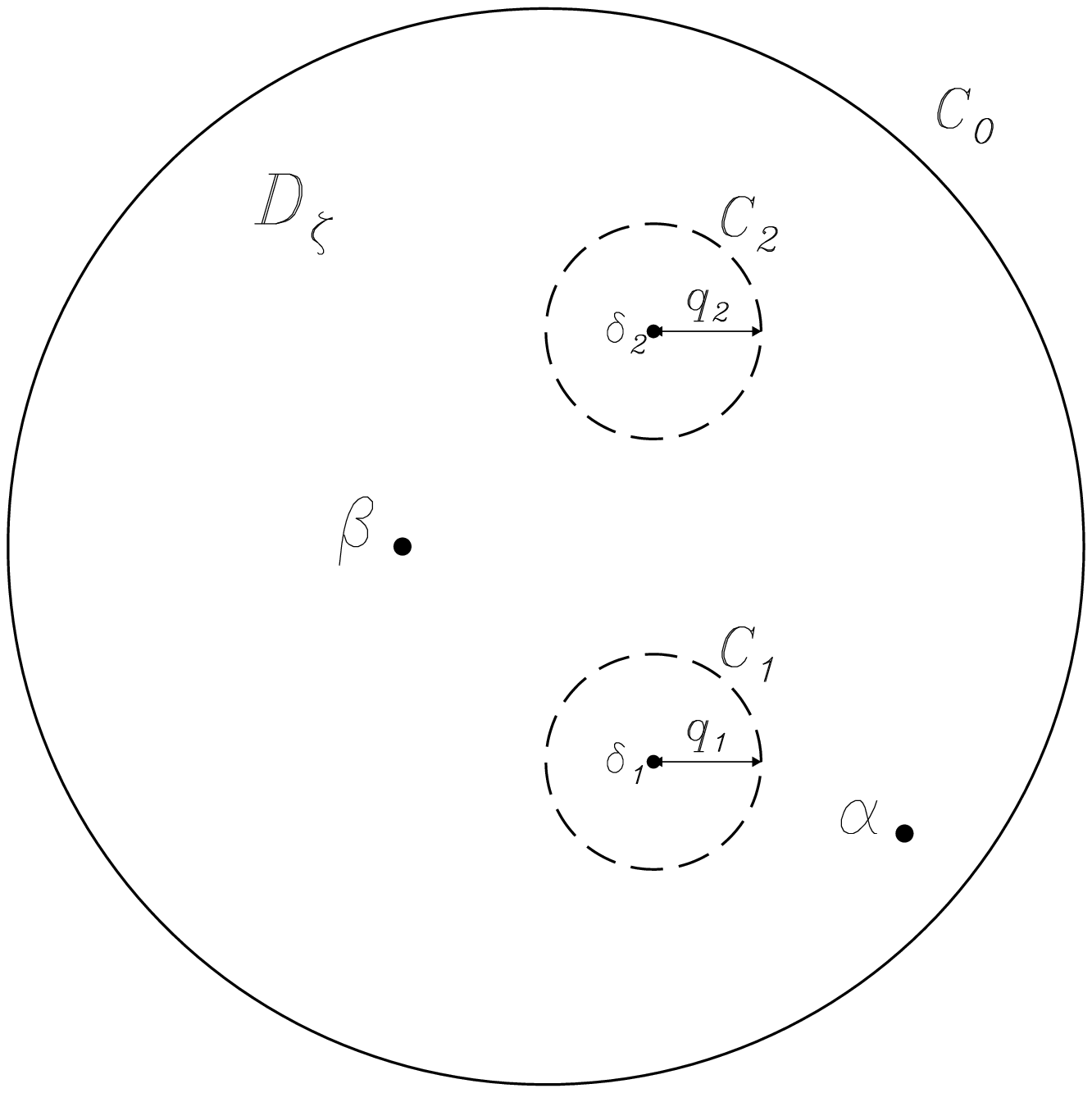}}
}
  \caption{(a) Schematic of the flow configuration. The thick dashed
    lines correspond to the auxiliary plates present in the Kasper
    Wing system (with $M=2$), but not in the single-plate system (with
    $M=0$). (b) Schematic of a typical pre-image domain.}
  \label{config}
\end{figure}

Let $\mathcal{D}_{z}$ denote the domain exterior to $M+1$ thin plates,
where $M=0$ corresponds to the single-plate case and $M=2$ {to}
the Kasper Wing case. A schematic of the configurations is shown in
figure \ref{config}(a).  To evaluate the velocity field $V(z)=u-iv$
induced at a point $z=x+iy$ {in $\mathcal{D}_{z}$}, the complex
potential of the system is first constructed in a conformally
equivalent pre-image circular domain.  This circular domain is
labelled $\mathcal{D}_{\zeta}$ and consists of the interior of the
unit disk with $M$ excised non-overlapping smaller disks. The boundary
of the unit circle will be labeled $C_{0}$ and those of the $M$
excised disks {$C_{k}$ for $k=1,\dots,M$}.  Complex coordinates in the
{transformed} domain will be {denoted} $\zeta$ (and thus $z
= z(\zeta)$).  Let the centres and radii of the $M$ excised disks be
respectively {$\delta_{k}\in\mathbb{C}$ and $q_{k}\in\mathbb{R}$} for
{$k=1,\dots,M$}. An example illustrating such a pre-image domain is shown
in figure \ref{config}(b). In what follows, overlines will be used to
denote complex conjugation and for some complex function $f(\zeta)$,
its conjugate function (Schwarz conjugate) is defined as
$\bar{f}(\zeta) := \overline{f(\bar{\zeta})}$ {(where ``$:=$''
  means that the expression on the left-hand side is defined by the
  one on the right-hand side)}.

{Let $\alpha, \beta \in \mathcal{D}_{\zeta}$, respectively, denote the location of
a point vortex and the point that is mapped to infinity in $\mathcal{D}_z$, i.e., $z(\beta)=\infty$.}
The flow under consideration consists of the following components (for
each of which the complex potential at a point $\zeta$ is stated):
\begin{enumerate}
     \item a potential flow which tends to a uniform {flow}
       inclined at angle ${\chi_0}$ as $|z|\rightarrow\infty$
       \begin{equation}\label{fc2}
        W_{U}(\zeta,\beta)=Ua\left[e^{i{\chi_0}}\frac{\partial}{\partial\bar{\beta}}-e^{-i{\chi_0}}\frac{\partial}{\partial\beta}\right]\log\left(\frac{\omega(\zeta,\beta)}{|\beta|\omega(\zeta,\bar{\beta}^{-1})}\right),
       \end{equation}
       where $a$ is a scaling constant such that
       $|\mathrm{d}W/\mathrm{d}z|\rightarrow U$ as
       $|z|\rightarrow\infty$ \citep[see][]{nelson:trapped-vortices};
     \item {flow due to a} single point vortex
       \begin{equation}\label{fc1}
         W_{V}(\zeta;\alpha)=-\frac{i\kappa}{2\pi}\log\left(\frac{\omega(\zeta,\alpha)}{|\alpha|\omega(\zeta,\bar{\alpha}^{-1})}\right),
       \end{equation}
       where $\kappa$ denotes the {circulation} of the point vortex;
     \item {flow due to a} point vortex with {circulation
       $-\kappa$ at the point $\beta$ (to remove the circulation around the disk $C_0$ owing to the point vortex at $\alpha$)}:
       \begin{equation}\label{fc4}
        W_{\infty}(\zeta,\beta)=\frac{i\kappa}{2\pi}\log\left(\frac{\omega(\zeta,\beta)}{|\beta|\omega(\zeta,\bar{\beta}^{-1})}\right);
       \end{equation}
     \item {flows due to prescribed circulations around each of
         the {disks}}
       \begin{equation}\label{fc3}
        W_{\Gamma}(\zeta,\beta)=\sum_{k=0}^{M}\frac{i\Gamma_{k}}{2\pi}\log\left(\frac{\omega(\zeta,\beta)}{\omega(\zeta,\bar{\theta}_{k}(\beta^{-1}))}\right),
       \end{equation}
       where $\Gamma_{k}$, $k=0,\dots,M$, is the desired circulation around the {disk $C_k$}.

\end{enumerate}
In {relations} \eqref{fc1}--\eqref{fc3}, $\omega(\zeta,\cdot)$ is
a special function known as the \textit{Schottky-Klein prime function}
associated with $\mathcal{D}_{\zeta}$ and $\{\theta_{k}|k=0,\dots,M\}$
are the related M\"{o}bius maps.  Both these functions will be
{described} below in section \ref{sec:SK}. The total complex potential in the 
\textit{pre-image} domain is then the sum of these components and is
{thus} given by
\begin{equation}
{ W(\zeta)=W_{\mathrm{V}}(\zeta;\alpha)+W_{U}(\zeta,\beta)+W_{\Gamma}(\zeta,\beta)+W_{\infty}(\zeta,\beta).}
\end{equation}

The velocity field in the \textit{physical} domain is retrieved via
\citep[see, e.g.,][]{saffman-vdynamics}
\begin{equation}\label{vz}
 V(z)= \left\{
  \begin{array}{ll}
   (z_{\zeta})^{-1} W_{\zeta}, & z\neq z(\alpha) \\
   (z_{\zeta})^{-1}\left[\widetilde{W}_{\zeta}
                                +(i\kappa/4\pi)z_{\zeta\zeta}/z_{\zeta}\right], & z=z(\alpha) \end{array} \right.,
\end{equation}
where {the subscript $\zeta$} represents the derivative with
respect to $\zeta$ and $\widetilde{W}_{\zeta}$ {is the velocity
  at the location of the point vortex in the pre-image domain with the
  self-induction term removed} {\citep[see, e.g.,][]{saffman-vdynamics}}. To compute (\ref{vz}) it is required
to know the form of the conformal mapping {$z = z(\zeta)$} from
$\mathcal{D}_{\zeta}$ to $\mathcal{D}_{z}$. For the domains under
consideration, when $M=0$,
\begin{equation}\label{cmap1}
 z(\zeta)=\frac{1}{2}\left(\zeta+\frac{1}{\zeta}\right)
\end{equation}
and when $M=2$,
\begin{equation}\label{cmap2}
 z(\zeta)=1+2S\frac{\omega(\zeta,\gamma)\omega(\zeta,\bar{\gamma}^{-1})}{\omega(\zeta,\beta)\omega(\zeta,\bar{\beta}^{-1})},
\end{equation}
where, owing to a degree of freedom in the Riemann mapping theorem,
$\beta$ can be chosen to lie anywhere in $\mathcal{D}_{\zeta}$, but it
will be required to compute {the values of $S$ and $\gamma$
  corresponding to the desired} configuration in $\mathcal{D}_{z}$.

The map given in \eqref{cmap1} is the celebrated \textit{Joukowski
map} which maps the unit disk ($|\zeta|=1$) to a thin slit located on the real axis ($\Im(z)=0$)
between $-1\leq\Re(z)\leq1$ and the interior of the disk to the region
exterior to this slit. Note that, for this map, the point mapped to
infinity in $\mathcal{D}_{z}$ corresponds to $\beta=0$.  The form of
the map appearing in \eqref{cmap2} is known as a \textit{radial slit
  map} (a subset of the more general group of conformal slit maps).
Such maps have a wide range of applications in applied mathematics
\citep{crowdy:conformal_slit_maps}.  Here, the map \eqref{cmap2} will
again map the unit disk to a thin slit located on the real axis between $-1\leq z\leq1$.
Additionally, the two excised disks will be mapped to sections of
``radial rays'' emanating from the point $z=1$. The region interior to
the unit disk and exterior to the two excised disks will be mapped to
the region exterior to the three radial thin plates. It should be
noted that the appearance of the Schottky-Klein prime function in
(\ref{cmap2}) is a consequence of the type of conformal map required
here and is not directly related to its appearance in
(\ref{fc1})--(\ref{fc3}). Further details regarding the relation
between these maps and the configurations desired will be given
shortly.

With the velocity field throughout $\mathcal{D}_{z}$ determined,
{the instantaneous position of the free point vortex, which we
  will denote $\mathbf{X}(t) := [x \ y]^{T}=[\Re(z) \ \Im(z)]^{T}$, is
  governed by} the non-linear dynamical system 
\begin{equation}\label{nlds}
 \frac{\mathrm{d}}{\mathrm{d}t}\mathbf{X}=\mathbf{F}(\mathbf{X})= \left[
  \begin{array}{ll} \phantom{-}\Re[V(z)] \\ -\Im[V(z)] \end{array} \right].
\end{equation}
First, stationary solutions are sought such that, in addition to the
point vortex being in equilibrium, the Kutta conditions prescribing
the flow separation at the trailing edge of each plate are satisfied.
This choice is motivated by the original study {of}
\citet{saffman:attachedvortex} {{on} the single-plate configuration}
and more discussion regarding this choice can be found in
\citet{nelson:trapped-vortices,nelson:robustness}. Imposing Kutta
conditions at each trailing edge means that {altogether} $M+3$
equations must be satisfied, i.e., the Kutta conditions at $(M+1)$
plates and two conditions for the equilibrium coordinates
{$\Re(z_{\alpha})$ and $\Im(z_{\alpha})$} of the point vortex.  The
unknowns in the problem thus are
{$\{\Re(z_{\alpha}),\Im(z_{\alpha}),\kappa,\Gamma_{0},\dots,\Gamma_{M}\}$}
giving $M+4$ real numbers. Solutions are therefore expected to trace
out one-parameter continuous loci. Labeling the position of the
trailing edge {of each plate} $z_{k}$ for {$k=0,\dots,M$}, stationary
solutions with a point vortex at $z_{\alpha}$ are characterized by
\begin{subequations}
\label{stabeq1}
\begin{alignat}{1}
   &\Re[V(z_{\alpha};{\chi_0},\kappa,\Gamma_{0},\dots,\Gamma_{M})]=0, \label{stabeq1a} \\
   &\Im[V(z_{\alpha};{\chi_0},\kappa,\Gamma_{0},\dots,\Gamma_{M})]=0, \label{stabeq1b} \\
   &|V(z_{k};{\chi_0},\kappa,\Gamma_{0},\dots,\Gamma_{M})|=0, \quad {k=0,\dots,M}, \label{stabeq1c}
\end{alignat}
\end{subequations}
where {the angle} ${\chi_0}$ can be chosen freely and the
unknowns {$z_{\alpha},\kappa,\Gamma_{0},..,\Gamma_{M}$} are solved
for. For the single-plate case ($M=0$), equations (\ref{stabeq1}) can
be solved exactly.  However, closed-form solutions are not possible
for the Kasper Wing system ($M=2$) and a numerical {approach} is
required. As in \citet{nelson:trapped-vortices,nelson:robustness}, a
Brownian ratchets scheme is used to solve equations \eqref{stabeq1} to
within {a prescribed} numerical tolerance. Further details
regarding such Brownian ratchets schemes can be found in
\citet{newton:bratchets}.

\subsection{The Schottky-Klein prime function}
\label{sec:SK}

In this section the aforementioned Schottky-Klein prime function and
the associated M\"{o}bius maps are briefly introduced. For further
details regarding this function and its properties {we refer the
  reader to,} e.g., \citet{crowdy:mconn}. One manner in which this
special function can be evaluated is via a classic product formula
\citep{baker}. Within $\mathcal{D}_{\zeta}$, for each of the $M$
excised circles {$C_{k}$, $k=1,\dots,M$}, we construct the
M{\"o}bius map
\begin{equation}\label{mobius}
{\theta_{k}(\zeta)=\delta_{k}+\frac{q_{k}^{2}\zeta}{1-\bar{\delta}_{k}\zeta}.}
\end{equation}
The Schottky-Klein prime function is then given by
\begin{equation}\label{skprod}
 \omega(\zeta,\gamma)=(\zeta-\gamma)\prod_{\vartheta\in\Theta''}\frac{[\vartheta(\zeta)-\gamma][\vartheta(\gamma)-\zeta]}{[\vartheta(\zeta)-\zeta][\vartheta(\gamma)-\gamma]},
\end{equation}
where the product is taken over all mappings $\vartheta$ belonging to
a special subset $\Theta''$ of the full Schottky group $\Theta$. 
(The full, or classical, Schottky group $\Theta$ is defined to be the infinite 
free group of mappings generated by compositions of the $M$ basic M\"{o}bius maps 
$\{\theta_{k}|k=1,\dots,M\}$ and their inverses $\{\theta^{-1}_{k}|k=1,\dots,M\}$
and including the identity map).
This subset contains all mappings, \textit{excluding} the identity
\textit{and} all inverse mappings, thus, for example, if the map
$\theta_{1}\theta_{2}^{-1}$ is included in the set,
$\theta_{2}\theta_{1}^{-1}$ must be excluded. Two important properties
of the Schottky-Klein prime function are
\begin{enumerate}
\item $\omega(\zeta,\gamma)$ has a simple zero at $\zeta=\gamma$,
\item $\omega(\cdot,\cdot)$ is such that the complex potential has
  constant imaginary part on all circles {$C_k$, $k=0,1,\dots,M$}.
\end{enumerate}
Formula (\ref{skprod}) is useful for concisely stating
$\omega(\cdot,\cdot)$ and can be applied in some practical situations.
In general, however, {expression} \eqref{skprod} fails to
converge or requires an impractically long time to compute (for
example, when two circles are close together or when the connectivity
is high). A more robust method is to compute $\omega(\cdot,\cdot)$ via
a Fourier-Laurent expansion, details of which are presented in
\citet{crowdy:skcomp}. Briefly, the algorithm works by introducing the
function $X(\cdot,\cdot)$ {defined as}
\begin{equation}
 X(\zeta,\gamma)=(\zeta-\gamma)^{2}\hat{X}(\zeta,\gamma),
\end{equation}
in which
\begin{equation}\label{f-l}
 \hat{X}(\zeta,\gamma)=A\left[1+\sum_{k=1}^{M}\sum_{m=1}^{\infty}\left(\frac{c_{m}^{(k)}q_{k}^{m}}{(\zeta-\delta_{k})^m}+
                               \frac{d_{m}^{(k)}Q_{k}^{m}}{(\zeta-\delta'_{k})^m}\right)\right],
\end{equation}
where $\delta'_{k}$ is the centre of the circle $C'_{k}$ obtained from
reflecting $C_{k}$ about $|\zeta|=1$, $Q_{k}$ is related to the radius
of $C'_{k}$, the constants \{$c_{m}^{(k)},d_{m}^{(k)}\; | \;
k=1,\dots,M, \; m=1,2,\dots$\} are determined numerically (for some
truncated $m$) and $A$ is a normalisation coefficient chosen such that
\begin{equation}
 \lim_{\zeta\rightarrow\gamma}\frac{X(\zeta,\gamma)}{(\zeta-\gamma)^{2}}=1.
\end{equation}
The Schottky-Klein prime function is then retrieved via
\begin{equation}
 X^{2}(\zeta,\gamma)=\omega(\zeta,\gamma),
\end{equation}
where the branch of the square root is taken such that
$\omega(\zeta,\gamma)$ behaves like $(\zeta-\gamma)$ as
$\zeta\rightarrow\gamma$.

\subsection{{Linear stability analysis of point-vortex equilibria}}
\label{sec:stab}

Linear stability analysis of the single plate and Kasper Wing systems
is {carried out} by adding a small perturbation $z'=x'+iy'$ to
the point-vortex equilibrium {$z_{\alpha}$ and performing
  linearization}. The evolution of this perturbation is governed by
the system
\begin{equation}
 \frac{\mathrm{d}}{\mathrm{d}t}\mathbf{X}'=\mathbf{A}\mathbf{X}',
\label{eq:X'}
\end{equation}
where $\mathbf{X}'(t)=[x'(t) \ y'(t)]^{T}$ and $\mathbf{A}$ is given by \citep[see][]{nelson:robustness}
\begin{equation}
 \mathbf{A}=\left[\begin{array}{cc} a & \phantom{-}b \\ c & -a \end{array} \right],
\end{equation}
with
\begin{subequations}
\begin{alignat}{1}
  &a=\Re\left[\left.\frac{\partial V}{\partial z}\right|_{\bar{z}_{\alpha}}\right], \\
  &b=-\Im\left[\left.\frac{\partial V}{\partial z}\right|_{\bar{z}_{\alpha}}\right]+
    \Im\left[\left.\frac{\partial V}{\partial \bar{z}}\right|_{z_{\alpha}}\right], \\
  &c=-\Im\left[\left.\frac{\partial V}{\partial z}\right|_{\bar{z}_{\alpha}}\right]-
    \Im\left[\left.\frac{\partial V}{\partial \bar{z}}\right|_{z_{\alpha}}\right].
\end{alignat}
\end{subequations}
Eigenvalue analysis of the matrix $\mathbf{A}$ reveals that
the system is
\begin{enumerate}
\item unstable when $a^{2}+bc>0$ ({corresponding to purely} real
  eigenvalues),
\item neutrally stable when $a^{2}+bc<0$ ({corresponding to}
  purely imaginary eigenvalues).
\end{enumerate}
{We emphasize that when the linearization has purely imaginary
  eigenvalues and there are no eigenvalues with positive real parts,
  as is the case here, then such linearization is inconclusive as
  regards the stability properties of the equilibrium and one must
  account for the effect of nonlinearity through suitable invariant
  manifold reductions \citep{p07d}.}  We {also} note that, since
system \eqref{nlds} is Hamiltonian, the two eigenvalues may be either
purely real and of opposite signs, or form a conjugate imaginary pair
\citep{n01}.  It should also be noted that when the system is
perturbed {away from the} equilibrium, the Kutta conditions
{\eqref{stabeq1c}} are no longer satisfied. For these conditions
to be satisfied throughout any time-dependent evolution, a method for
shedding vorticity from the rear tip of each plate is required. This
topic will be revisited in \S \ref{stochastic}.

\subsection{Configurations and analysis}\label{configs}

{We now introduce the configurations which will be subject to
  further study and analyze their properties in the uncontrolled
  setting.}  In all configurations, the main plate (which $C_{0}$ is
mapped to) will lie {in the region defined by $-1 \le x \le 1$
  and $y=0$} and, when $M=2$, the centres of the auxiliary plates
(which $C_{1}$ and $C_{2}$ are mapped to) will lie at $z=1+0.4e^{\pm
  i\phi}$. Three values of {the angle $\phi$} will be used,
{{these being} $\pi/12$, $\pi/6$ and $5\pi/12$,} and the length of the
auxiliary plates will be set to $0.1$. This means that $C_{1}$ and
$C_{2}$ are mapped to straight segments with endpoints, respectively,
at $z=1+(0.4\pm0.05)e^{i\phi}$ and $z=1+(0.4\pm0.05)e^{-i\phi}$.  The
angle of attack {of the oncoming flow} is set to
${\chi_0}=0.1$.

Owing to the configurations chosen in $\mathcal{D}_{z}$, when $M=2$,
$\gamma$ must be set unity. The constant $\beta$ can be chosen freely
and will be set to $-0.4$ in all results that follow.  {In the
  construction of} $\mathcal{D}_{\zeta}$ and the conformal map, the
remaining unknowns are $S,\delta_{1},q_{1},\lambda_{1}$ and
$\lambda_{2}$, where $\lambda_{1}$ and $\lambda_{2}$ are the angles
{(with respect to $\delta_1$)} of the points on $C_{1}$ that map
to the auxiliary plate edges in $\mathcal{D}_{z}$, which leaves six
real unknowns. We note that due to the symmetry of the configurations
considered here, $\delta_{2}=\overline{\delta_{1}}$ and $q_{2}=q_{1}$
(see \citet{nelson:robustness} for details regarding non-symmetric
Kasper Wing configurations).  {The six unknowns can be found from
  the} six real{-valued} equations obtained from
\begin{subequations}
\label{eq:cmp}
\begin{alignat}{1}
 &z(-1)=-1, \\
 &|z(\delta_{1}+q_{1}e^{i\lambda_{1}})-1|=0.35, \\
 &|z(\delta_{1}+q_{1}e^{i\lambda_{2}})-1|=0.45, \\
 &\mathrm{Arg}[z(\delta_{1}+q_{1}e^{i\lambda_{2}})]=\phi, \\
 &\left|\frac{\mathrm{d}z}{\mathrm{d}\zeta}(\delta_{1}+q_{1}e^{i\lambda_{1}})\right|=0, \\
 &\left|\frac{\mathrm{d}z}{\mathrm{d}\zeta}(\delta_{1}+q_{1}e^{i\lambda_{2}})\right|=0.
\end{alignat}
\end{subequations}
These equations are solved via Newton's iteration and solutions for
three different values of $\phi$ are given in {table} \ref{tab:configs}.
The scaling constant $a$ (appearing in (\ref{fc2})) must be set to
\begin{equation}
 a=1/2
\end{equation}
when $M=0$, and
\begin{equation}
 a=2S\frac{\omega(\beta,\gamma)\omega(\beta,\bar{\gamma}^{-1})}
{\omega(\beta,\bar{\beta}^{-1})},
\end{equation}
when $M=2$. Loci of solutions will be parametrised in terms of their
vertical coordinate in the physical domain. Thus, moving along any
given locus will correspond to an increasing $\Im(z_{\alpha})$.
Additionally, {since they are more relevant from the physical
  point of view}, only loci stemming from the rear tip of the main
plate will be considered \citep[see][]{nelson:trapped-vortices}.

\begin{table}
\begin{center}
  \caption{Solutions of equations \eqref{eq:cmp} for three different values of $\phi$.}
    \vspace{1pc}
\begin{tabular}{ l | c | c | c | c | c } 
     \Bmp{1.0cm} \centering $\phi$ \Emp
   & \Bmp{1.0cm} \centering $S$ \Emp
   & \Bmp{1.0cm} \centering $\delta_{1}$ \Emp 
   & \Bmp{1.0cm} \centering $q_{1}$  \Emp 
   & \Bmp{1.0cm} \centering $\lambda_{1}$ \Emp
   & \Bmp{1.0cm} \centering $\lambda_{2}$ \Emp \\ \cline{1-6}
   $\pi/12$  & $0.2243$ &  $0.0243-0.0562i$ & $0.0130$ & $6.2484$ & $3.0932$ \rule[-5pt]{0pt}{25pt} \\ 
   $\pi/6$  & $0.2242$ &  $0.0209-0.1109i$ & $0.0135$ & $6.2105$ & $3.0530$ \rule[-5pt]{0pt}{25pt} \\ 
   $5\pi/12$  & $0.2241$ &  $-0.0038-0.2900i$ & $0.0159$ & $6.1191$ & $2.9396$ \rule[-5pt]{0pt}{25pt}
\label{tab:configs}
\end{tabular}
\end{center}
\end{table} 

\begin{figure}
 \centering
\mbox{
  \subfigure[]{\label{lssaff}\includegraphics[width=0.475\textwidth]{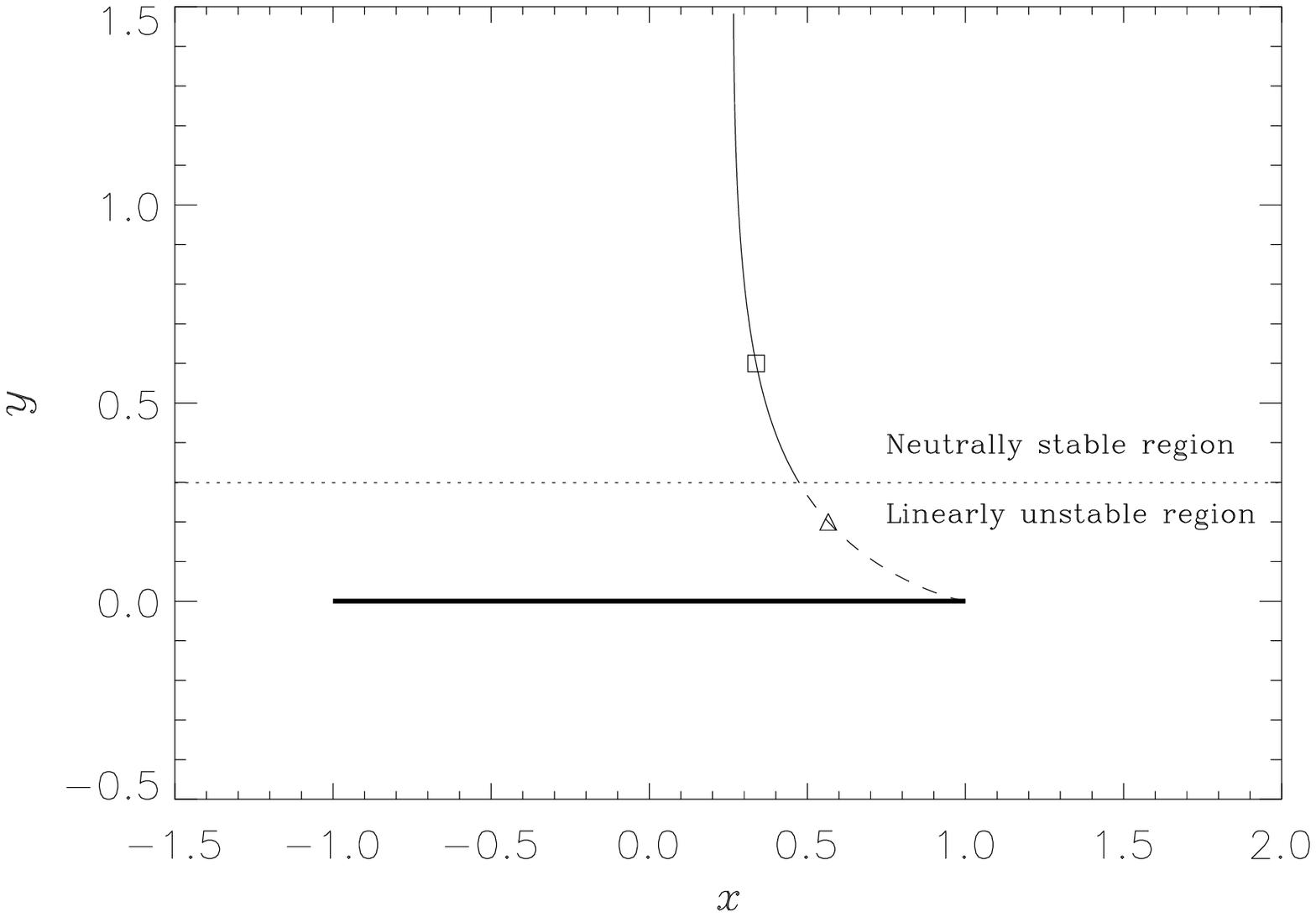}}\qquad
  \subfigure[]{\label{lskw0}\includegraphics[width=0.475\textwidth]{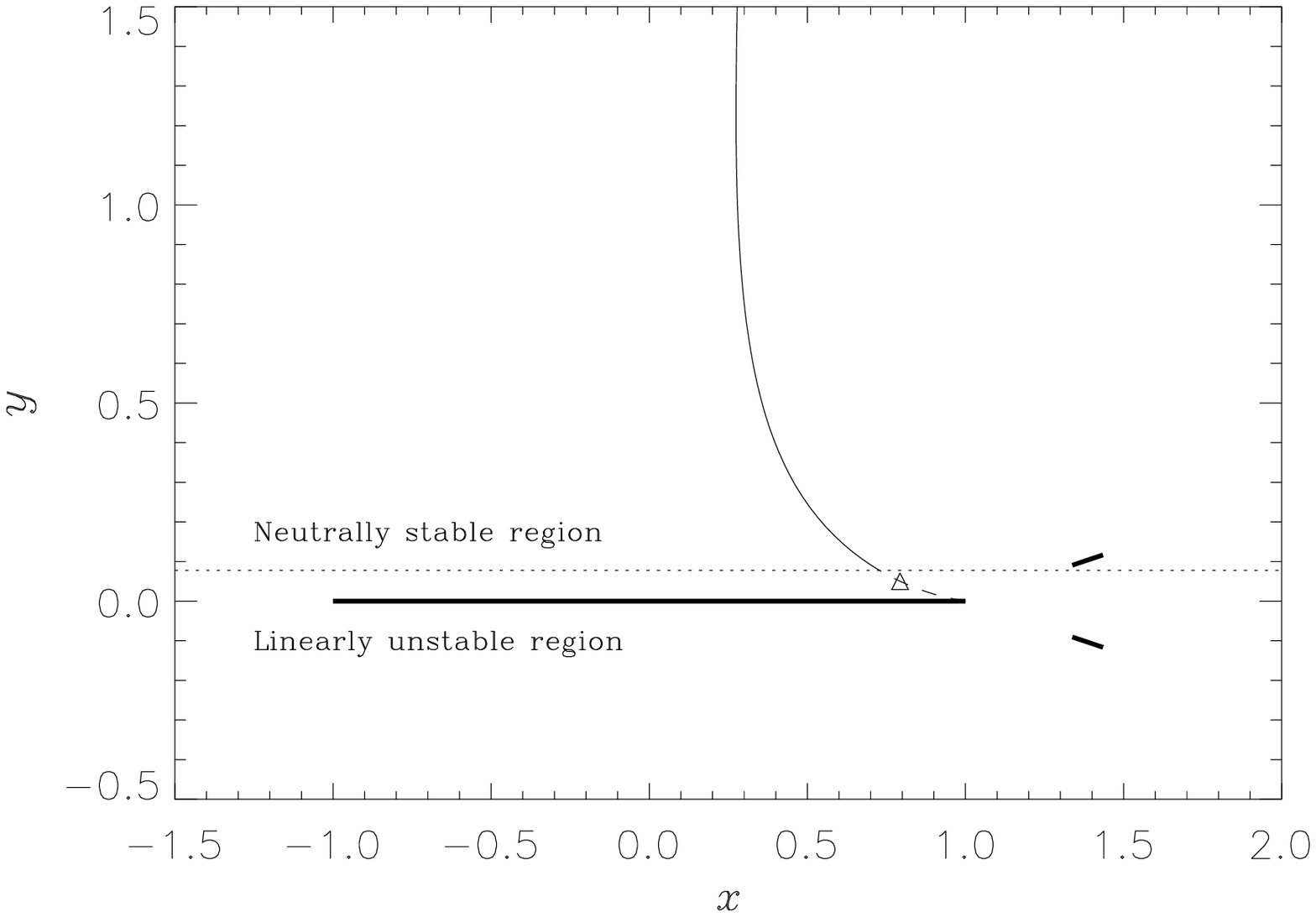}}}
\mbox{
  \subfigure[]{\label{lskw1}\includegraphics[width=0.475\textwidth]{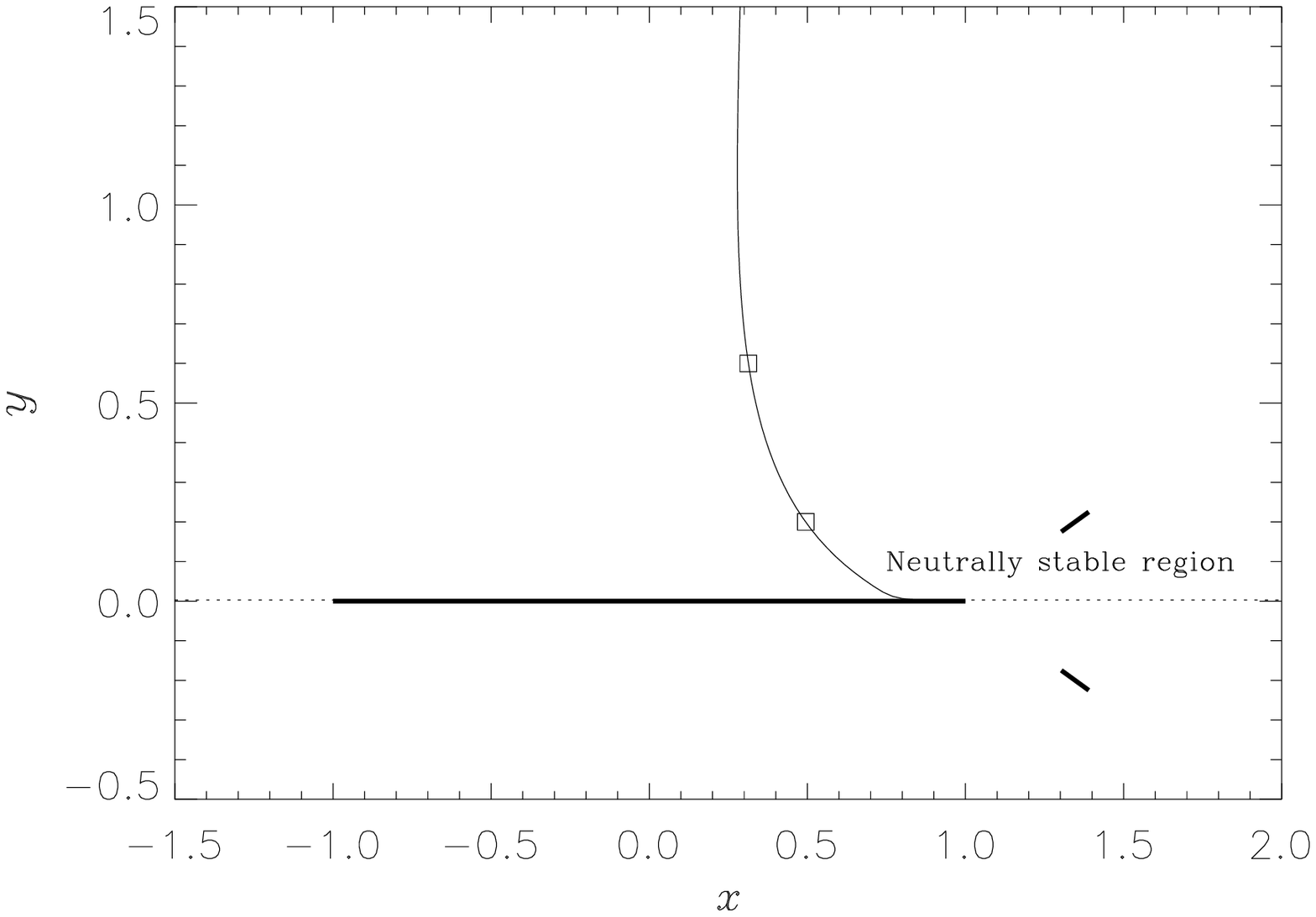}}\qquad
  \subfigure[]{\label{lskw2}\includegraphics[width=0.475\textwidth]{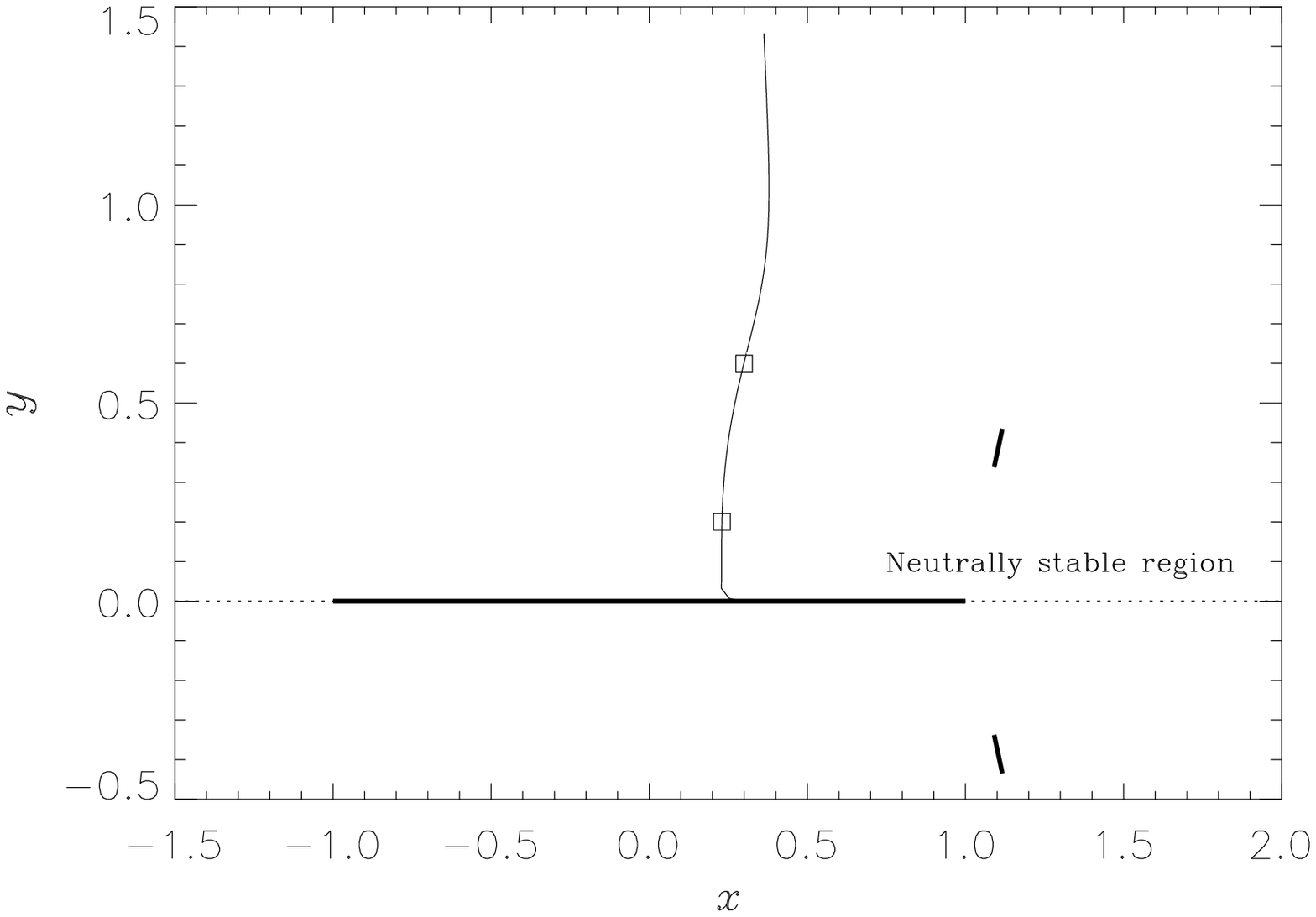}}
}
  \caption{Rear tip equilibria and their linear stability for (a)
    $M=0$ (b) $M=2$, $\phi=\pi/12$, (c) $M=2$, $\phi=\pi/6$ and (d)
    $M=2$, $\phi=5\pi/12$. Thick solid lines represent plates,
    {whereas} thin solid and dashed curves represent, {respectively,
      the neutrally stable and unstable parts of the loci}. The dotted
    horizontal lines {separate the sections of the equilibrium
      loci} characterized by linear instability and neutral stability.
    Square and triangle symbols represent, respectively, the locations
    of {neutrally} stable and unstable equilibria that will be subject
    to further analysis (information regarding these these equilibria
    is summarized below in {table} \ref{tab:cases}).}
  \label{linstab}
\end{figure}

Figures \ref{linstab}(a)--\ref{linstab}(d) show {the equilibrium loci
  emanating from the rear tip of the main plate of the configurations
  just introduced and we also identify parts of the loci characterized
  by different stability properties.}  For the configuration {with
  $M=0$}, it is seen that a linearly unstable region exists close to
the plate (for ${0 <}\Im(z_{\alpha})\lessapprox 0.3$). Small
perturbations to equilibria with $z_{\alpha}$ {further away from the
  main plate} exhibit neutral stability.  When $M=2$ {and}
$\phi=\pi/12$, the shape of the locus and stability properties are
relatively similar to the case {with $M=0$}, but the unstable region
has shrunk substantially so that now {only} equilibria lying in the
region {with} $\Im(z_{\alpha})\lessapprox 0.075$ exhibit linear
instability. Again, above this unstable region small perturbations
exhibit neutral stability. For the configuration {with $M=2$ and
  $\phi=\pi/6$, the unstable} region shrinks even further so that
linear instability is only witnessed when $\Im(z_{\alpha})\lessapprox
0.003$.  However, when $M=2$ {and} $\phi=5\pi/12$, the shape of the
locus and the corresponding stability properties of equilibria change
quite dramatically.  Equilibria stemming from the rear tip of the main
plate remain extremely close to the plate prior to a sharp rise away
from it when $\Re(z_{\alpha})$ is roughly $0.2$ {(given the finite
  graphical resolution, this latter region cannot be discerned in
  figure \ref{linstab}(d))}.  The neutrally stable region now also
extends down to the plate and unstable equilibria are {present only in
  a small neighbourhood of the rear tip.}

\begin{figure}
\begin{center}
\includegraphics[width=0.7\textwidth]{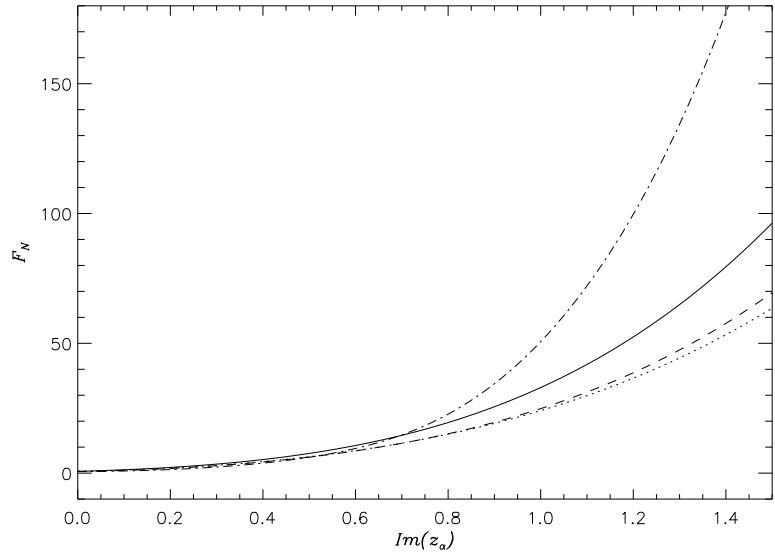}
\caption{Lift $F_{N}$ against $\Im(z_{\alpha})$. The solid curve
  represents the single plate configuration ({with} $M=0$), the
  dotted {curve} the configuration {with $M=2$ and $\phi=\pi/12$},
  the dashed {curve} the configuration {with $M=2$ and $\phi=\pi/6$}
  and the dot-dash {curve} the configuration {with $M=2$ and
    $\phi=5\pi/12$}.}
\label{fig:lift}
\end{center}

\end{figure}

For each configuration, the lift force experienced by the main plate
for increasing $\Im(z_{\alpha})$ is shown in figure \ref{fig:lift}.
These forces are computed {using the {Blasius} formula}
\citep{crowdy:cylinders}
\begin{equation}\label{liftf}
{F_{x}-iF_{y}=\frac{i}{2}\oint_{C_0}\left(\frac{\mathrm{d}W}{\mathrm{d}z}\right)^{2}\mathrm{d}z,}
\end{equation}
{where $F_x$ and $F_y$ are the force components acting in the
  directions $x$ and $y$} with integration carried out
{numerically} using a trapezoidal rule.  The lift, $F_{N}$, is
then retrieved through taking the component of (\ref{liftf}) in the
direction normal to the oncoming flow. For small values of
$\Im(z_{\alpha})$ the lift experienced by the main plate is similar
for all {flow} configurations.  As $\Im(z_{\alpha})$ increases,
the lift in the configuration {with $M=2$ and {$\phi=\pi/12$}} grows
slowest, {whereas for vortex equilibria furthest away from the
  plate} the configuration {with $M=2$ and $\phi=5\pi/12$}
experiences the highest lifts. 
{In regard to the data shown in figure \ref{fig:lift}, we can
  conclude that in all cases with auxiliary plates the lift $F_N$
  increases continuously with the inclination angle $\phi$ from values
  lower than in the single-plate configuration ($M=0$) to higher
  values. In particular, when the inclination angle is close to $\phi = 3\pi/10$,
  the lift is comparable to its value in the single-plate
  configuration.  Although the configurations with small
  inclination angles $\phi$ have smaller lift than the single-plate
  configuration, they can still be useful since, as shown in figure
  \ref{linstab}, they are characterized by more favourable stability
  properties. In addition, differences in lift become more evident only for
  vortex equilibria located further away from the main plate, i.e.,
  with larger $\Im(z_{\alpha})$, which are less relevant from the
  application point of view.}

\begin{figure}
 \centering
\mbox{
  \subfigure[]{\label{trajuc1}\includegraphics[width=0.475\textwidth]{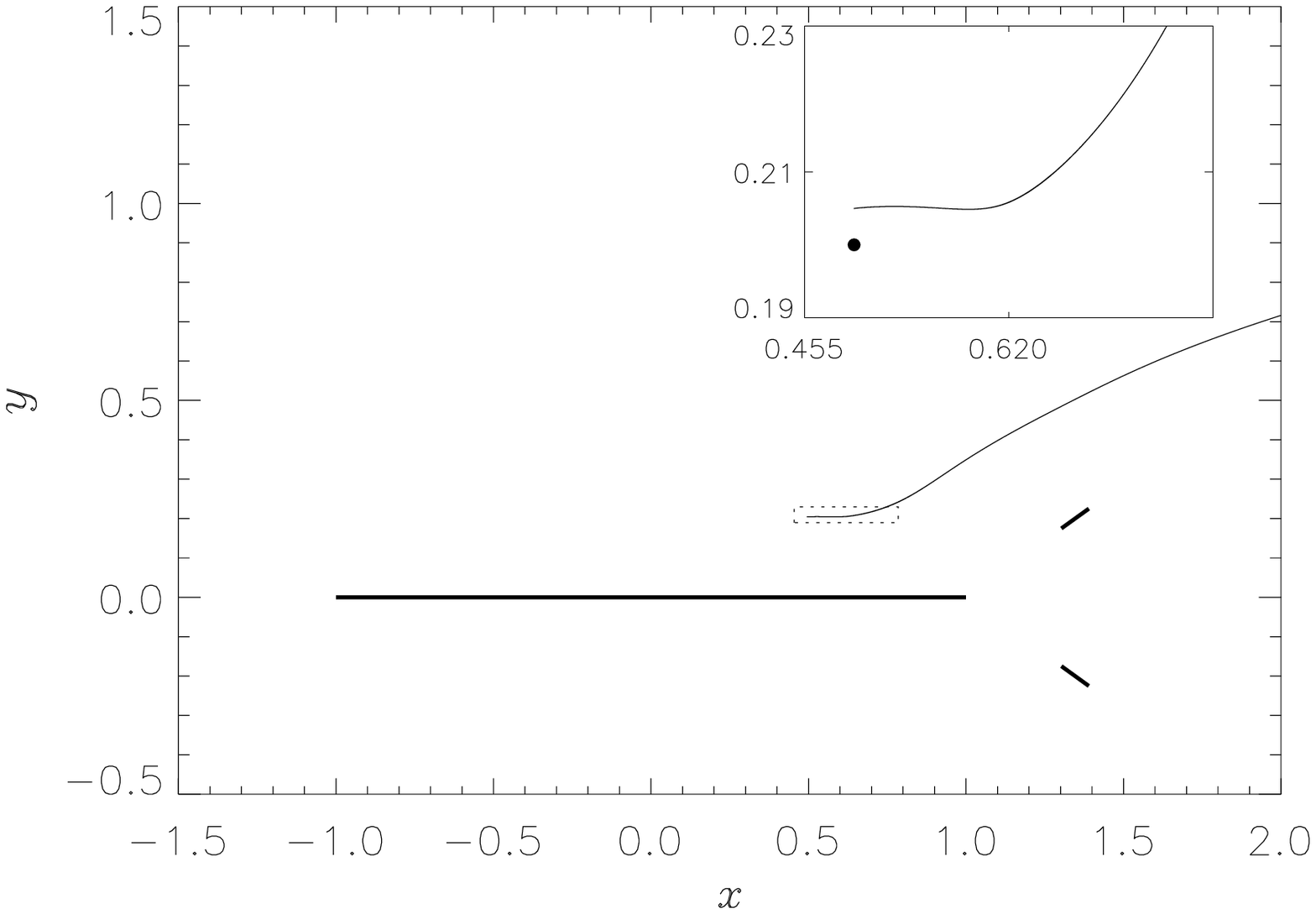}}\qquad
  \subfigure[]{\label{trajuc2}\includegraphics[width=0.475\textwidth]{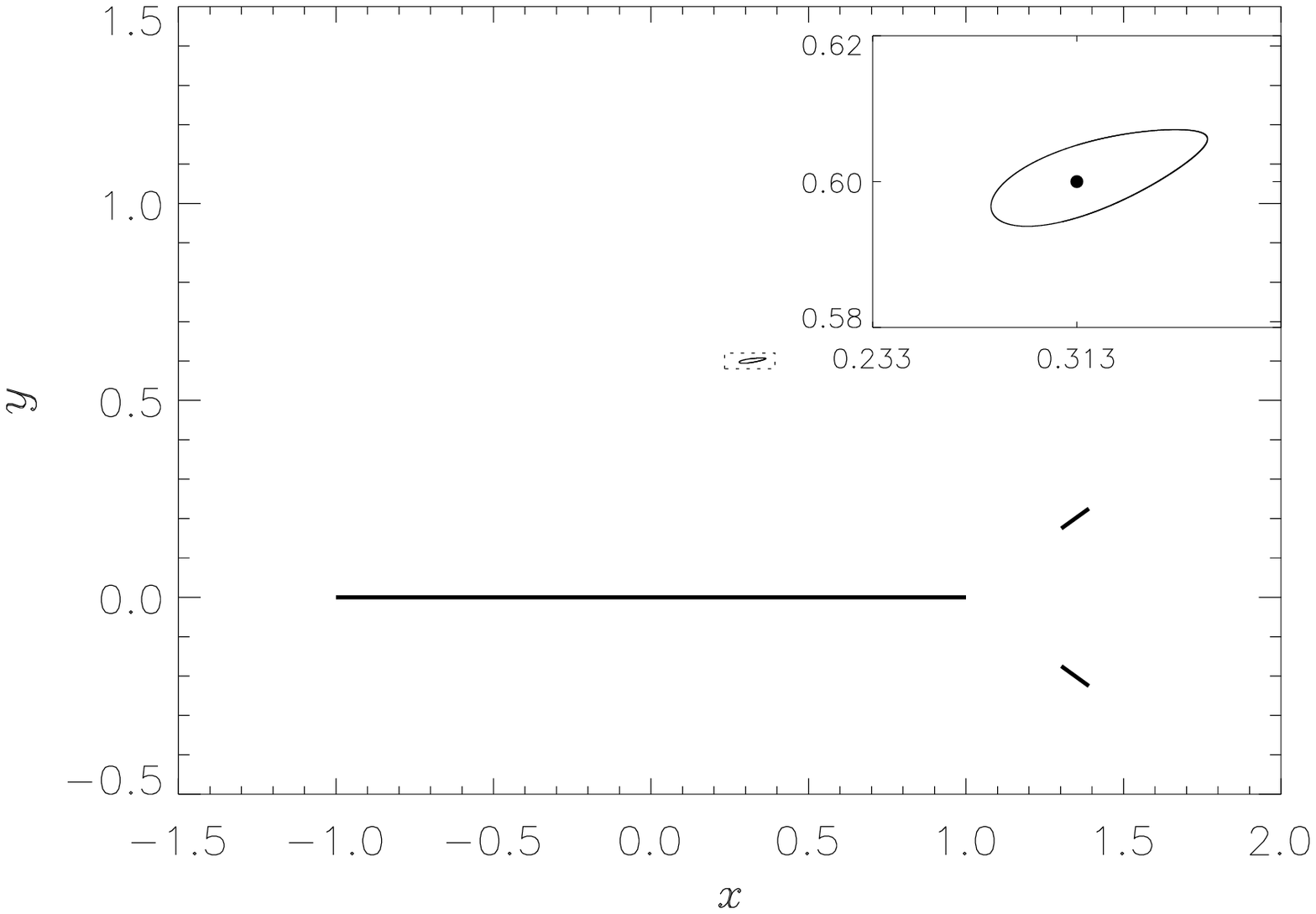}}
}
  \caption{Trajectories (represented by the solid {curves}) of the
    {solutions of} system \eqref{nlds} in (a) Case \#3 and (b)
    Case \#4 (see {table} \ref{tab:cases}) in which the equilibria
    $z_{\alpha}$ are perturbed to $z_{\alpha}+\delta$, where
    $\delta=0.005i$, corresponding to the time interval $0\leq t \leq
    50$. The dotted rectangles represent the regions magnified in the
    insets.  In the magnified regions the solid circle {marks}
    the unperturbed equilibrium.}
  \label{trajuc}
\end{figure}

\begin{figure}
 \centering
  \subfigure[case \#1]{\label{sl-c1}\includegraphics[width=0.3\textwidth]{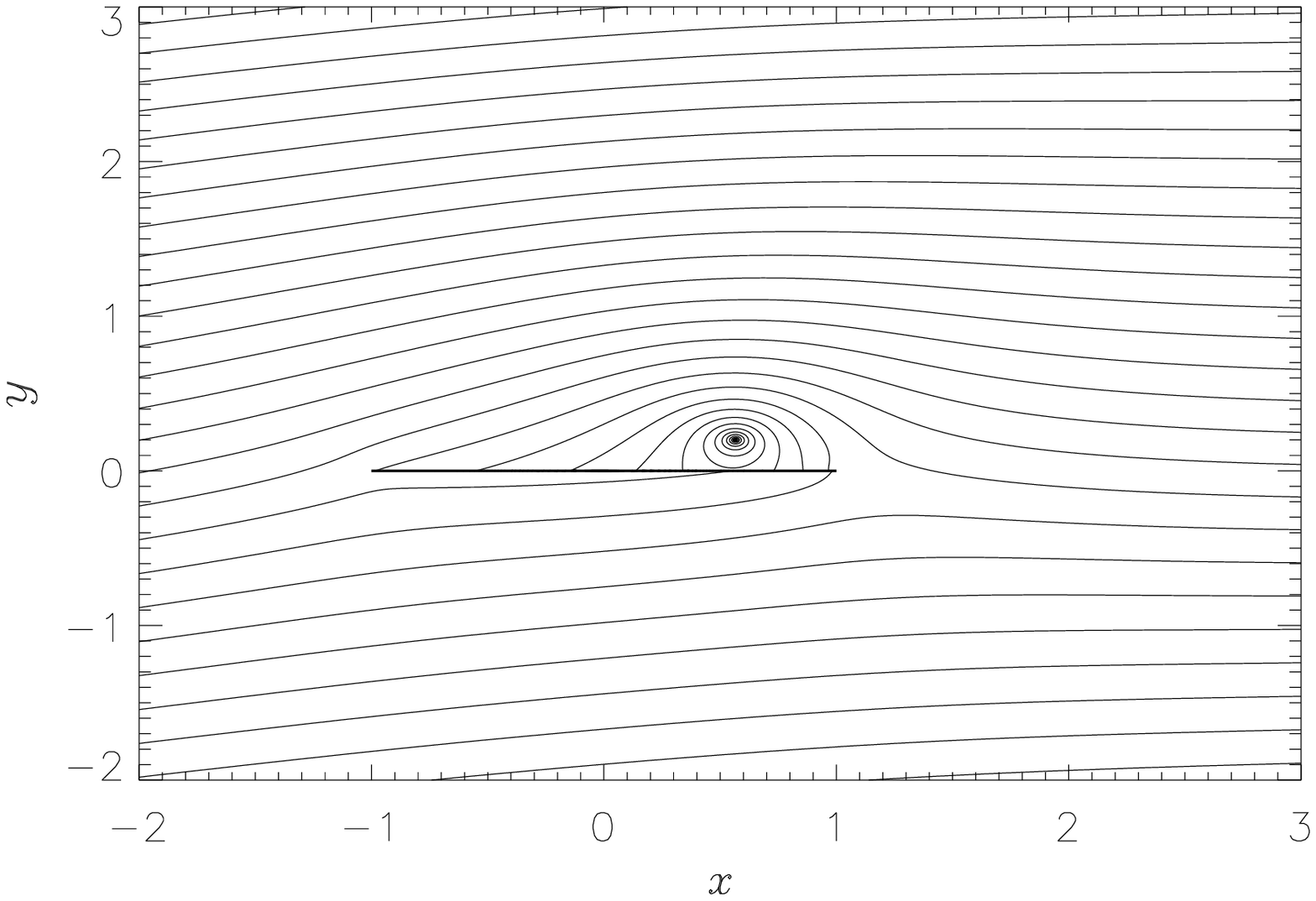}}\qquad
  \subfigure[case \#2]{\label{sl-c2}\includegraphics[width=0.3\textwidth]{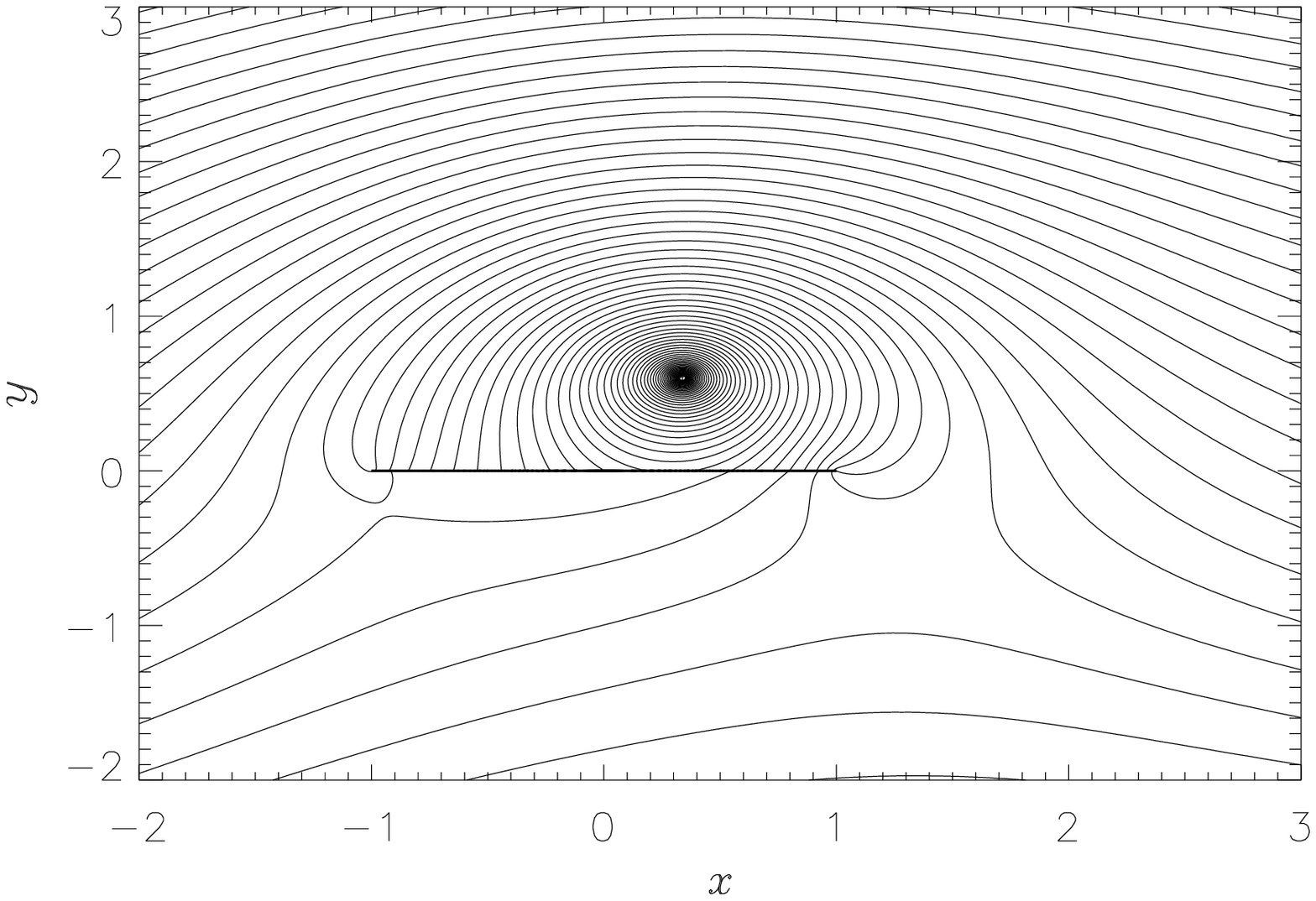}}\qquad
  \subfigure[case \#3]{\label{sl-c3}\includegraphics[width=0.3\textwidth]{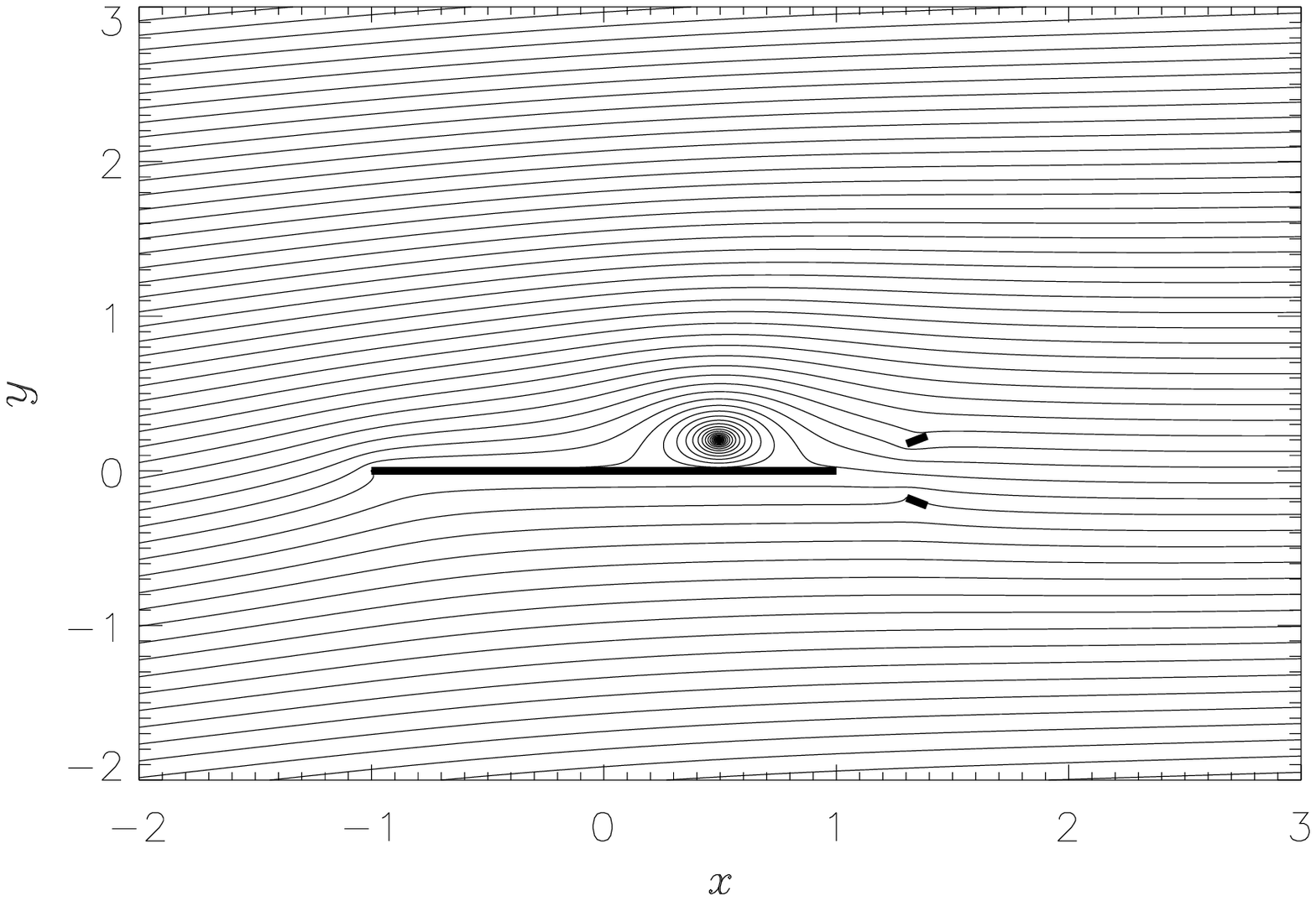}}\qquad
  \subfigure[case \#4]{\label{sl-c4}\includegraphics[width=0.3\textwidth]{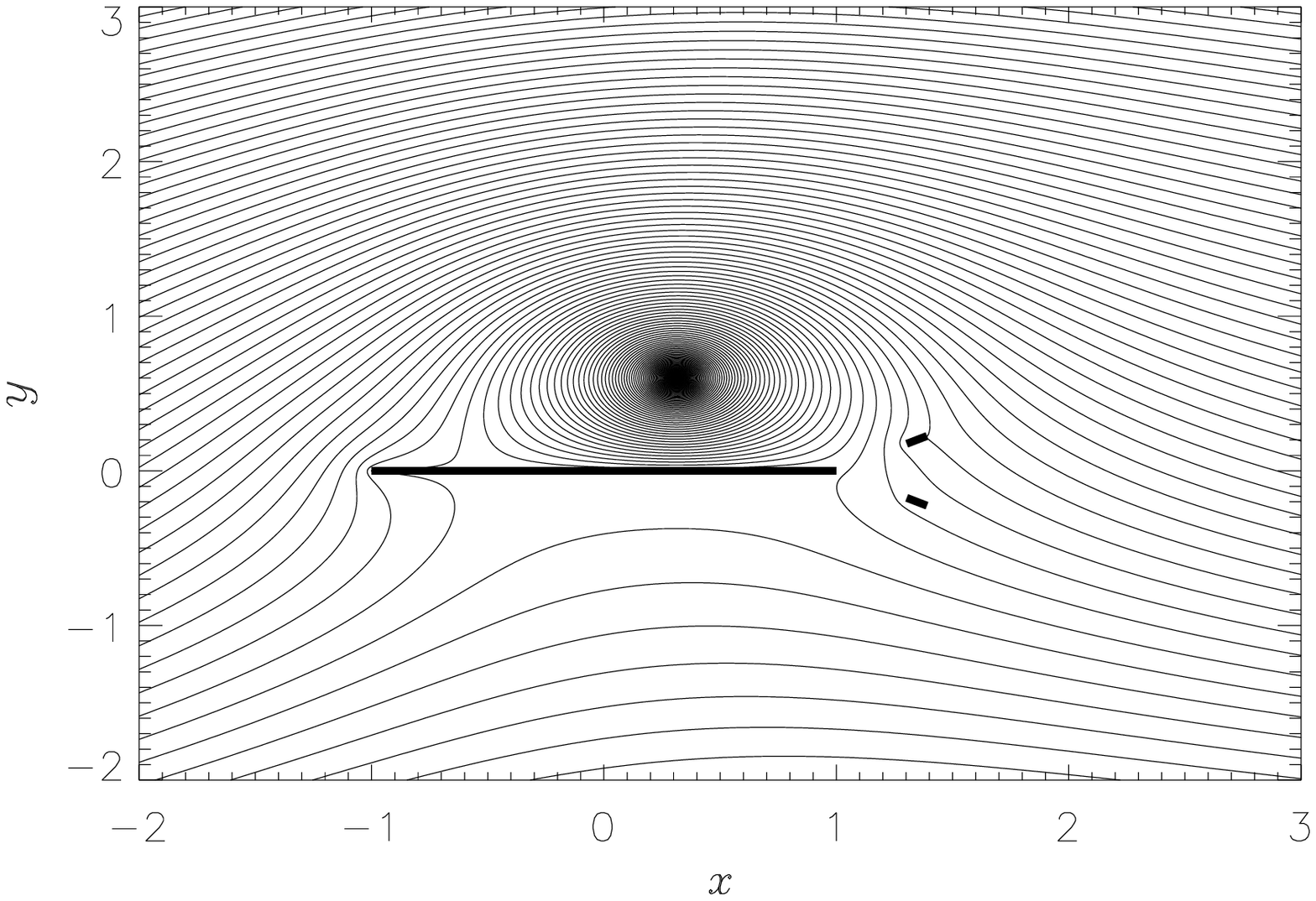}}\qquad
  \subfigure[case \#5]{\label{sl-c5}\includegraphics[width=0.3\textwidth]{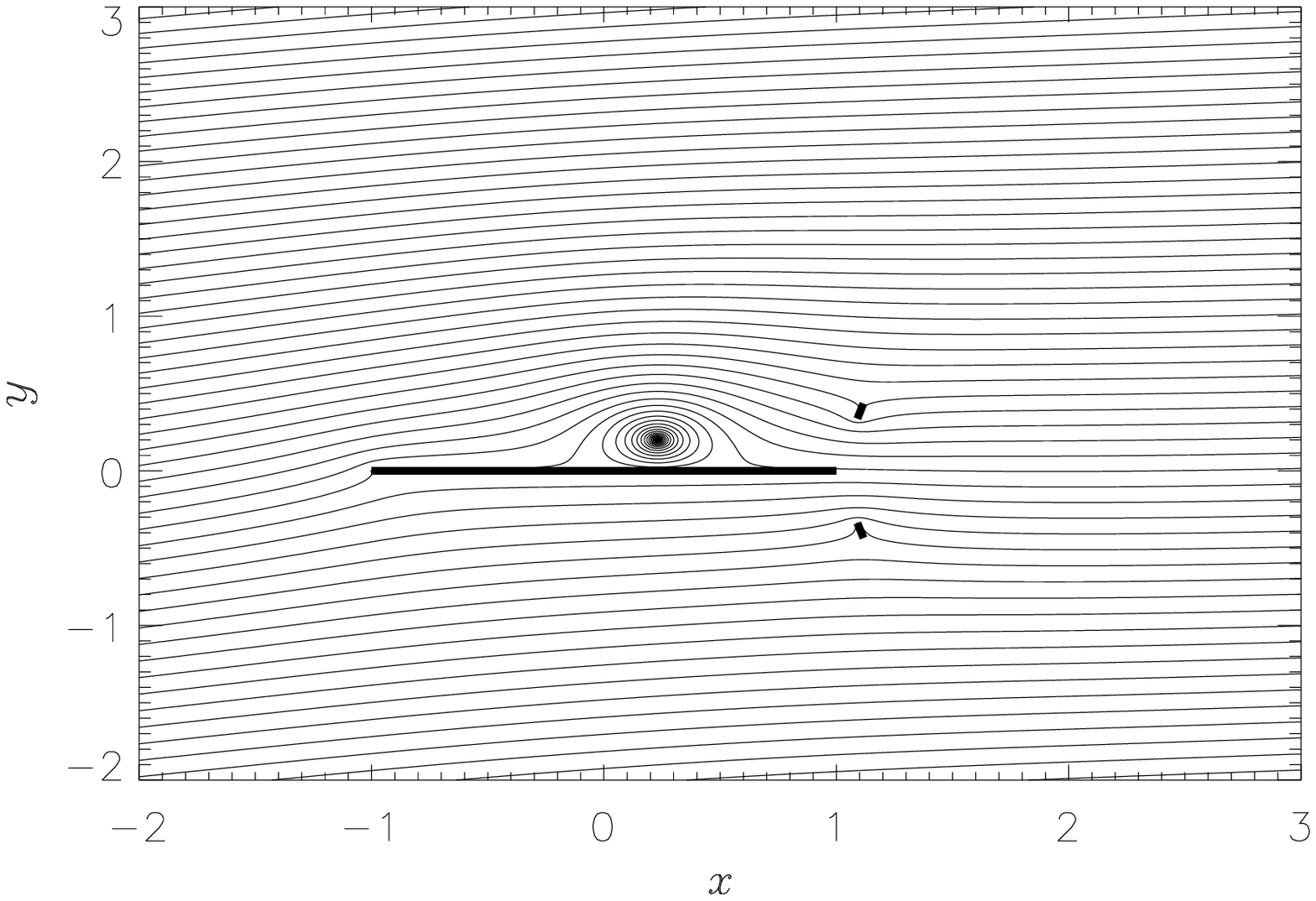}}\qquad
  \subfigure[case \#6]{\label{sl-c6}\includegraphics[width=0.3\textwidth]{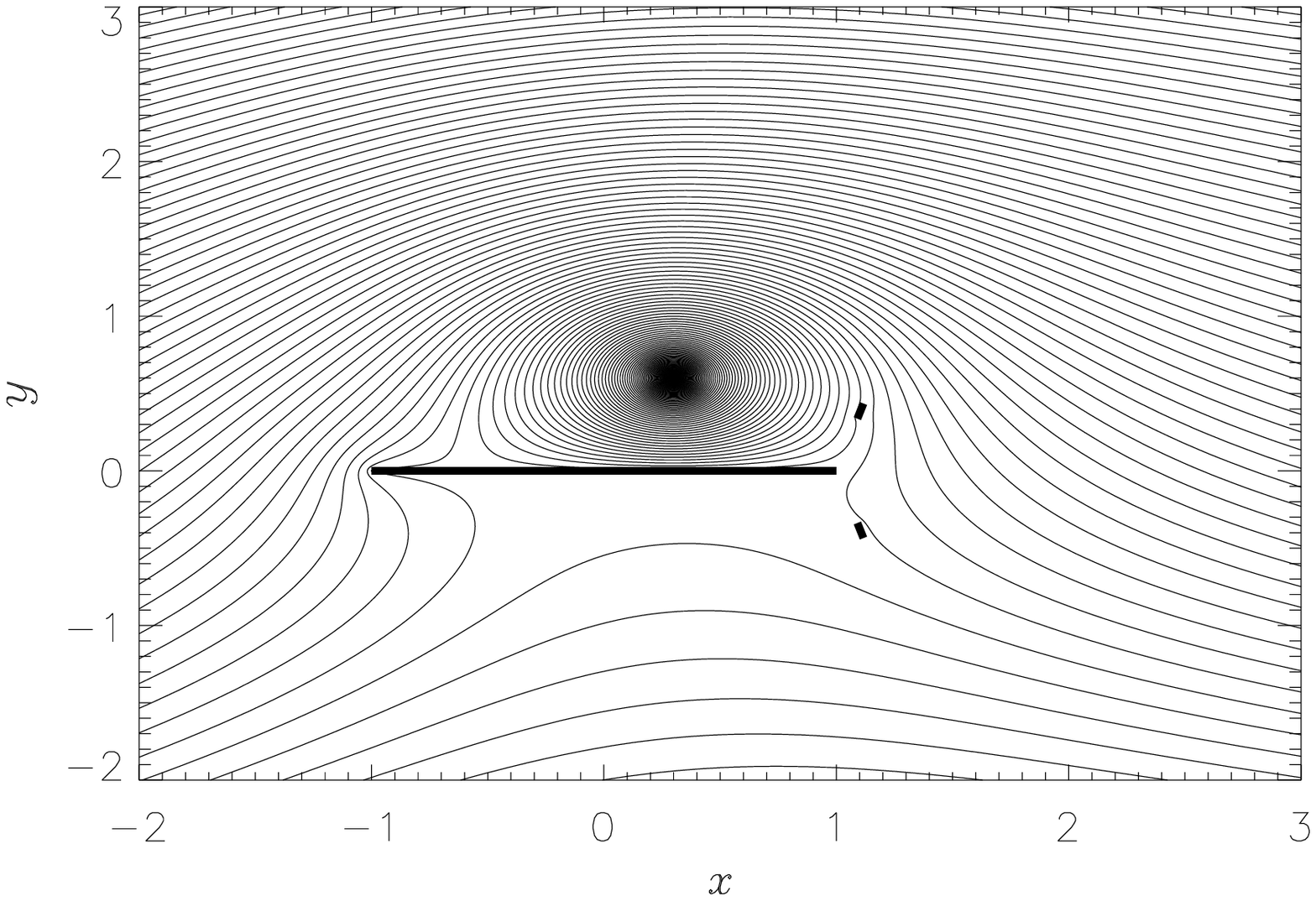}}\qquad
  \subfigure[case \#7]{\label{sl-c7}\includegraphics[width=0.3\textwidth]{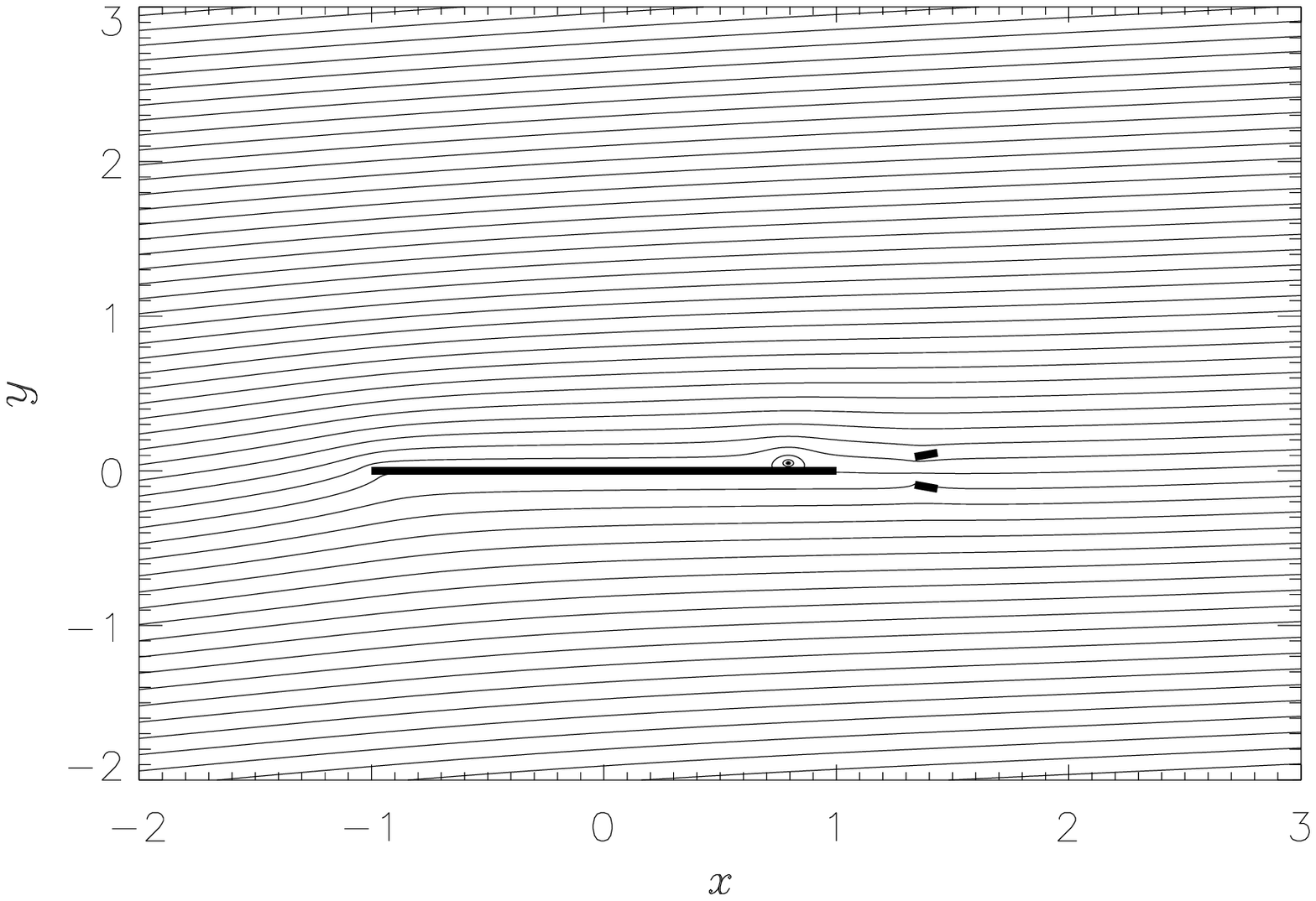}}
  \caption{(a)--(g) The streamline {patterns} $\psi=\Im[W]$
    {corresponding to the vortex equilibria from {table}
      \ref{tab:cases}}.}
  \label{sl-c1-7}
\end{figure}

Finally in this section, uncontrolled point-vortex trajectories are
presented {for configurations with $M=2$ and $\phi=\pi/6$}.
Evolutions are computed through numerical integration of {system}
\eqref{nlds} using Euler's {explicit} method with a time-step of
$\mathrm{d}t=0.001$ {(the choice of this method is motivated by its
  straightforward extension to the stochastic case considered in
  \S\ref{sec:random})}. Figures \ref{trajuc}(a) and \ref{trajuc}(b)
show the {trajectories} of the vortex when it is perturbed by $0.005i$
away from the equilibria, respectively, at $\Im(z_{\alpha})\approx0.2$
and $\Im(z_{\alpha})\approx0.6$.  {Responses to these perturbations}
show good agreement with the linear theory for
$\Im(z_{\alpha})\approx0.6$; that is, the vortex follows a closed
orbit over a long period of time.  However, when
$\Im(z_{\alpha})\approx0.2$, the vortex quickly escapes showing that
nonlinear effects become important even for very small perturbations.
Again, it should be noted that during these evolutions the Kutta
conditions are not enforced. {We emphasize that, from the practical
  point of view, neutrally stable equilibria are not desirable,
  because neutral stability in the linearised setting does not imply
  the stability of the nonlinear system, an effect already observed in
  figure \ref{trajuc}(a). In addition, vortices moving along closed
  trajectories circumscribing the equilibrium give rise to fluctuating
  loads on the airfoils which degrade their performance.}  To close
this section, {in {table} \ref{tab:cases}} we collect the information
about the different cases which were discussed here and which will be
further studied in the controlled setting below. {These cases have
  been selected to be representative of the different behaviours in
  the controlled setting (cf.~\S\ref{nres} and \S\ref{stochastic})}.
Note that due to the proximity of the vortex to the main plate in case
\#7, this case it is not necessarily as physically important as the
others considered. It is however included in this study for
completeness so that a Kasper Wing configuration demonstrating linear
instability is represented {as well}. It is also noted that this
case will not be analysed in as much detail as the others considered.
{To complete the physical picture, the streamline patterns
  corresponding to the seven equilibrium configurations from {table}
  \ref{tab:cases} are presented in figure \ref{sl-c1-7}. We see that
  the cases corresponding to equilibrium locations further away from
  the main plate feature larger recirculation regions.}

\begin{table}
\begin{center}
  \caption{Summary information about the different flow cases
      which will be analysed in the subsequent sections.}
    \vspace{1pc}
\begin{tabular}{ l | c | c | c | c | c | c | c | c } 
     \Bmp{1.0cm} \centering Case \Emp
   & \Bmp{1.0cm} \centering $M$ \Emp
   & \Bmp{1.0cm} \centering $\phi$ \Emp 
   & \Bmp{1.0cm} \centering $z_{\alpha}$  \Emp  
   & \Bmp{1.0cm} \centering $\kappa$  \Emp 
   & \Bmp{1.0cm} \centering $\Gamma_0$  \Emp 
   & \Bmp{1.0cm} \centering $\Gamma_1$  \Emp 
   & \Bmp{1.0cm} \centering $\Gamma_2$  \Emp 
   & \Bmp{2.0cm} \centering stability \Emp \\ \cline{1-9}
\#1  & 0 &  {-} & $0.566+0.200i$ & $-3.112$ & $0.926$ & - & - & unstable  \rule[-5pt]{0pt}{25pt} \\ 
\#2  & 0 &  {-} & $0.338+0.599i$ & $-14.272$ & $3.647$ & - & - & neutrally stable   \rule[-5pt]{0pt}{25pt} \\
\#3  & 2 &  $\pi / 6$ & $0.495+0.199i$ & $-2.993$ & $1.204$ & $0.199$ & $-0.134$ & neutrally stable  \rule[-5pt]{0pt}{25pt} \\ 
\#4  & 2 &  $\pi / 6$ & $0.313+0.600i$ & $-12.829$ & $3.935$ & $0.281$ & $0.063$ & neutrally stable   \rule[-5pt]{0pt}{25pt} \\
\#5  & 2 &  $5 \pi / 12$ & $0.230+0.200i$ & $-2.954$ & $1.591$ & $0.322$ & $-0.296$ & neutrally stable  \rule[-5pt]{0pt}{25pt} \\ 
\#6  & 2 &  $5 \pi / 12$ & $0.300+0.599i$ & $-13.496$ & $3.886$ & $0.128$ & $-0.037$ & neutrally stable   \rule[-5pt]{0pt}{25pt} \\
\hdashline
\#7  & 2 &  $\pi / 12$ & $0.793+0.050i$ & $-0.681$ & $-0.204$ & $0.092$ & $-0.090$ & unstable   \rule[-5pt]{0pt}{25pt}
\label{tab:cases}
\end{tabular}
\end{center}
\end{table} 

\section{Control-oriented characterization of the flow model}
\label{ctr-char}

In the present study {the intensity of} a point sink-source
located on the upper {boundary of the} main plate is used as the
control variable. Other choices of flow actuation are also possible
and, for example, circular cylinder rotation was used in the
{work of} \citet{protas:lfs}. The effect of the sink-source is
represented by the addition of a new term to {system}
\eqref{nlds} which is {now} written as
\begin{equation}
 \frac{\mathrm{d}}{\mathrm{d}t}\mathbf{X}=\mathbf{F}(\mathbf{X})+m\mathbf{b}(\mathbf{X}),
 \label{eq:F}
\end{equation}
where $m$ is the sink-source strength and 
\begin{equation}
 \mathbf{b}(\mathbf{X})=\left[\begin{array}{ll} \phantom{+}\Re[\mathrm{d}W_{S}(z)/\mathrm{d}z] \\ -\Im[\mathrm{d}W_{S}(z)/\mathrm{d}z] \end{array} \right],
 \label{ss:b}
\end{equation}
in which $W_{S}(z)$ represents the potential induced {at the
  point $z$} by the unit sink-source. First, we note that since the
actuator is chosen to lie on {{\em upper}} side of the main plate, its
location in the {pre-image} domain, labelled $\zeta_a$, will be
restricted to $\zeta_a=\exp(-i\sigma)$ for $0<\sigma<\pi$.  Then, in
the physical domain, $z_a = z(\zeta_a)$ {is such that $-1<\Re(z_a)<1$
  and $\Im(z_a) = 0$.  Hereafter $x_a := \Re(z_a)$ will denote the
  location of the sink-source actuator.}  The complex potential owing
to the point sink-source singularity in the pre-image domain is given
by
\begin{equation}
 W_{s}(\zeta)=\frac{1}{2\pi}\log\left[\frac{\omega(\zeta,\zeta_{a})\omega(\zeta,\bar{\zeta}_{a}^{-1})}{\omega(\zeta,\beta)\omega(\zeta,\bar{\beta}^{-1})}\right].
\label{eq:W_s}
\end{equation}
Introducing
\begin{equation}\label{eq:W_S}
 W_S(z(\zeta)) = W_s(\zeta),
\end{equation}
the derivatives appearing in \eqref{ss:b} can now be evaluated.
Re-deriving the linearised system with the additional term
{representing the flow actuation, cf.~\eqref{ss:b},} gives
\begin{equation}
 \frac{\mathrm{d}}{\mathrm{d}t}\mathbf{X}'=\mathbf{A}\mathbf{X}'+m\mathbf{B},
\label{eq:Xc}
\end{equation}
where $\mathbf{B}$ is a $2\times{1}$ matrix obtained from evaluating
\eqref{ss:b} at the equilibrium $z_{\alpha}$.

In order to formulate a meaningful control problem, it is required to
identify a physical objective that the control algorithm will seek to
achieve. This objective will be expressed in terms of system outputs,
i.e., certain measurable quantities that characterize the system
evolution and the system input, i.e., the control strength $m$.
Taking some equilibrium {configuration} as the ``base'' state, we
choose elimination of perturbations to the pressure difference between
two points on the main plate boundary resulting from perturbations to
this base state as the control objective. The two measurement points
{labelled $z_{+}$ and $z_{-}$} are chosen to have the same real
coordinate and lie on the upper and lower sides of the plate, {so
  that} in $\mathcal{D}_z$, $z_{+} = z_{-} \in {[-1,1] \subset}
\mathbb{R}$. Introducing the inverse map from the physical to
pre-image domain as
\begin{equation}\label{eq:zeta}
 \zeta = \zeta(z),
\end{equation}
we have the relation $\zeta(z_{-})=\overline{\zeta(z_{+})}$, where both
points will lie on the unit circle $C_{0}$. In a potential flow the
pressure {$p_m$} at a given boundary point can be calculated from the
Bernoulli equation as
$p_{m}=p_{0}+\frac{1}{2}(|V_{0}|^{2}-|V_{m}|^{2})$, where $p_{0}$ and
$V_{0}$ are the pressure and complex velocity at some arbitrary point
in the flow domain, and $V_{m}$ is the complex velocity at the
boundary point.  Thus, the pressure difference across the main plate
can be calculated as $\Delta
p=\frac{1}{2}(|V(z_{+})|^{2}-|V(z_{-})|^{2})$.  Choosing this
quantity as an output of system \eqref{eq:F} gives the following
output equation
\begin{equation}
  h(z_{\alpha}) := \frac{1}{2}\left[|V(z_{+})+mD_{+}|^2-|V(z_{-})+mD_{-}|^2\right],
 \label{eq:h}
\end{equation}
where $V(z_{+/-})$ is obtained from evaluating \eqref{vz} at $z_{+/-}$ and
\begin{equation}
 D_{+}=\left.\frac{\mathrm{d}W_{S}}{\mathrm{d}z}\right|_{z_{+}}, \phantom{---}
 D_{-}=\left.\frac{\mathrm{d}W_{S}}{\mathrm{d}z}\right|_{z_{-}}.
\end{equation}
Linearising for small $m$ then gives 
\begin{equation}
 h(z_{\alpha}) \cong \frac{1}{2}\left[|V(z_{+})|^2-|V(z_{-})|^2\right]+mD,
\end{equation}
where 
\begin{equation}
 D=\frac{1}{2}\left[\overline{V(z_{+})}D_{+}+V(z_{+})\overline{D_{+}}-\overline{V(z_{-})}D_{-}-V(z_{-})\overline{D_{-}}\right],
\end{equation}
is a scalar representing the direct effect of the control on the
measurement (i.e., the control-to-measurement map) in the linearised
regime.  This particular choice of the observation operator $h$ is
motivated by practical considerations, as measurements of pressure
differences across a wing are relatively easy to implement in a
laboratory experiment (i.e., either by direct measurement using
instrumentation, or {indirectly} from the airspeed distribution
using basic physical principles).  When considering the evolution of
small perturbations $\mathbf{X}'$ around the equilibrium, equation
\eqref{eq:h} can again be linearised, {this time with respect to
  $\mathbf{X}$,} which yields
\begin{equation}
 h(z_{\alpha}+z') \cong
 h(z_{\alpha}) + \mathbf{C}\mathbf{X}',
 \label{eq:hl}
\end{equation}
where the linearised observation operator $\mathbf{C}$ is given by a
$2\times{1}$ matrix
\begin{equation} 
 \mathbf{C} = 
 \left[
 \left.\frac{\partial \Delta p}{\partial x}\right|_{z_{\alpha}} \phantom{--}
 \left.\frac{\partial \Delta p}{\partial y}\right|_{z_{\alpha}}
 \right],
 \label{eq:C}
\end{equation}
in which
\begin{subequations}
 \begin{alignat}{1}
 \frac{\partial}{\partial x} &= \frac{\partial}{\partial z} + \frac{\partial}{\partial \bar{z}}, \\
 \frac{\partial}{\partial y} &= i\left(\frac{\partial}{\partial z} - \frac{\partial}{\partial \bar{z}}\right).
 \end{alignat}
\end{subequations}
{Hereafter $x_m := \Re(z_{+/-})$ will denote the location of the
  sensor.}  Since our linearised model reproduces the actual dynamics
only approximately, the difference between its predictions and the
actual flow behaviour can be regarded as disturbances which can be
accounted for by introducing a stochastic variable $w$ referred to as
the ``system (plant) noise''. It affects the linearised system
dynamics via a $[2\times 1]$ matrix $\mathbf{G}$ and the linearised
system output via a scalar value $H$. Furthermore, we assume that the
pressure measurement may be additionally contaminated with noise
$\MN$, where $\MN$ is a stochastic process.  With these definitions in
place, the linearised model can now put in the standard state--space
form \citep[see][]{OptC}
\begin{subequations}
\label{eq:Xs}
\begin{alignat}{3}
\frac{\d}{\d t} \X' &= \A \X' + && m \B + && \G w, 
\label{eq:Xsa} \\
             Y  &= \C \X' + && mD  + && H w + \MN.
\label{eq:Xsb} 
\end{alignat}
\end{subequations}
Before designing a controller for system \eqref{eq:Xs}, it must first
be verified that this is in fact feasible given the internal structure
of the system with its inputs and outputs. This can be done by
analysing the controllability and observability of system
\eqref{eq:Xs}.  {\em Controllability} is characterized by the number
of modes $\N_c$ that can be affected by the control authority
available.  The difference between the system dimension (2 in the
present case) and $\N_c$ gives the number of uncontrollable modes. For
the system under consideration, $\N_c$ is calculated according to
\citep[see][]{OptC}
\begin{equation}
\N_c := \rank\left[ \B \ \ \A\B \right].
\label{eq:Nc}
\end{equation}
Evaluating {condition} \eqref{eq:Nc} for each of the cases \#1--7
gives $\N_c=2$ for all actuator locations $x_a$, meaning that {in
  general} the matrix pair $\{\A,\B\}$ is completely controllable and
both modes present in the system can be controlled in all cases.  In a
similar spirit, {\em observability} is characterized by the number of
modes $\N_o$ that can be reconstructed based on the measurements
available and the difference between the system dimension and $\N_o$
gives the number of unobservable modes.  For the linear time-invariant
system \eqref{eq:Xs}, in each of the cases \#1--7, $\N_o$ is
calculated as
\begin{equation}
\N_o := \rank\left[ \C^T \ \ \A^T \C^T \right] = 2
\label{eq:No}
\end{equation}
which means that the matrix pair $\{\A,\C\}$ is completely observable
{for {\em almost} all sensor locations $x_m$} in all cases {(as
  will be discussed below, observability may be in fact lost for
  certain isolated sensor locations $x_m$ forming a zero-measure
  subset of $[-1,1]$)}.

This section is concluded with a brief discussion regarding the
optimal placement of the actuator and sensor, i.e., the best choices
of $x_a$ {in \eqref{eq:W_s}} and $x_m$ in \eqref{eq:h}. This is an
important issue from the implementation point of view, as a judicious
choice of $x_a$ will {maximize the control authority} and a
judicious choice of $x_m$ will maximize the information that can be
extracted from the available measurements.  Decomposing $\X'$ in terms
of the right {(column)} eigenvectors $\bxi_1$ and $\bxi_2$ of
$\A$ as $\X'=\sum_{k=1,2}{\rho_{k}} \, \bxi_{k}$ (for
${\rho_{k}}\in\mathbb{R}$), an equation for the linearised
dynamical system \eqref{eq:Xsa} can be expressed as
\begin{equation}
 \frac{\d}{\d t} \sum_{k=1,2}{\rho_{k}}\,\bxi_{k} = \A \sum_{k=1,2}{\rho_{k}}\,\bxi_{k} + m \B + \G w.
 \label{eq:ctr1}
\end{equation}
Taking the inner product of \eqref{eq:ctr1} with the left
{(row)} eigenvector $\bpsi_k$ {of $\A$} then yields
\begin{equation}
\begin{split}
 \frac{\d}{\d t} \rho_{k} &= \lambda_{k}\,\rho_k + m (\bpsi_k\B) + (\bpsi_k\G) w \\
                             &= \lambda_{k}\,\rho_k + m b_{k} + (\bpsi_k\G) w,
 \label{eq:ctr2}
\end{split}
\end{equation}
for $k=1,2$ {where $\lambda_1$ and $\lambda_2$ are the
  eigenvalues of $\A$}. The quantities $b_1 := \bpsi_1\B$ and $b_2 :=
\bpsi_2\B$ are referred to as {``modal control residuals''}
\citep{bl98} and give a quantitative measure of the sensitivity of the
{eigenmodes $\bxi_1$ and $\bxi_2$} to the control {input
  represented by} $\B$. When $b_k = 0$, $k=1,2$, this implies
uncontrollability of the corresponding mode.  On the other hand, when
$b_k$ is large {relative to other terms in \eqref{eq:ctr2}}, the
control is capable of leaving a large imprint on the corresponding
mode. In a similar manner, the linearisation of the pressure
perturbation \eqref{eq:Xsb} can be expressed as
\begin{equation}
\begin{alignedat}{2}
Y & =  \C &&\sum_{k=1,2} \lambda_{k} \bxi_k + mD + Hw + \MN \\
     & =            &&\sum_{k=1,2} \lambda_{k} c_k    + mD + Hw + \MN.
\end{alignedat}
\label{eq:Ya}
\end{equation}
The quantities $c_1 := \C\bxi_1$ and $c_2 := \C\bxi_2$, referred to as
the ``modal observation residuals'', are therefore related to
observability of the eigenmodes. Again, when $c_k = 0$, $k=1,2$, this
implies unobservability of the corresponding mode.  On the other hand,
when $c_k$ is large {relative to other terms in \eqref{eq:Ya}},
the corresponding mode leaves a large imprint on the measurement.

\begin{figure}
  \centering
\mbox{
  \subfigure[]{\label{ctr-m0-02}\includegraphics[width=0.475\textwidth]{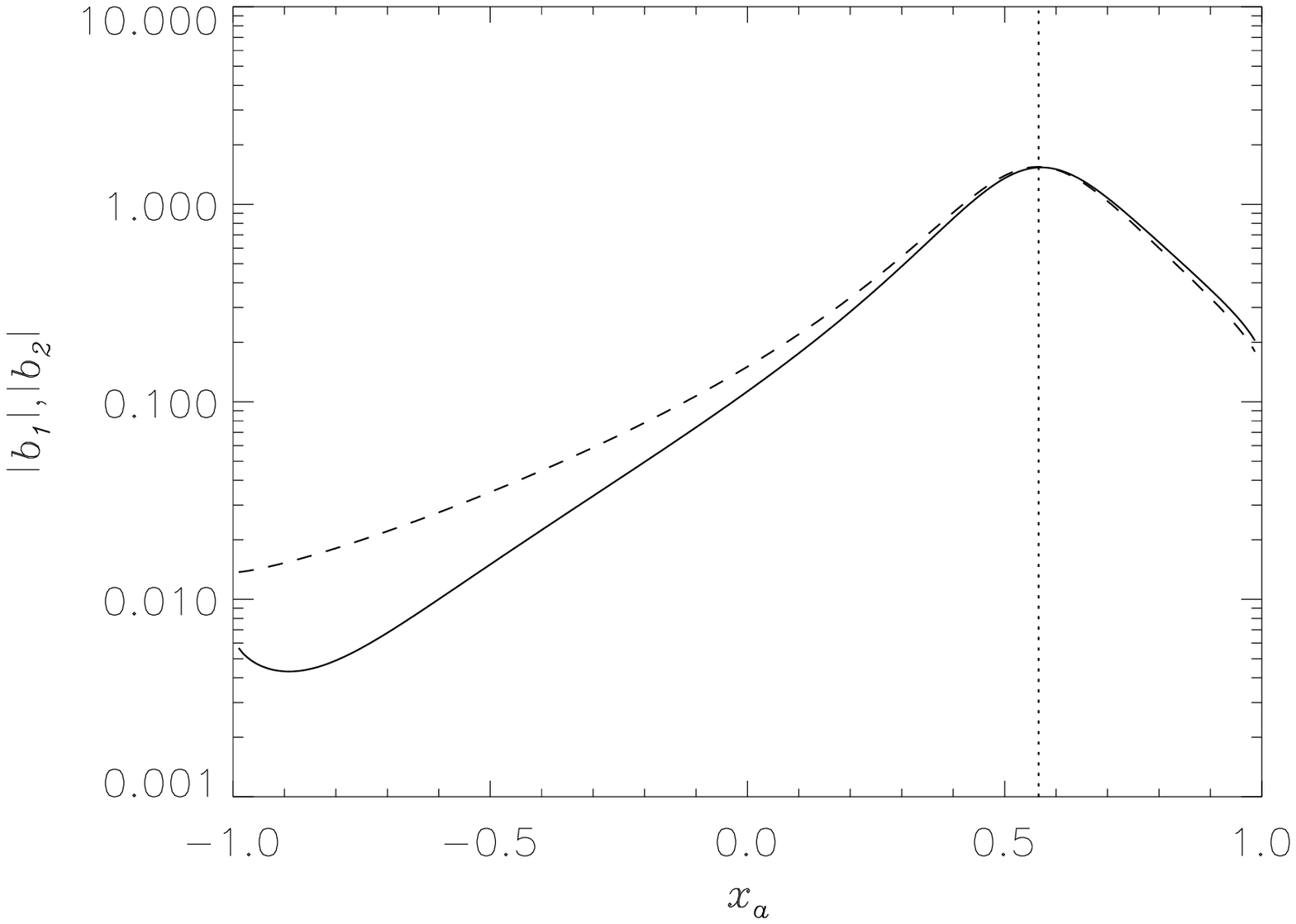}}\qquad
  \subfigure[]{\label{obs-m0-02}\includegraphics[width=0.475\textwidth]{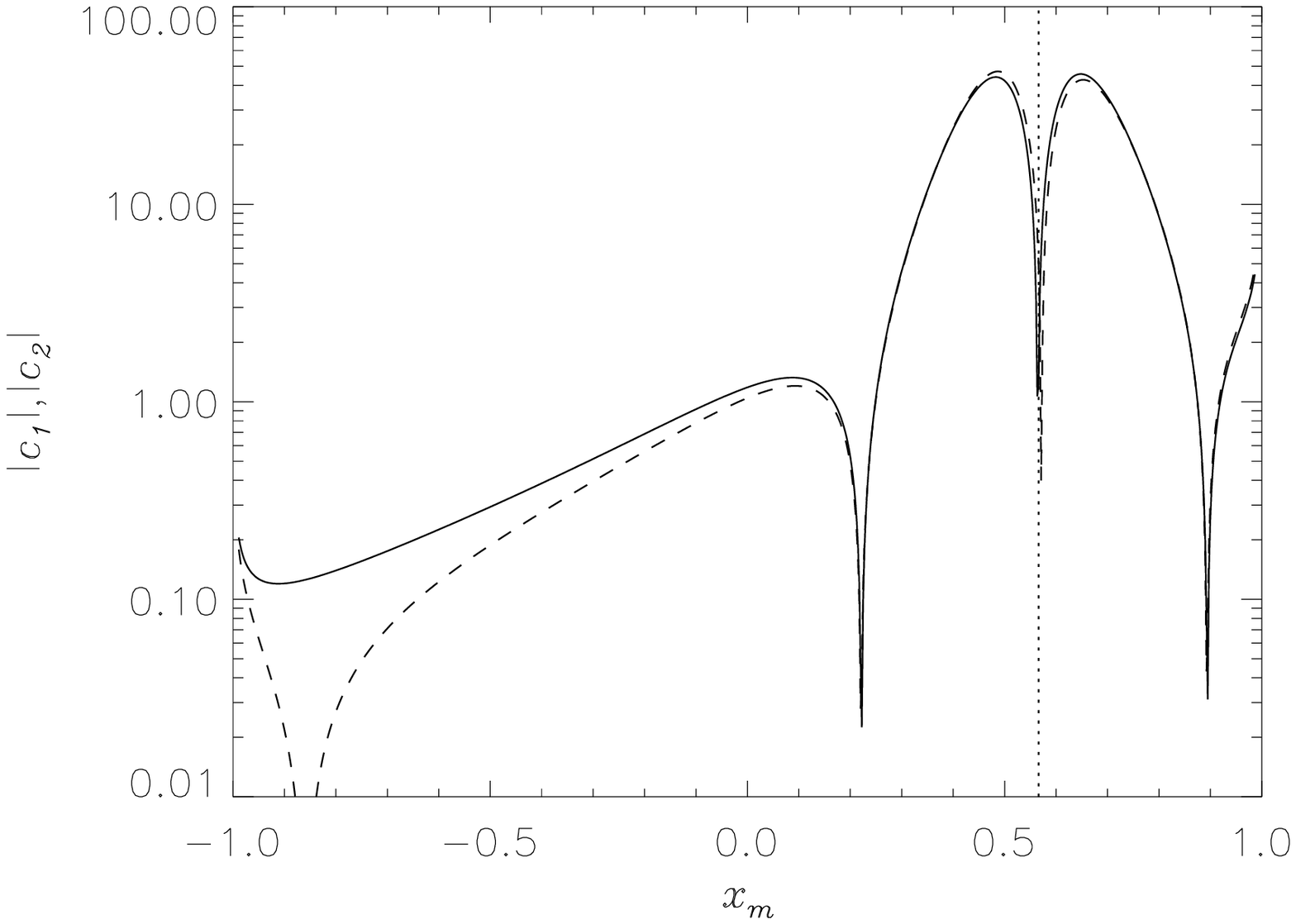}}
}
  \caption{(a) Dependence of the absolute value of the control
    residuals $|b_{1}|$ (solid curve) and $|b_{2}|$ (dashed curve) on
    the actuator {location} $x_a$ in case \#1. The maxima of
    {the residuals} $|b_{1}|$ and $|b_{2}|$ occur at $x_a=0.571$
    and $x_a=0.564$, respectively.  The dotted vertical line indicates
    the $x$-coordinate of the point vortex equilibrium. (b) Same as
    (a), but for the observability residuals $|c_1|$ and $|c_2|$.  The
    {maxima of the residuals} $|c_{1}|$ and $|c_{2}|$ occur at
    $x_m=0.648$ and $x_m=0.487$, respectively.}
  \label{ctrobs-m0-02}
\bigskip\bigskip\bigskip
\mbox{
  \subfigure[]{\label{ctr-m0-06}\includegraphics[width=0.475\textwidth]{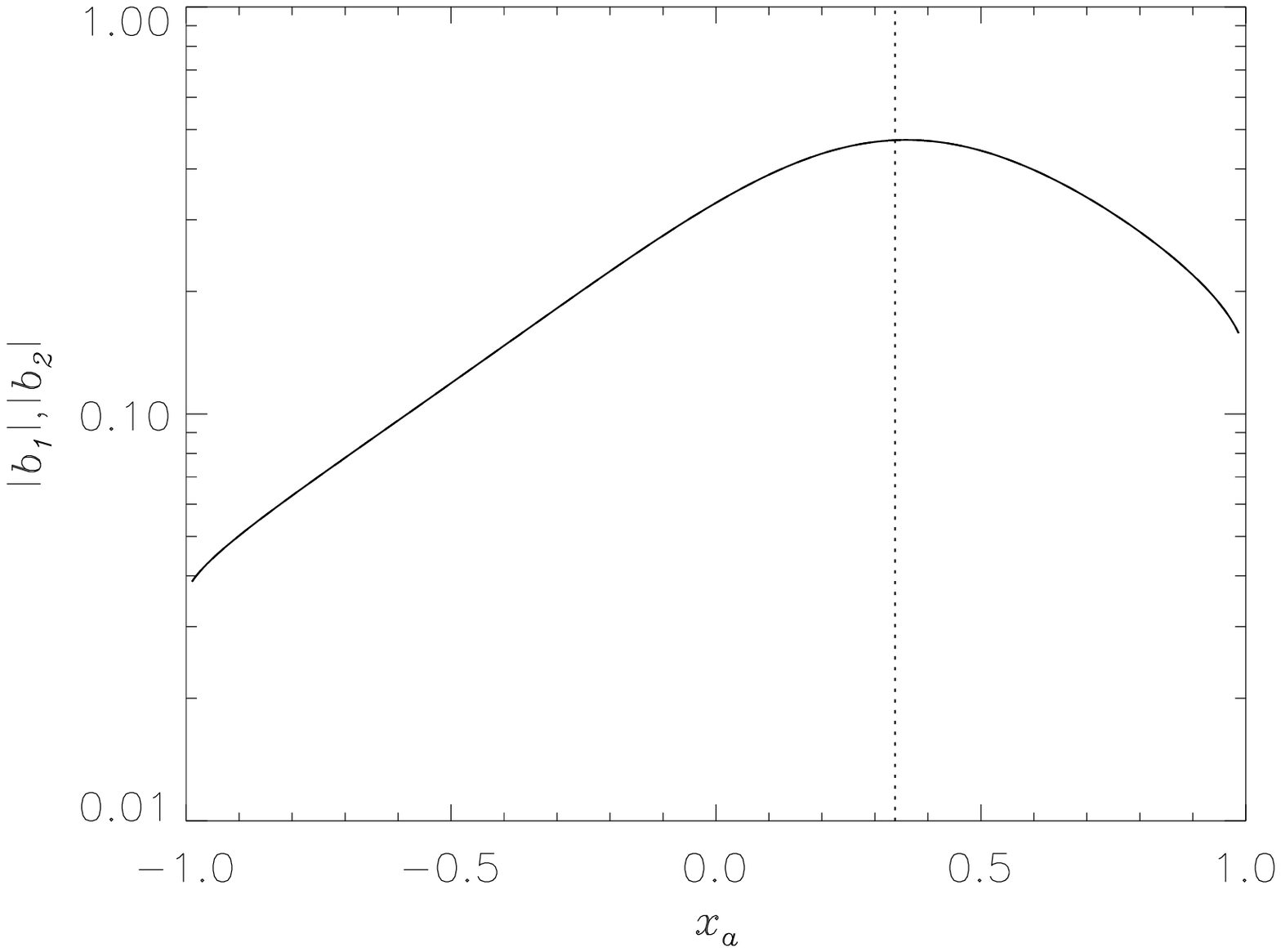}}\qquad
  \subfigure[]{\label{obs-m0-06}\includegraphics[width=0.475\textwidth]{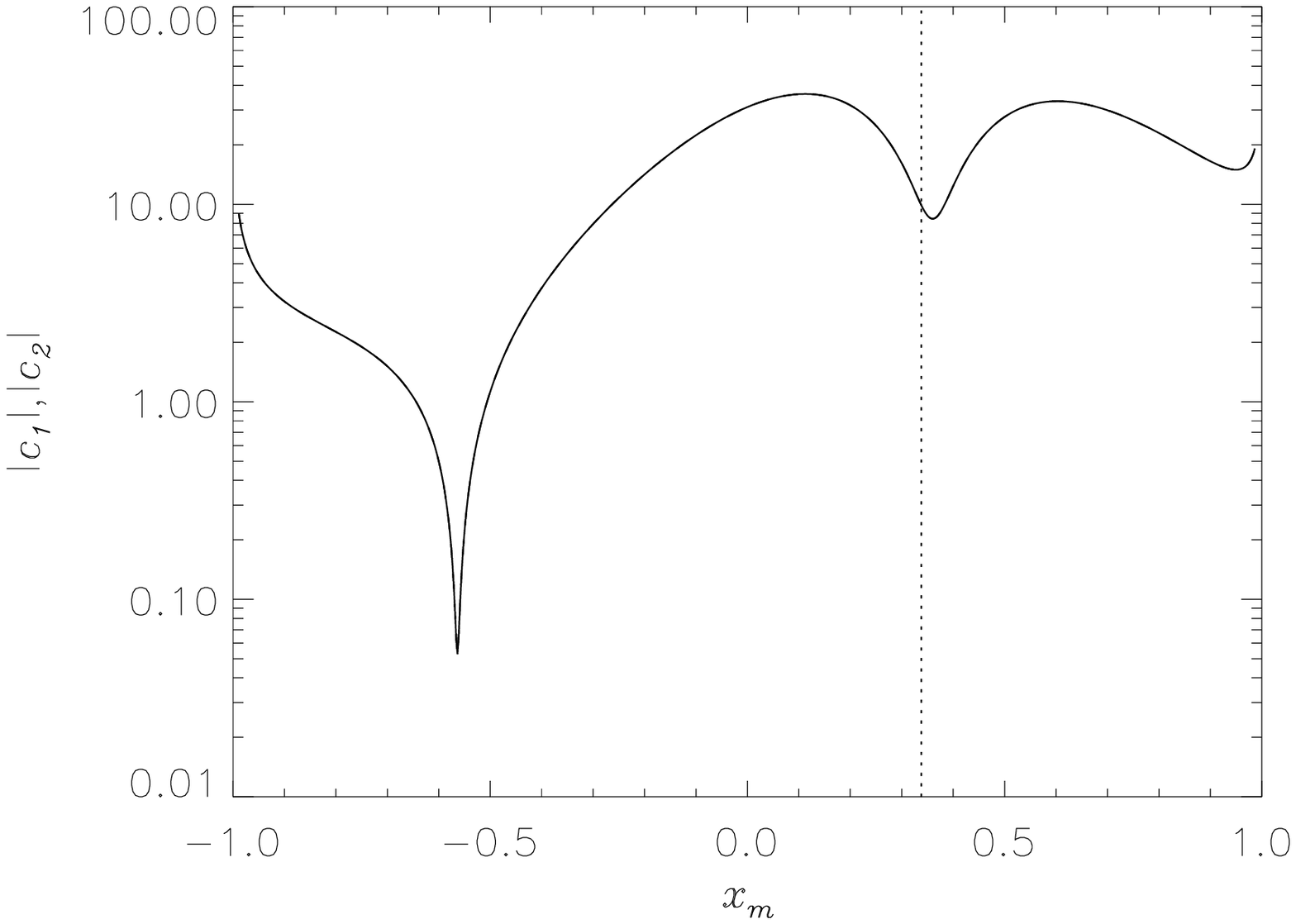}}
}
  \caption{(a) Dependence of the absolute value of the control
    residuals $|b_{1}|$ and $|b_{2}|$ on the actuator {location} $x_a$
    in case \#2 {(the values of $|b_{1}|$ and $|b_{2}|$ are
      essentially the same in this configuration)}.  The maximum value
    of the control residuals occurs at $x_a=0.358$. The dotted
    vertical line indicates the $x$-coordinate of the point vortex
    equilibrium. (b) Same as (a), but for the observability residual.
    Here the maximum value occurs at $x_m=0.113$.}
  \label{ctrobs-m0-06}
\end{figure}

\begin{figure}
 \centering
\mbox{
  \subfigure[]{\label{ctr-m2a30-02}\includegraphics[width=0.475\textwidth]{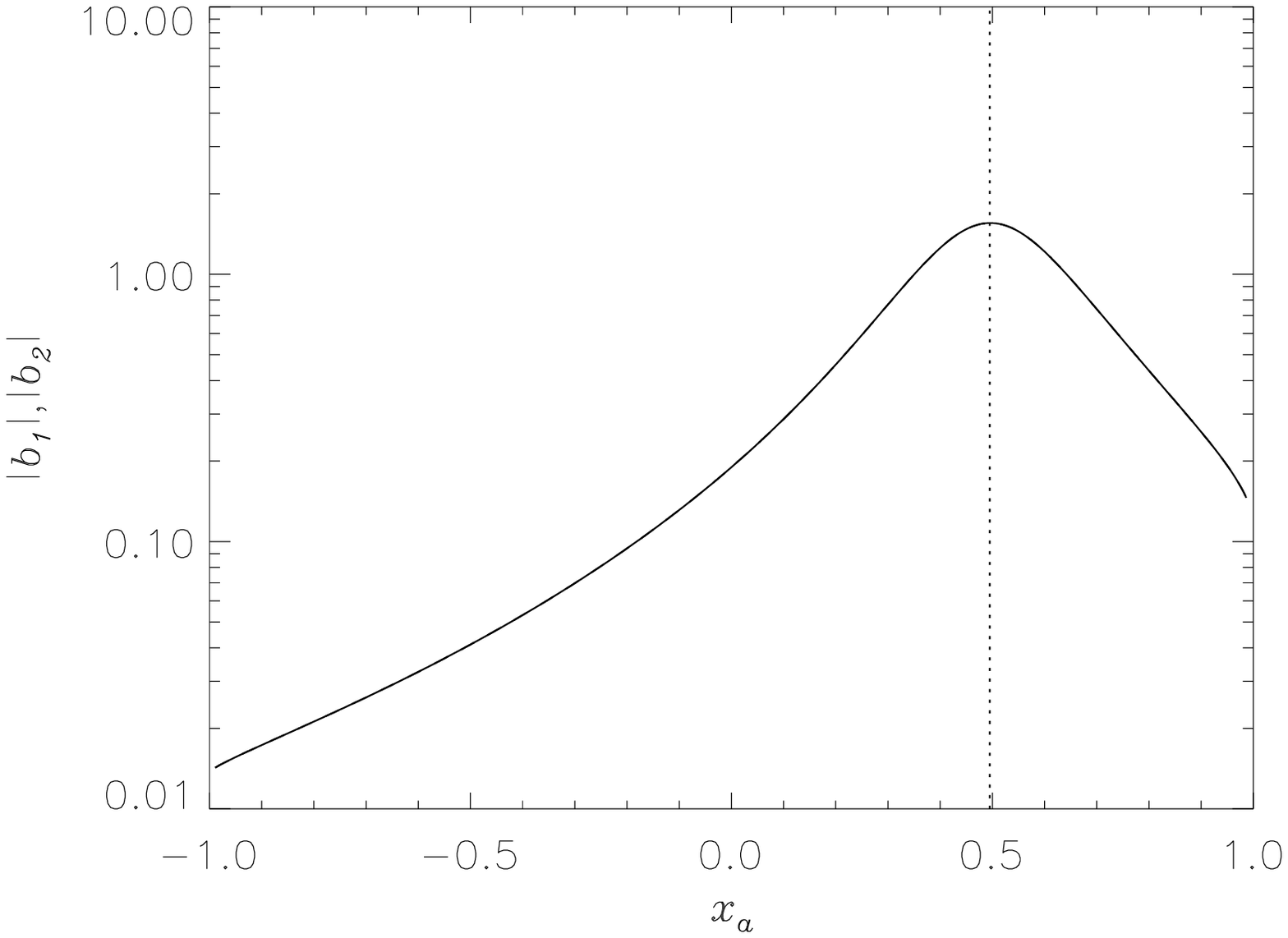}}\qquad
  \subfigure[]{\label{obs-m2a30-02}\includegraphics[width=0.475\textwidth]{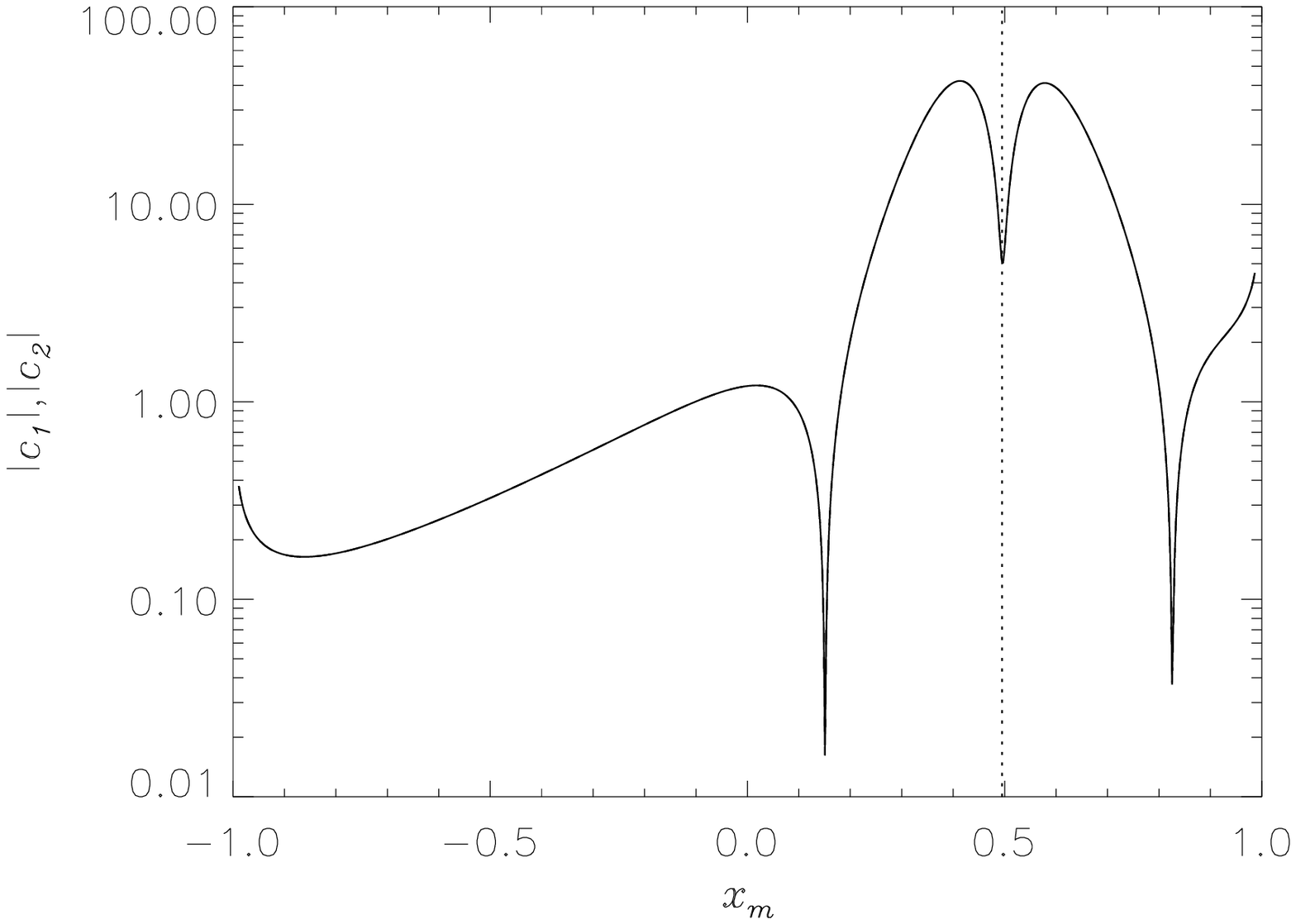}}
}
  \caption{(a) Dependence of the absolute value of the control
    residuals $|b_{1}|$ and $|b_{2}|$ on the actuator {location}
    $x_a$ in case \#3 {(the values of $|b_{1}|$ and
    $|b_{2}|$ are essentially the same in this configuration)}.  The
    maximum value of the control residuals occurs at $x_a=0.497$. The
    dotted vertical line indicates the $x$-coordinate of the point
    vortex equilibrium. (b) Same as (a), but for the observability
    residual. Here the maximum value occurs at $x_m=0.413$.}
  \label{ctrobs-m2a30-02}
\bigskip\bigskip\bigskip
\mbox{
  \subfigure[]{\label{ctr-m2a30-06}\includegraphics[width=0.475\textwidth]{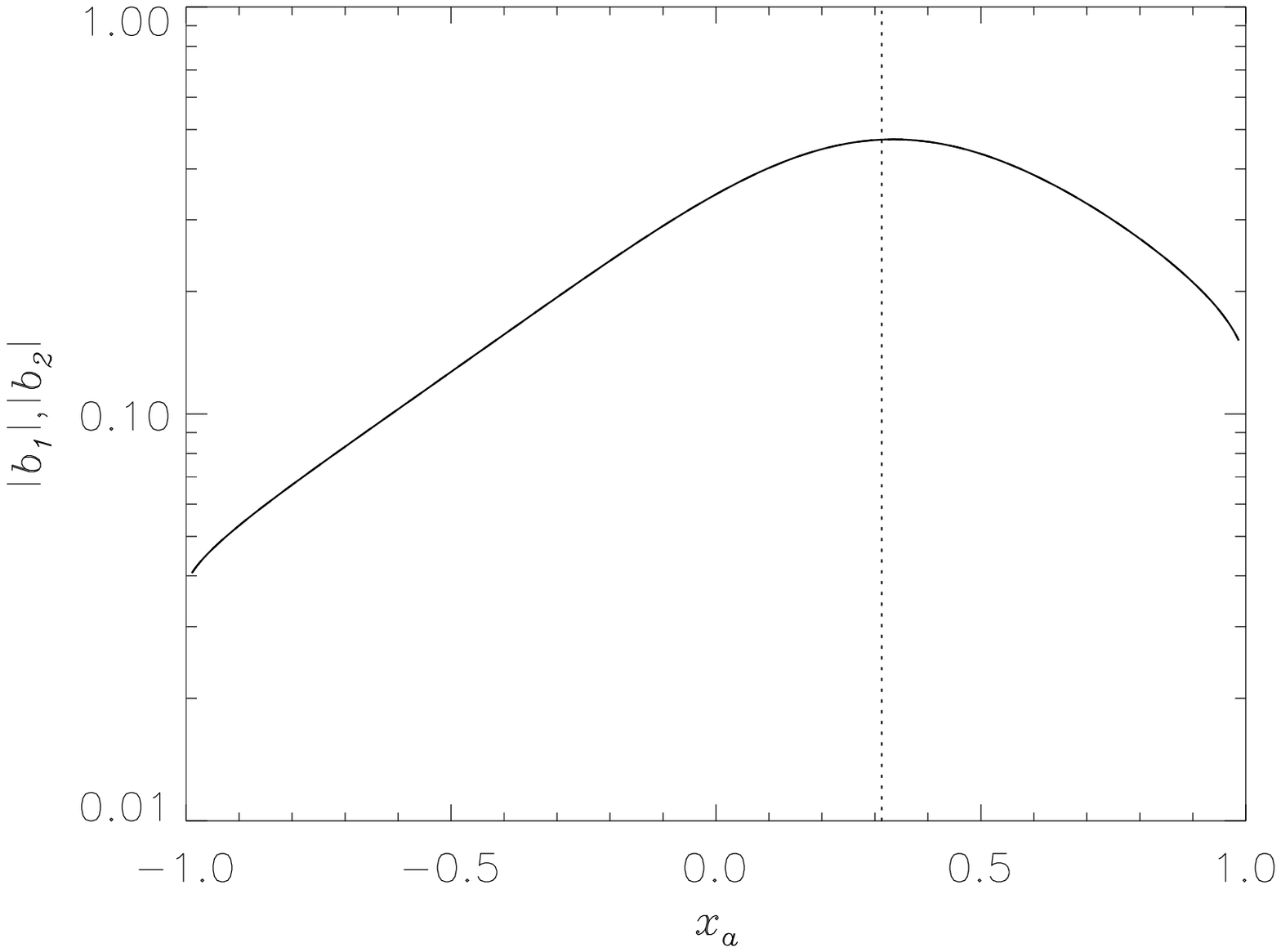}}\qquad
  \subfigure[]{\label{obs-m2a30-06}\includegraphics[width=0.475\textwidth]{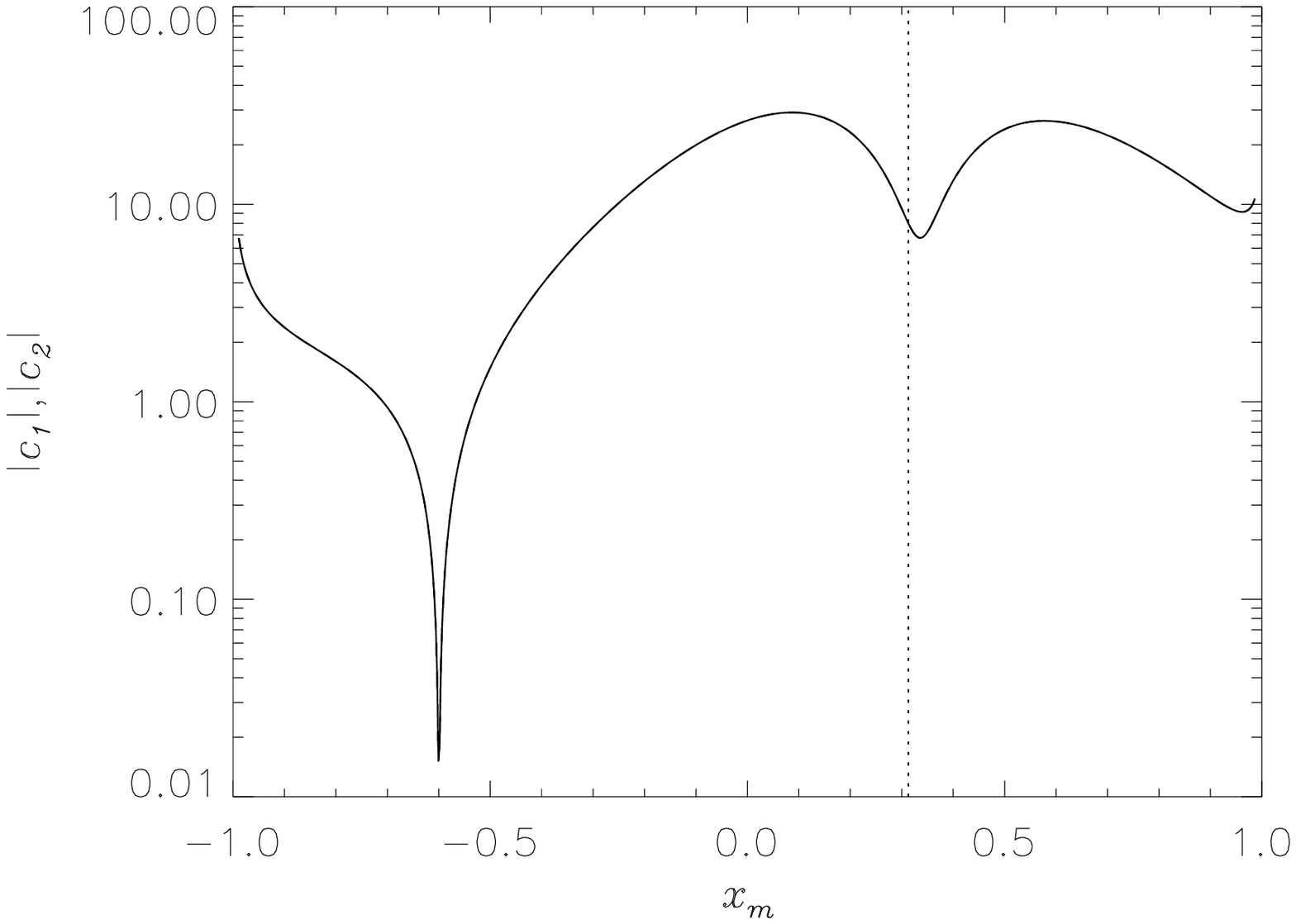}}
}
  \caption{(a) Dependence of the absolute value of the control
    residuals $|b_{1}|$ and $|b_{2}|$ on the actuator {location} $x_a$
    in case \#4 {(the values of $|b_{1}|$ and $|b_{2}|$ are
      essentially the same in this configuration)}.  The maximum value
    of the control residuals occurs at $x_a=0.334$. The dotted
    vertical line indicates the $x$-coordinate of the point vortex
    equilibrium. (b) Same as (a), but for the observability residual.
    Here the maximum value occurs at $x_m=0.087$.}
  \label{ctrobs-m2a30-06}
\end{figure}

The dependence of the absolute values of the control {residual on
  $x_a$} and observability residual on $x_m$ is shown for cases \#1
and \#2 in figures \ref{ctrobs-m0-02} and \ref{ctrobs-m0-06}, and for
cases \#3 and \#4 in figures \ref{ctrobs-m2a30-02} and
\ref{ctrobs-m2a30-06}. To obtain these figures, {the coordinates
  $x_a, x_m \in [-1,1]$} were discretized {using} 800 equispaced
{grid} points and the residuals {were} evaluated at each
point.  Residuals for cases \#5--7 produced similar features to
those already presented {in figures
  \ref{ctrobs-m0-02}--\ref{ctrobs-m2a30-06}} and therefore, for
brevity, the full details of these are excluded.  In {most} cases the
residuals of both modes are very similar and, in fact, for cases
\#2--6 the residuals are (to numerical tolerance) essentially the
same. Only small differences {between the residuals of the two
  modes} are seen in cases \#1 and \#7. {We remark that while
  the control residuals are in all cases bounded away from zero, the
  observability residuals may vanish at some isolated points. This
  occurs when the observability residuals $c_1$ and $c_2$ change sign
  and is manifested by ``spikes'' visible in figures
  \ref{ctrobs-m0-02}(b)--\ref{ctrobs-m2a30-06}(b) (due to numerical
  resolution, these spikes are smeared and do not actually reach
  zero). We therefore conclude that observability may be lost for some
  isolated sensor locations $x_m$.}  In the results that follow,
{the actuator will be situated at the location $x_a$}
corresponding to $\mathrm{max}\{|b_1|,|b_2|\}$ and {the sensor
  will be situated at} $x_m$ corresponding {to the} maximum of
the observation residuals {$\mathrm{max}\{|c_1|,|c_2|\}$}. A
summary of the {optimal} locations of $x_a$ and $x_m$ for each
configuration is shown in {table} \ref{max-res-table}.

\begin{table}
\begin{center}
  \caption{Summary of the actuator and sensor placement in each case
    together with the corresponding {maximum} values of the
    controllability and observability residuals.}
    \vspace{1pc}
\begin{tabular}{ l | c | c | c | c } 
     \Bmp{2.0cm} \centering Case \Emp
   & \Bmp{2.0cm} \centering $x_a$ \Emp
   & \Bmp{2.2cm} \centering $\max\{{|b_1|,|b_2|}\}$ \Emp 
   & \Bmp{2.0cm} \centering $x_m$  \Emp 
   & \Bmp{2.2cm} \centering $\max\{{|c_1|,|c_2|}\}$ \Emp \\ \cline{1-5}
\#1  & $0.564$  & $1.546$  & $0.487$ & $47.072$   \rule[-5pt]{0pt}{25pt} \\ 
\#2  & $0.358$  & $0.471$  & $0.113$ & $36.190$   \rule[-5pt]{0pt}{25pt} \\
\#3  & $0.497$  & $1.554$  & $0.413$ & $42.117$   \rule[-5pt]{0pt}{25pt} \\ 
\#4  & $0.334$  & $0.473$  & $0.087$ & $29.174$   \rule[-5pt]{0pt}{25pt} \\
\#5  & $0.230$  & $1.567$  & $0.148$ & $40.431$   \rule[-5pt]{0pt}{25pt} \\ 
\#6  & $0.319$  & $0.472$  & $0.074$ & $32.301$   \rule[-5pt]{0pt}{25pt} \\
\hdashline
\#7  & $0.793$  & $6.309$  & $0.774$ & $138.500$   \rule[-5pt]{0pt}{25pt}
\label{max-res-table}
\end{tabular}
\end{center}
\end{table}

\section{LQG control design}
\label{sec:lqg}

\begin{figure}
\begin{center}
\includegraphics[width=.8\textwidth]{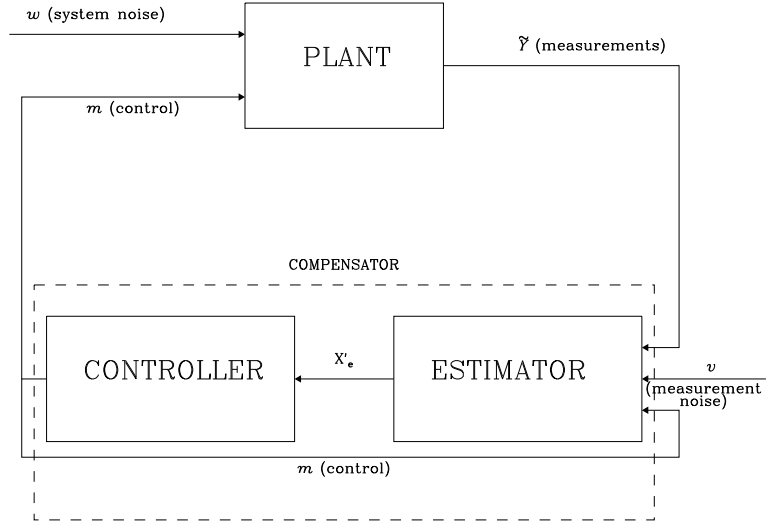}
\caption{Schematic of a compensator composed of an estimator and a
  controller.}
\label{fig:lqg}
\end{center}
\end{figure}

In this section the control algorithm is derived for the model system
{\eqref{eq:Xs}} based on {Linear} Optimal Control Theory
\citep{OptC}. The derivation follows closely the approach implemented
in \citet{protas:lfs}. The objective is to find a {\em feedback}
control law $m = - \K \X'$, where $\K$ is a $[2 \times 1]$ feedback
matrix, that will {asymptotically} stabilize system \eqref{eq:Xs}
while minimizing a performance criterion represented by the following
cost functional
\begin{equation}
  \J(m) := {\mathbb{E}\left[ \int_0^{\infty} (Y Q Y + m R m ) \d t \right]},
\label{eq:J}
\end{equation}
where {$\mathbb{E}$} denotes the expectation, whereas $Q \ge 0$ and $R>0$ are
given numbers. We note that the cost functional \eqref{eq:J} balances
the linearised system output $Y$ (i.e., the pressure difference across
the wing at the sensor location $(x_m,0)$, {cf.~\eqref{eq:Xsb},}
and the control effort, whereas the feedback control law provides a
recipe for determining the actuation (i.e., the strength $m$ of the
sink-source) based on the state of the linearised model (i.e., the
perturbation $\X'$ of the {equilibrium}). In practice, however,
the state $\X'$ of the model \eqref{eq:Xs} is not known. Instead,
noisy measurements $\tilde{Y}$ of the actual system (i.e., the
nonlinear single-plate or Kasper Wing system \eqref{eq:F}) are
available and can be used in an {\em estimation procedure} to
construct an estimate $\X'_e$ of the model state $\X'$. The evolution
of the state estimate $\X'_e$ is governed by the estimator system
\begin{subequations}
\label{eq:Xe}
\begin{alignat}{2}
\frac{\d}{\d t} \X'_e &= \A \X'_e + &&m\B + \L ( \tilde{Y} - Y_e ), 
\label{eq:Xea} \\
             Y_e  &= \C \X'_e + && mD,
\label{eq:Xeb} 
\end{alignat}
\end{subequations}
where $\L$ is a feedback matrix that will be chosen below in a manner
ensuring that the estimation error vanishes in the infinite time
horizon, i.e., that $\X'_e \rightarrow \X'$ as $t \rightarrow \infty$.
Thus, the estimator assimilates available observations into the system
model, so as to produce an evolving estimate of the system state.
Finally, the controller and the estimator can be combined to form a
{\em compensator} in which the feedback control is determined based on
the state estimate $\X'_e$ as
\begin{equation}
m = - \K \X'_e.
\label{eq:Gc}
\end{equation}
The flow of information in a compensator is shown schematically in
figure \ref{fig:lqg}.

The design of a Linear-Quadratic-Gaussian (LQG) compensator can be
accomplished using standard methods of Linear Control Theory
\citep[see, e.g.,][]{OptC} and is outlined below only briefly.
Assuming that all the stochastic variables are white and Gaussian, the
separation principle can be applied which means that the control and
estimation problems can be solved independently of each other. Based
on the above assumptions, solution of the control problem can be
further simplified by invoking the principle of certainty equivalence
stating that the optimal feedback matrix $\K$ for the stochastic
system \eqref{eq:Xs} with the cost function \eqref{eq:J} is exactly
the same as for the corresponding deterministic system obtained by
setting the stochastic disturbance $w$ to zero. The matrix $\K$ is
then determined via
\begin{equation}
\K = \frac{1}{R} \B^T \P
\label{eq:K}
\end{equation}
in which the $1\times{2}$ matrix $\P$ is a symmetric
positive--definite solution of the algebraic Riccati equation
\begin{equation}
\A^T \P + \P \A + {Q\, \C^T \C} - \frac{1}{R} \P \B \B^T \P = \0.
\label{eq:ric1}
\end{equation}
We note that the feedback matrix $\K$ will depend on the choice of the
output weight $Q$ and the control penalty $R$ in the cost
functional \eqref{eq:J}. The optimal estimator feedback matrix
needed in \eqref{eq:Xea} is given by
\begin{equation}
\L = {\frac{1}{\MM}} \SS \C^T,
\label{eq:L}
\end{equation}
where the matrix $\SS$ is a symmetric positive--definite solution of the algebraic
Riccati equation
\begin{equation}
\A \SS + \SS \A^T + W \G \G^T - {\frac{1}{\MM}} \SS \C^T \C \SS = \0,
\label{eq:ric2}
\end{equation}
in which the disturbance structure is assumed to be
{$\mathbb{E}[w(t)w(\tau)^T] = W \delta(t - \tau)$} and
{$\mathbb{E}[v(t)v(\tau)^T]=\MM\delta(t - \tau)$}.  Thus, the optimal
estimator feedback $\L$ depends on the covariances of the system and
measurement disturbances, $W$ and $\MM$, respectively, and yields an
estimator known as the Kalman filter. For the case of the simple
point-vortex models studied here{,} the algebraic Riccati equations
\eqref{eq:ric1} and \eqref{eq:ric2} can be solved using standard
techniques. As a matter of fact, equation \eqref{eq:ric1} represents a
system of three coupled quadratic equations {and} can be reduced
to a scalar quartic equation that, in {principle}, can be solved in
closed form.  However, the {analytic} expressions obtained are
{prohibitively complicated} and in practice it is much more convenient
to use a numerical solution provided by the control toolbox in {\tt
  Matlab} \citep{matlab:CStoolbox}.

The LQG compensator is an example of an $\HH_2$ controller / estimator
design in which disturbances are assumed Gaussian and uncorrelated
with the state and control.  Robustness of the compensator can be
enhanced by performing an $\HH_{\infty}$ controller / estimator design
where disturbances are allowed to have the worst--case form. In the
present study, however, the point-vortex model has a very
simple structure and robustness can be achieved by hand--tuning the
compensator.  Consequently, we do not pursue the $\HH_{\infty}$
compensator design here and refer the reader to the review paper
of \citet{b01} for a discussion of the utility of the $\HH_{\infty}$
design in the context of flow control problems.

\section{Numerical results: {deterministic setting}}
\label{nres}

In this section we present computational results in which LQG-based
{feedback} stabilization is added to the base flow. The
configurations to be examined were previously introduced in {table}
\ref{tab:cases}, labelled cases \#1--7.

Integration of \eqref{eq:F} was again carried out using Euler's
{explicit} method with a time step of $\mathrm{d}t=0.001$. This
time step was chosen as it is sufficiently small for a wide range of
cases.  Larger time steps are suitable in many situations, {except
  that} when the vortex passes close to a plate a small time step is
required.  In the solution of the estimation problem \eqref{eq:Xe} it
was assumed that the covariances of the plant and measurement
disturbances were given by
\begin{equation}\label{eq:pmd}
  W=1.0, \phantom{...} M=1.0.
\end{equation}
{Unless otherwise stated,} in the solution of the control problem
\eqref{eq:ric1} we chose
\begin{equation}\label{eq:cparam}
 R=1.0, \qquad Q = 1.0.
\end{equation}
Initially, small perturbations of the equilibrium {vortex positions}
will be considered to demonstrate that the control is successful in
each of the chosen cases. Following this, the range {of perturbations
  for} which the control {succeeds in stabilizing the equilibrium will
  be examined} in each case.

\begin{figure}
 \centering
\mbox{
  \subfigure[]{\label{traj0}\includegraphics[width=0.475\textwidth]{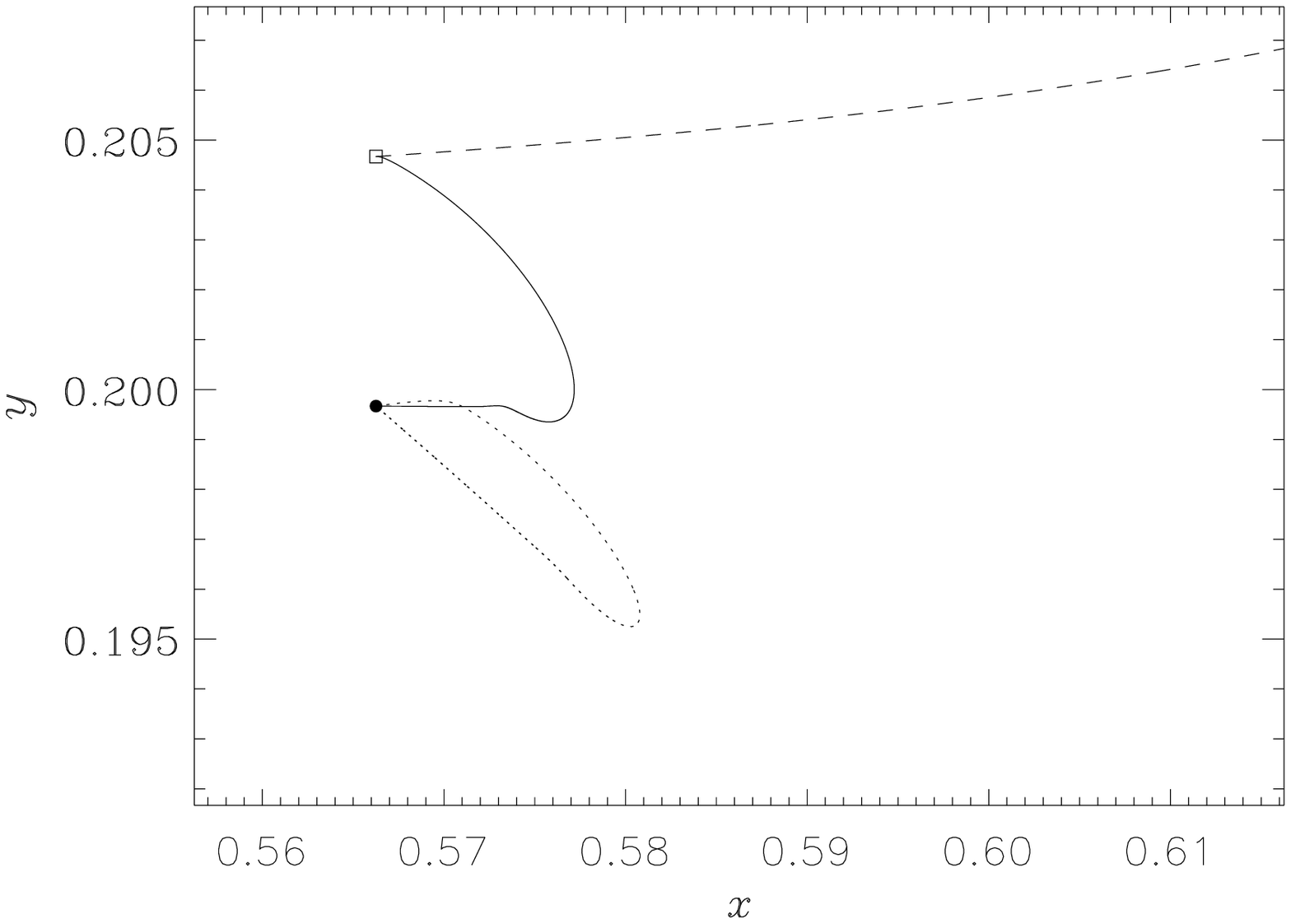}}\qquad
  \subfigure[]{\label{mhist0}\includegraphics[width=0.475\textwidth]{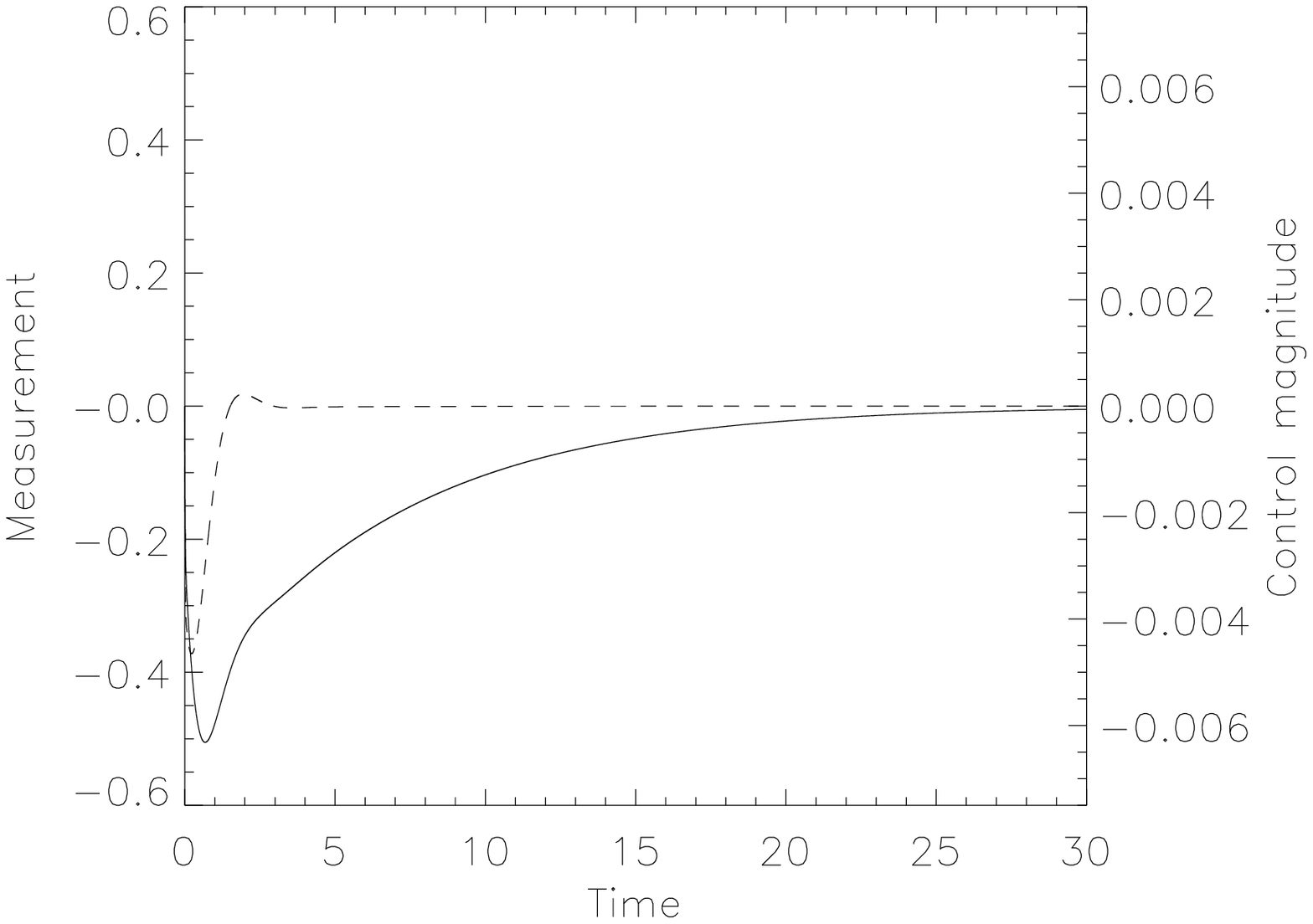}}
}
\mbox{
  \subfigure[]{\label{traj2}\includegraphics[width=0.475\textwidth]{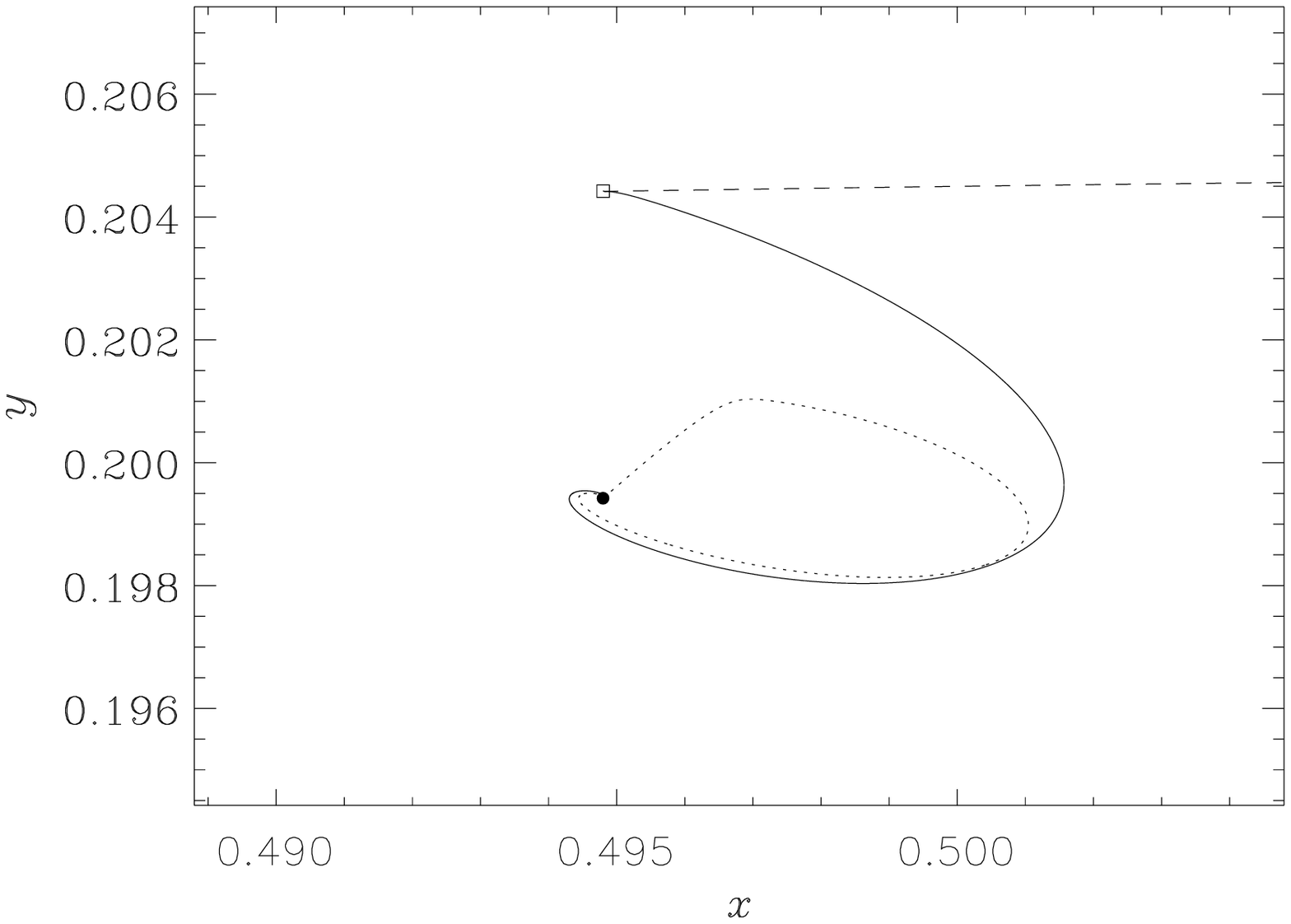}}\qquad
  \subfigure[]{\label{mhist2}\includegraphics[width=0.475\textwidth]{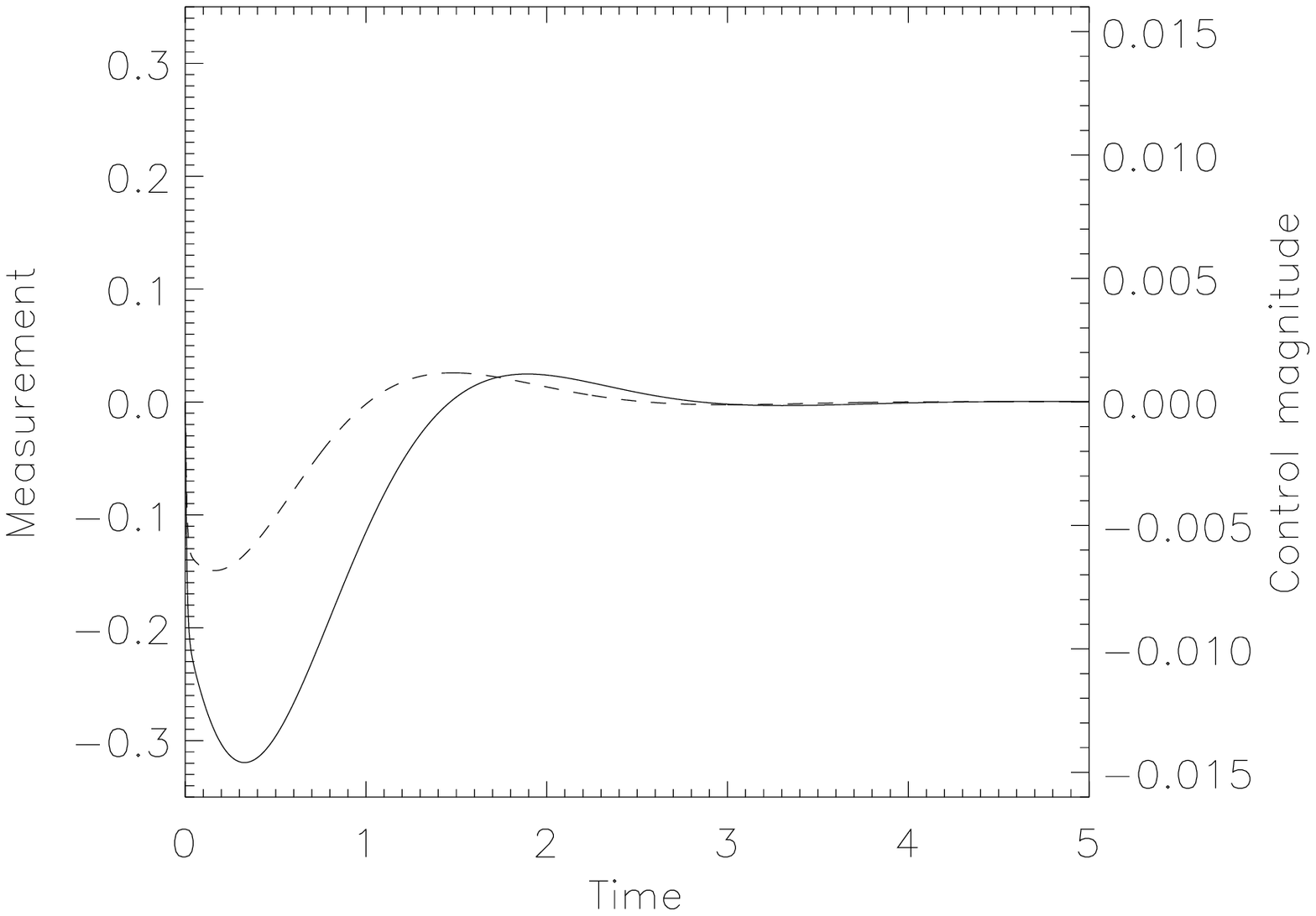}}
}
\mbox{
  \subfigure[]{\label{traj4}\includegraphics[width=0.475\textwidth]{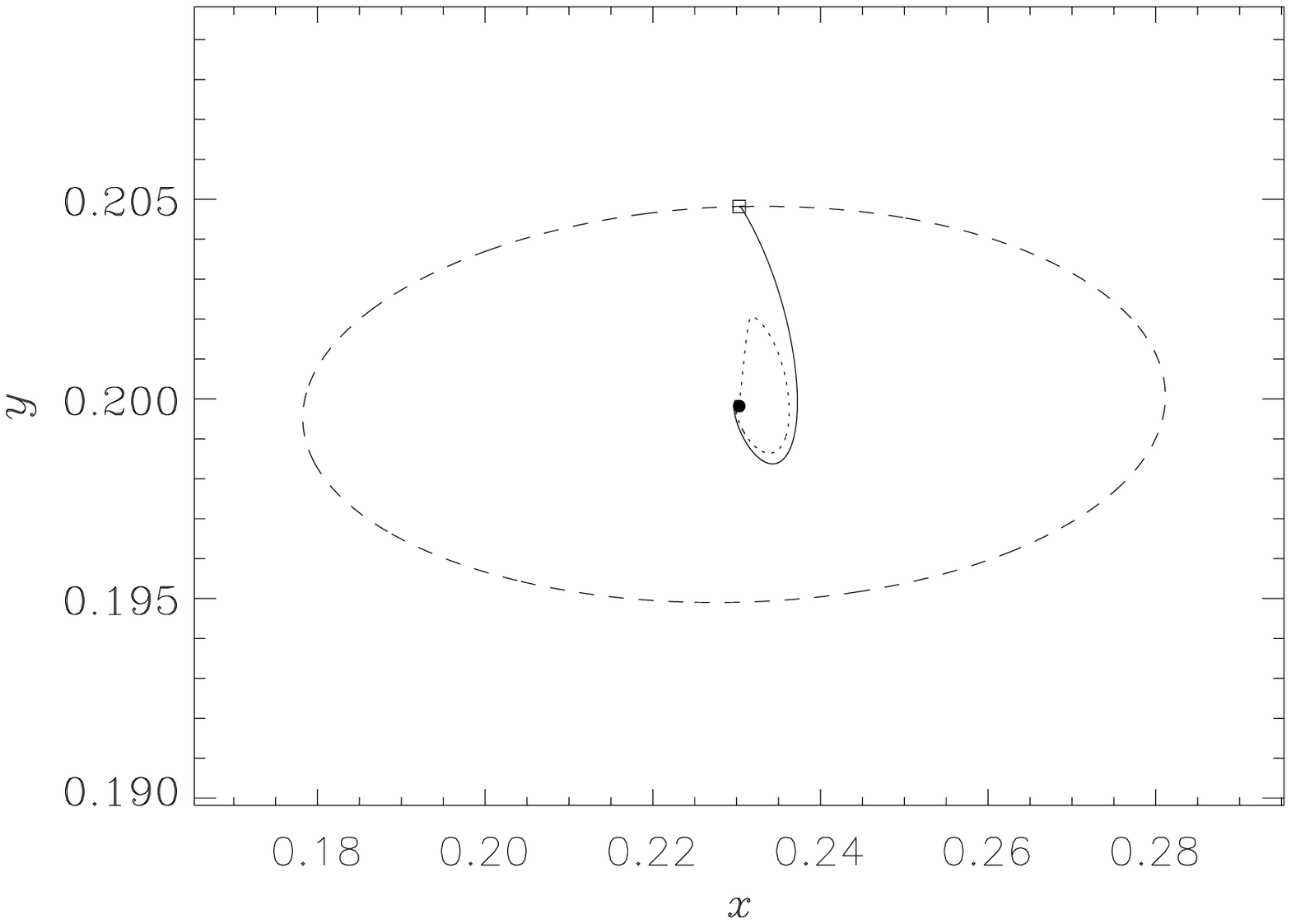}}\qquad
  \subfigure[]{\label{mhist4}\includegraphics[width=0.475\textwidth]{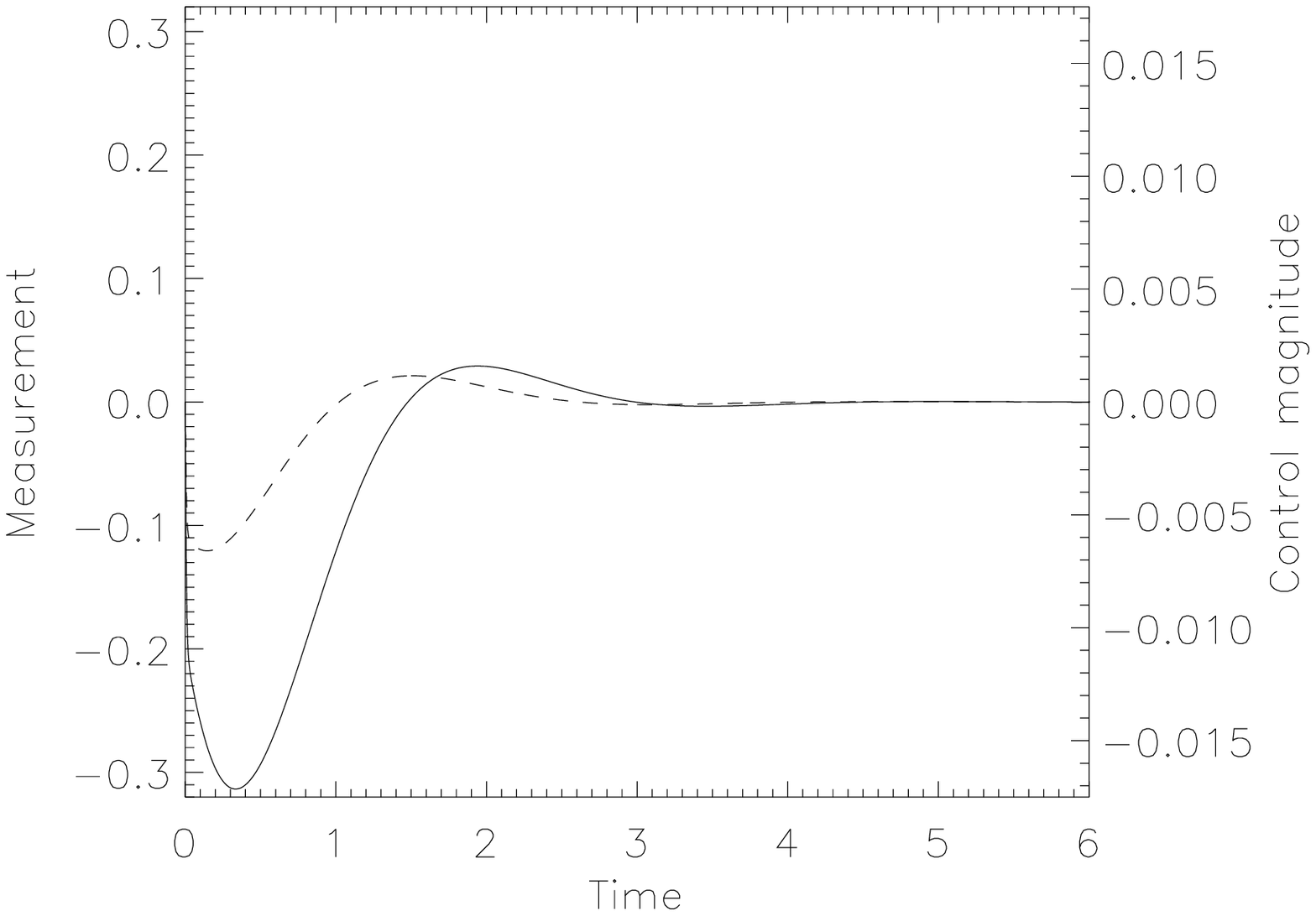}}
}
  \caption{{The first, second and third row correspond to cases \#1,
      \#3 and \#5:} (a), (c) \& (e) The uncontrolled vortex trajectory
    is represented by the dashed curve, the trajectory with LQG-based
    stabilization by the solid curve and the estimator trajectory by
    the dotted curve.  The solid circular symbol represents the
    unperturbed equilibrium position and the square the initial
    perturbed position. (b), (d) \& (f) The corresponding linearised
    measurement {$Y(t)$} (solid curve) and control {intensity $m(t)$}
    (dashed curve).}
  \label{y02-controlled}
\end{figure}

\begin{figure}
 \centering
\mbox{
  \subfigure[]{\label{traj1}\includegraphics[width=0.475\textwidth]{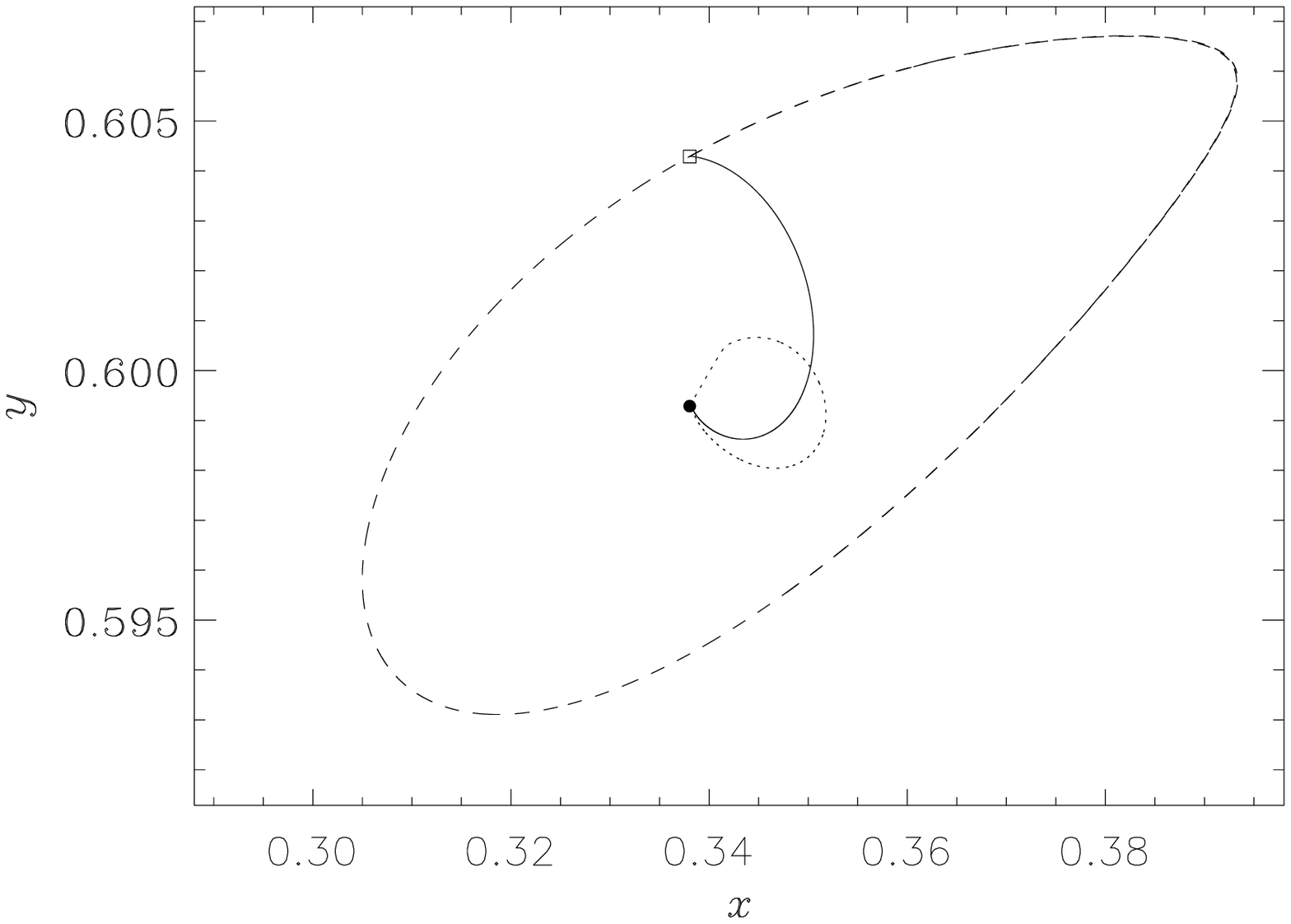}}\qquad
  \subfigure[]{\label{mhist1}\includegraphics[width=0.475\textwidth]{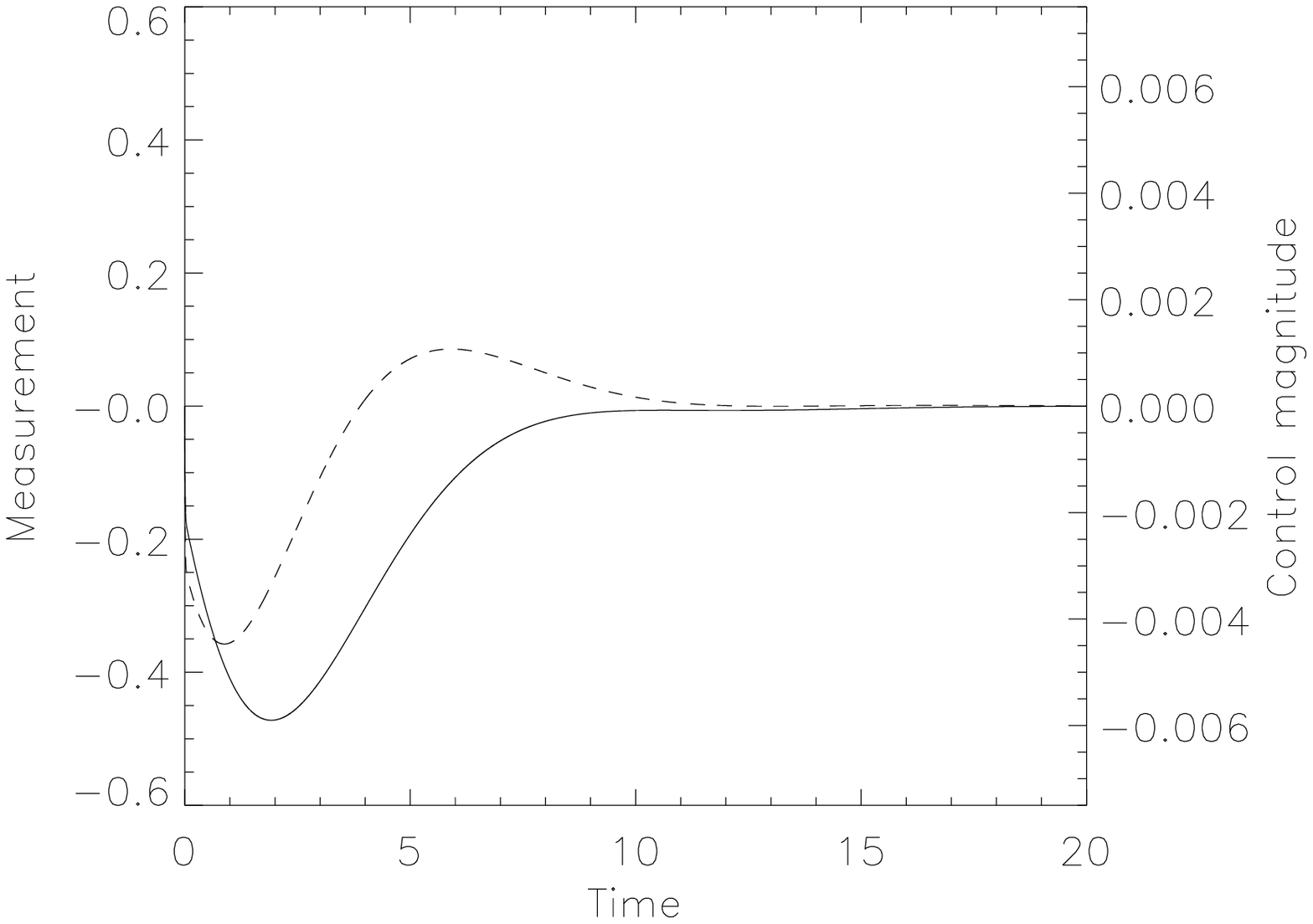}}
}
\mbox{
  \subfigure[]{\label{traj3}\includegraphics[width=0.475\textwidth]{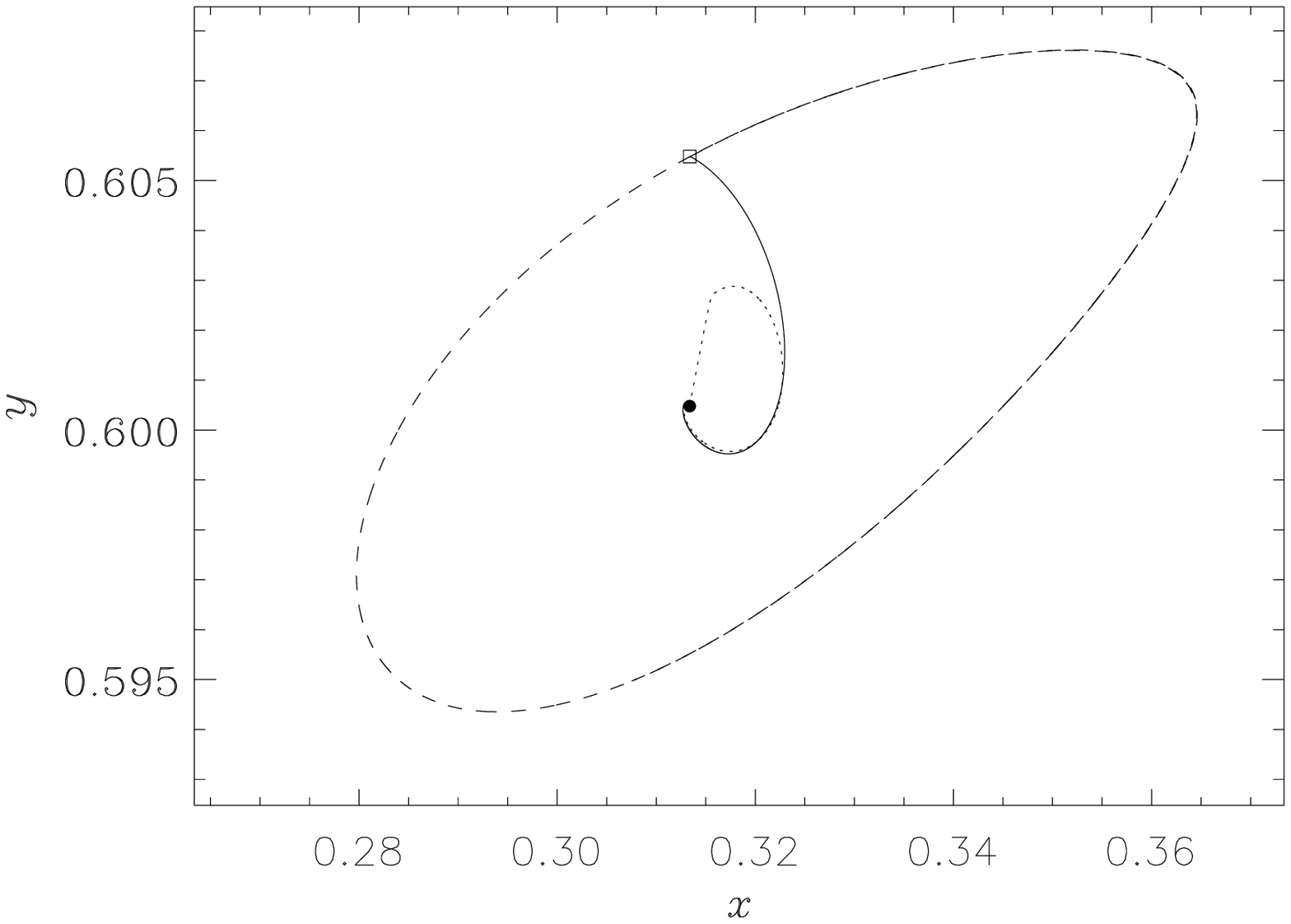}}\qquad
  \subfigure[]{\label{mhist3}\includegraphics[width=0.475\textwidth]{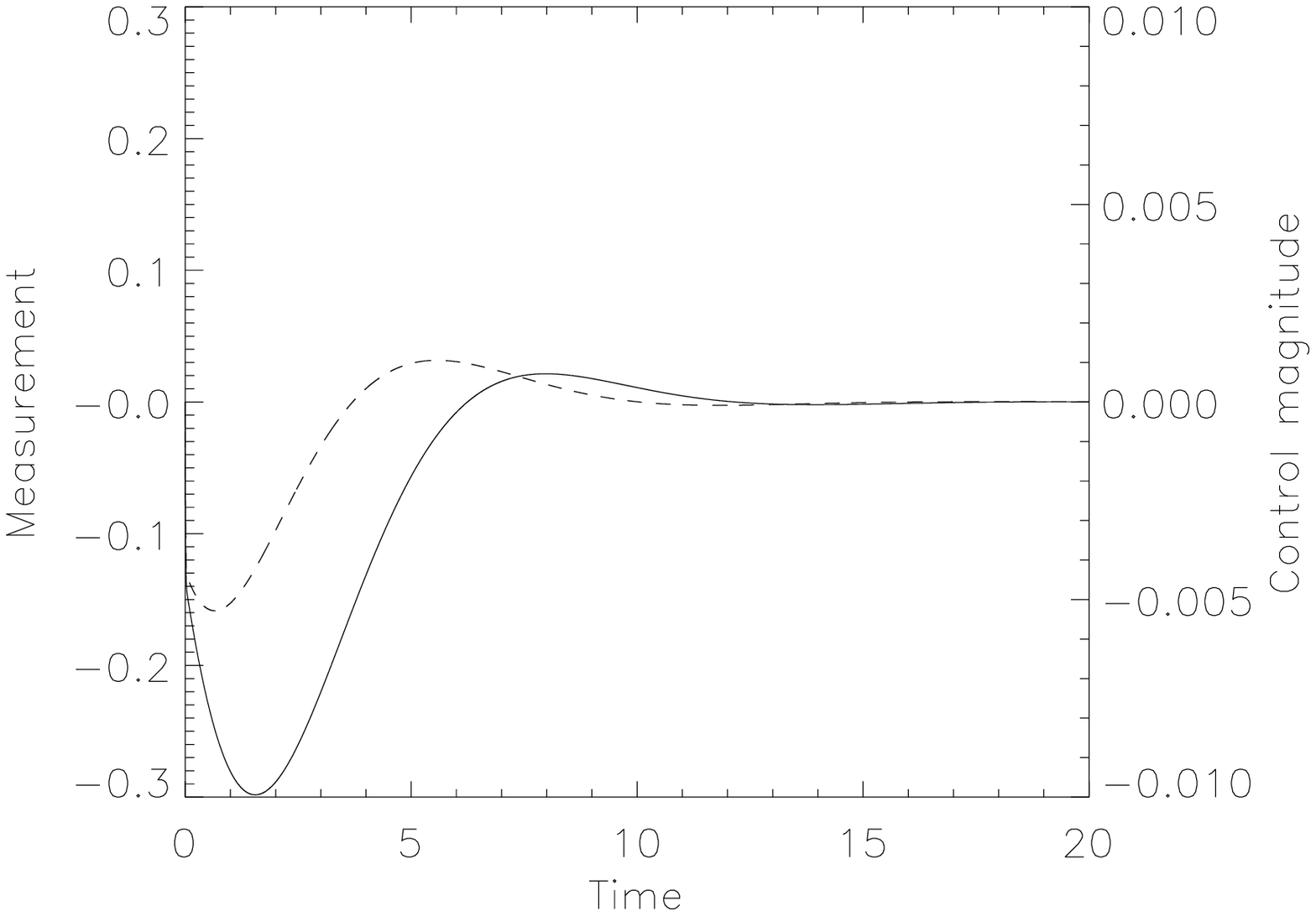}}
}
\mbox{
  \subfigure[]{\label{traj5}\includegraphics[width=0.475\textwidth]{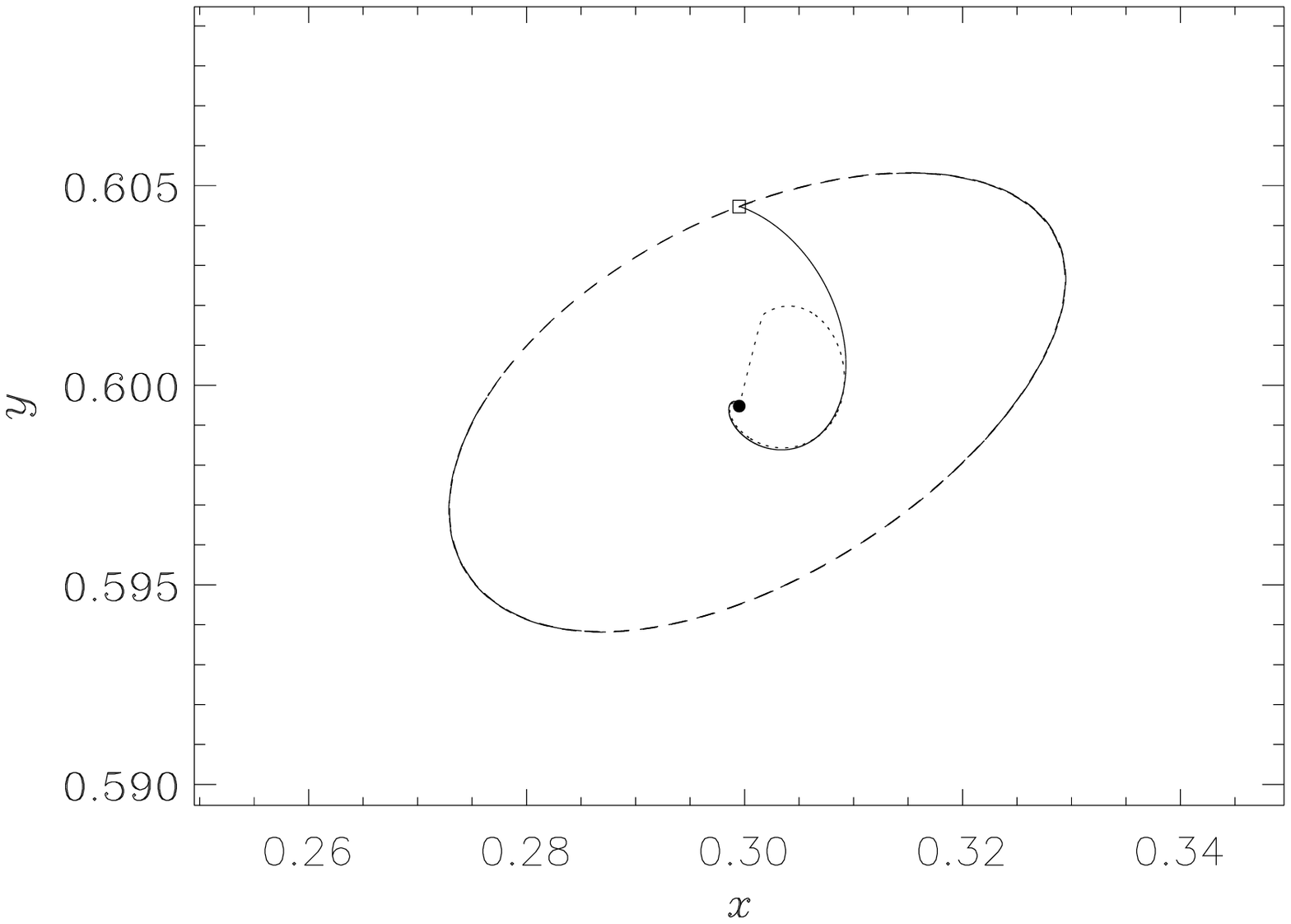}}\qquad
  \subfigure[]{\label{mhist5}\includegraphics[width=0.475\textwidth]{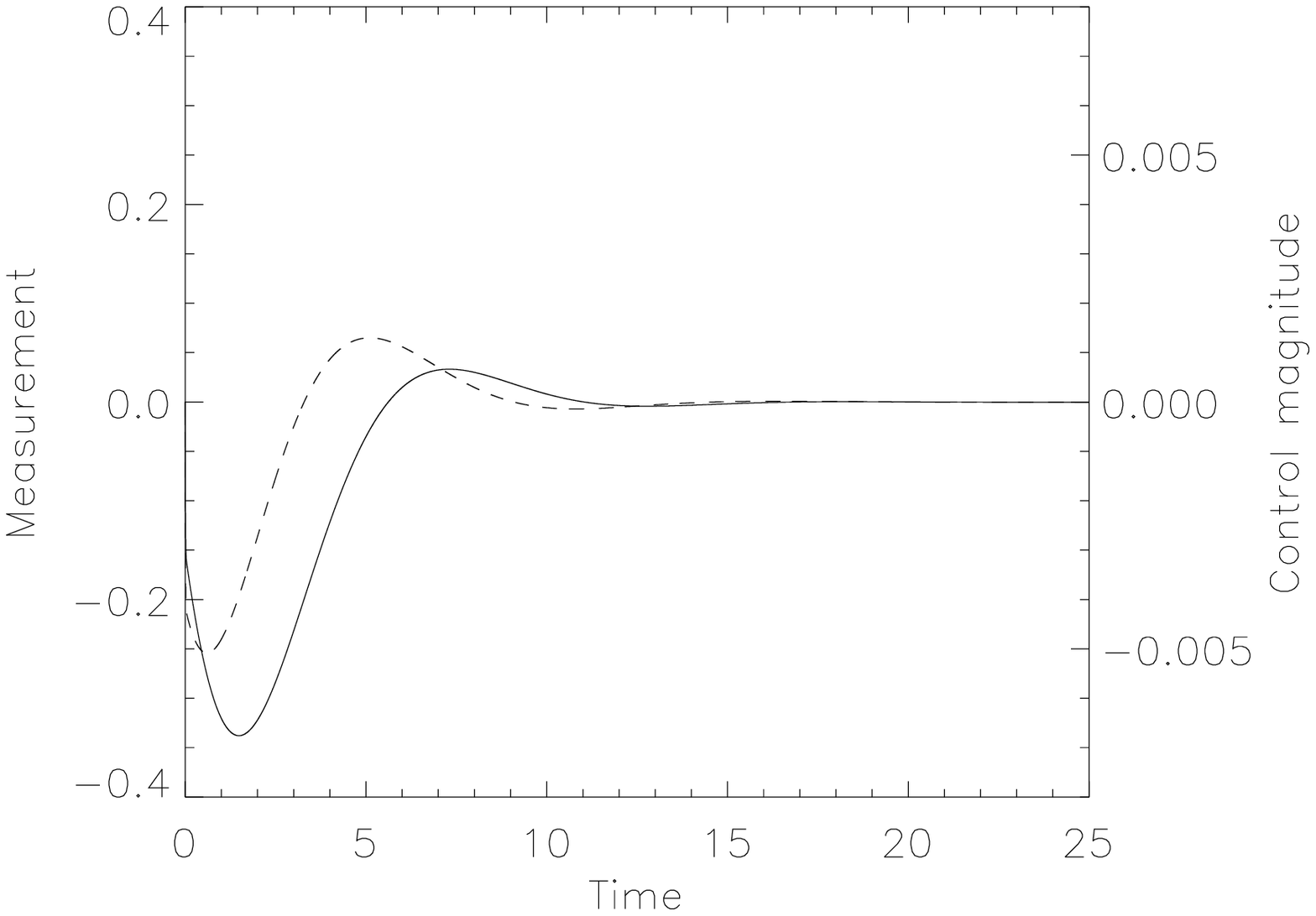}}
}
  \caption{{The first, second and third row correspond to cases \#2,
      \#4 and \#6:} (a), (c) \& (e) The uncontrolled vortex trajectory
    is represented by the dashed curve, the trajectory with LQG-based
    stabilization by the solid curve and the estimator trajectory by
    the dotted curve.  The solid circular symbol represents the
    unperturbed equilibrium position and the square the initial
    perturbed position. (b), (d) \& (f) The corresponding linearised
    measurement {$Y(t)$} (solid curve) and control {intensity $m(t)$}
    (dashed curve).}
  \label{y06-controlled}
\end{figure}

\begin{figure}
 \centering
\mbox{
  \subfigure[]{\label{traj6}\includegraphics[width=0.475\textwidth]{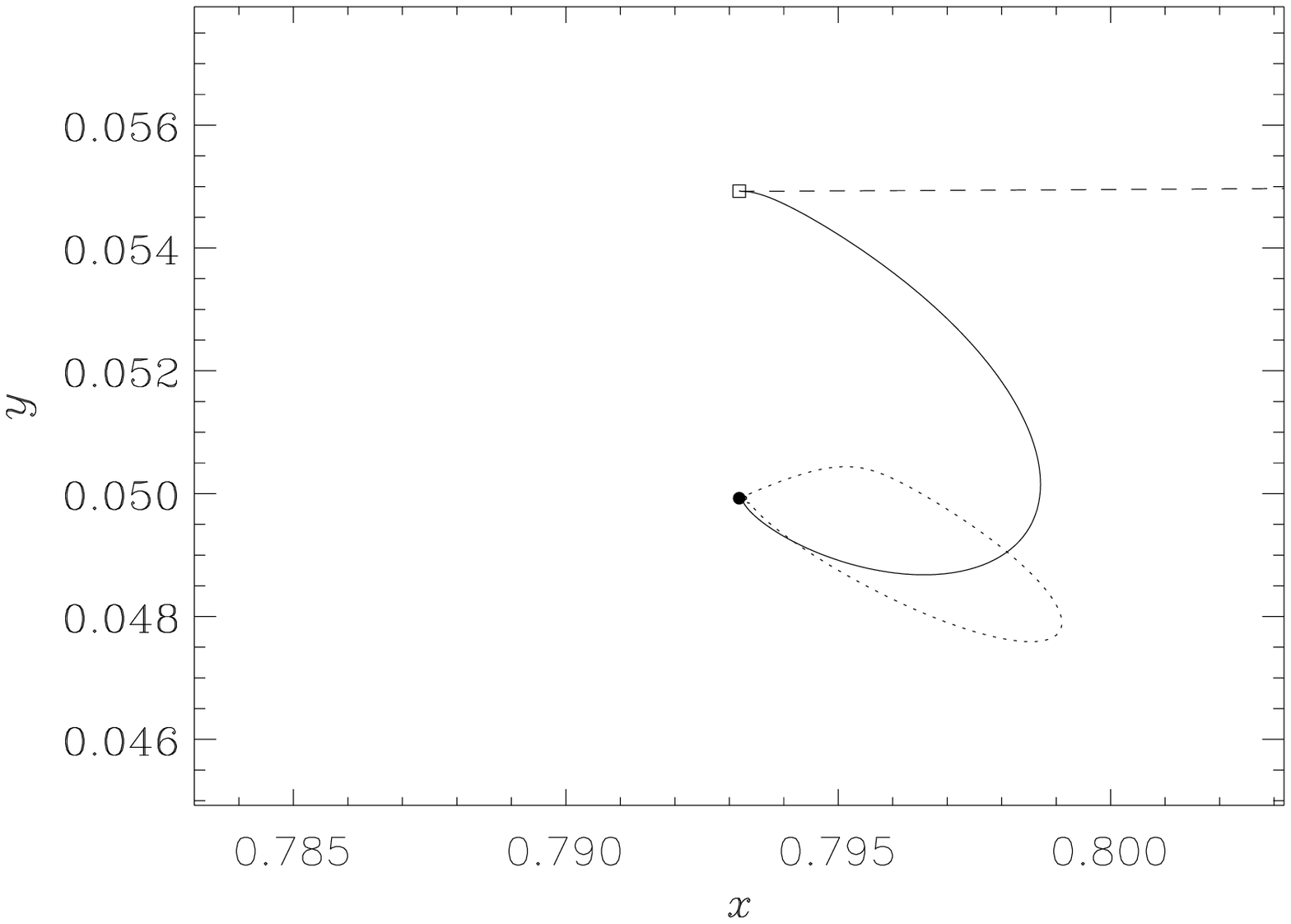}}\qquad
  \subfigure[]{\label{mhist6}\includegraphics[width=0.475\textwidth]{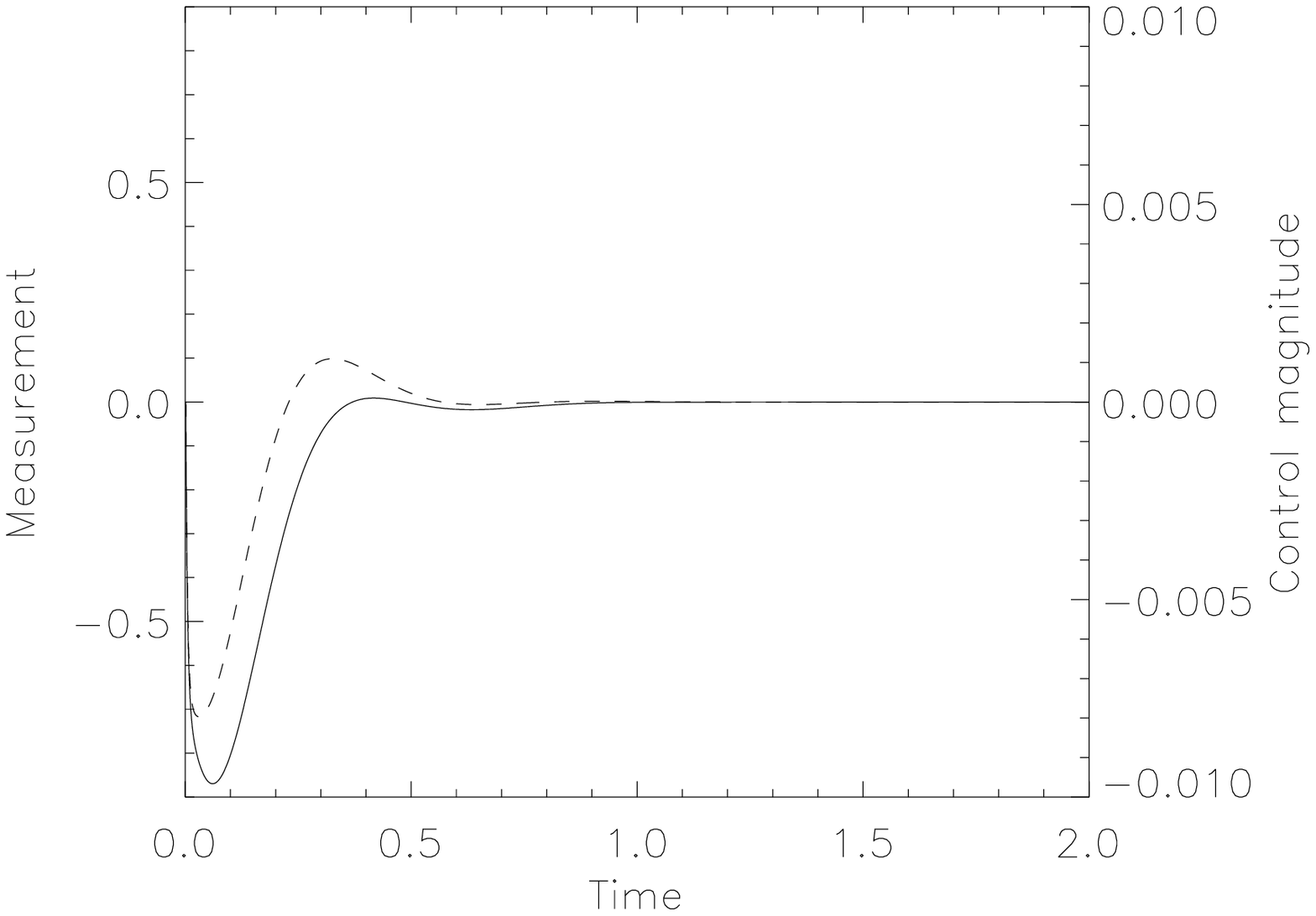}}
}
  \caption{Case \#7: (a) The uncontrolled vortex trajectory is
    represented by the dashed curve, the trajectory with LQG-based
    stabilization by the solid curve and the estimator trajectory by
    the dotted curve.  The solid circular symbol represents the
    unperturbed equilibrium position and the square the initial
    perturbed position. (b) The corresponding measurement
    {$Y(t)$} (solid curve) and control {intensity $m(t)$}
    (dashed curve).}
  \label{kw3-y005-controlled}
\end{figure}

\begin{figure}
 \centering
\mbox{
  \subfigure[]{\label{fig:pevol135}\includegraphics[width=0.475\textwidth]{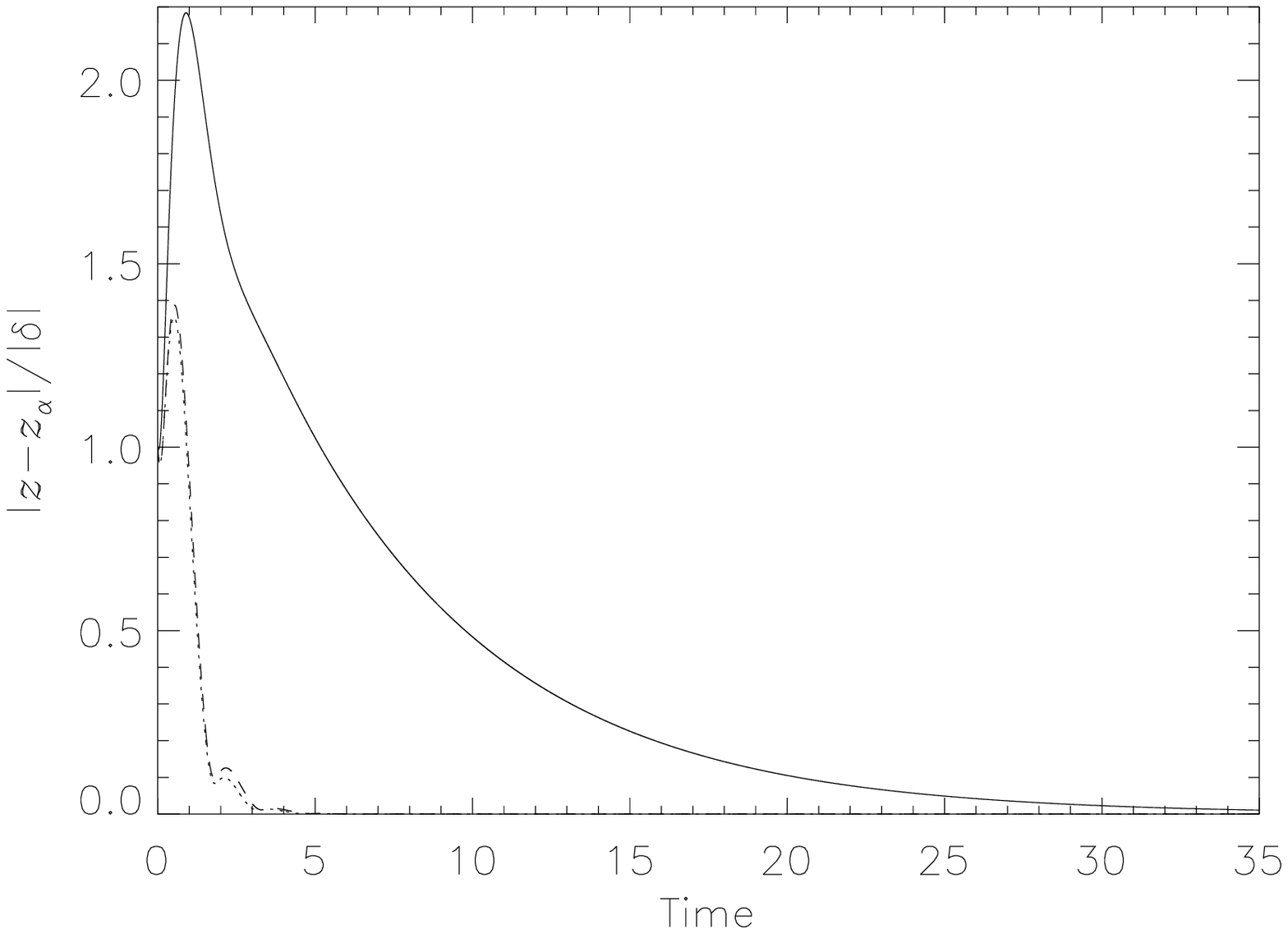}}\qquad
  \subfigure[]{\label{fig:pevol246}\includegraphics[width=0.475\textwidth]{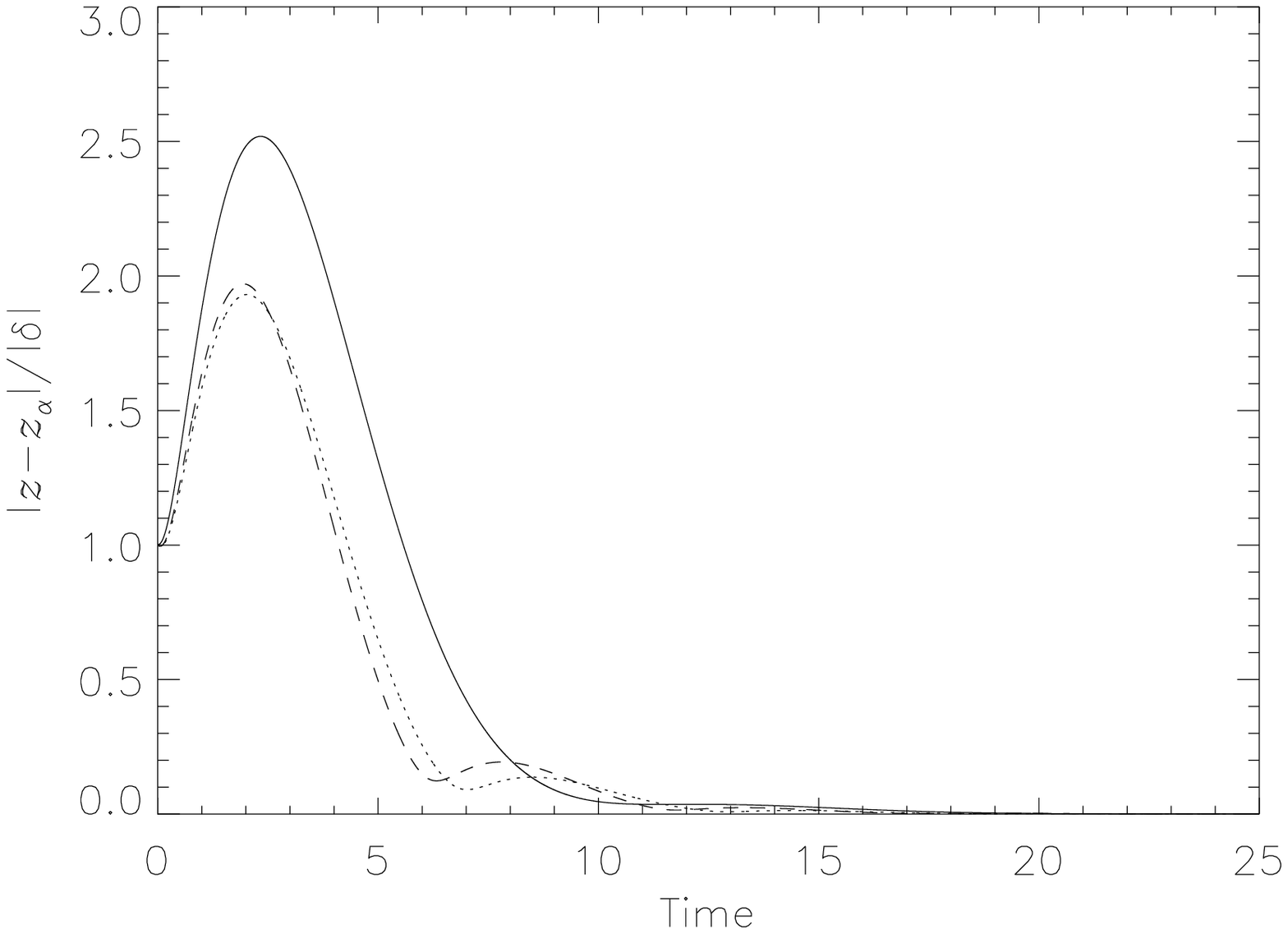}}
}
  \caption{Time evolution of the magnitude of the perturbation for:
    (a) case \#1 (solid curve), case \#3 (dotted curve) and case \#5
    (dashed curve), (b) case \#2 (solid curve), Case \#4 (dotted
    curve) and Case \#6 (dashed curve).}
  \label{fig:pevol}
\end{figure}

Figures \ref{y02-controlled}--\ref{kw3-y005-controlled} {correspond
  to} cases \#1--7 in which the vortex is {initially} perturbed by
$\delta=0.005i$ {away} from its equilibrium position $z_{\alpha}$ (see
{table} \ref{tab:cases}). For these small perturbations, and with LQG
stabilization added, the trajectories in all cases are stabilized and
the vortex position approaches $z_{\alpha}$ as $t\rightarrow\infty$.
However, the time frame over which stabilization occurs, {evident
  in} the time-varying control {intensity $m(t)$}, differs between
cases.  {As regards case \#3, we note that the trajectory in the
  uncontrolled configuration diverges (figure
  \ref{y02-controlled}(c)), even though the equilibrium is neutrally
  stable, cf.~table \ref{tab:cases}. This is because in this case the
  magnitude of the initial perturbation $\delta$ is already outside
  the range of validity of linearization.}  From a practical point of
view, it is pertinent to compare cases in which the vortex
{equilibrium is at} the same {elevation} above the plate, i.e., cases
\#1, \#3 and \#5, and cases \#2, \#4 and \#6.  Letting the current
vortex position be ${z(t;\delta)}$, figures \ref{fig:pevol}(a) and
\ref{fig:pevol}(b) show the time evolution of the magnitude of the
{normalized} perturbation (given by
$|{z(t;\delta)}-z_{\alpha}|/|\delta|$).  Figure \ref{fig:pevol}(a)
shows clearly that the two neutrally stable configurations (cases \#3
and \#5) are stabilized far quicker than the unstable configuration
(case \#1). In the neutrally stable cases \#2, \#4 and \#6, shown in
figure \ref{fig:pevol}(b), stabilization occurs over a similar
time-frame but it is notable that in the single-plate configuration
(case \#2) the perturbation {exhibits a far greater transient growth}
prior to successful stabilization. It is also noted that an initial
``kick'' is seen in the estimator trajectories {and} is
particularly evident in the trajectories shown in figure
\ref{y06-controlled} when the vortex has a larger circulation.  This
kick is due to the initial perturbation of the vortex {location} which
is not accounted for in the initial condition of the estimator
{($\X'_e(0) = \mathbf{0}$ in \eqref{eq:Xea})}.

\begin{figure}
 \centering
\mbox{
  \subfigure[case \#1]{\label{basin-m0-y02}\includegraphics[width=0.475\textwidth]{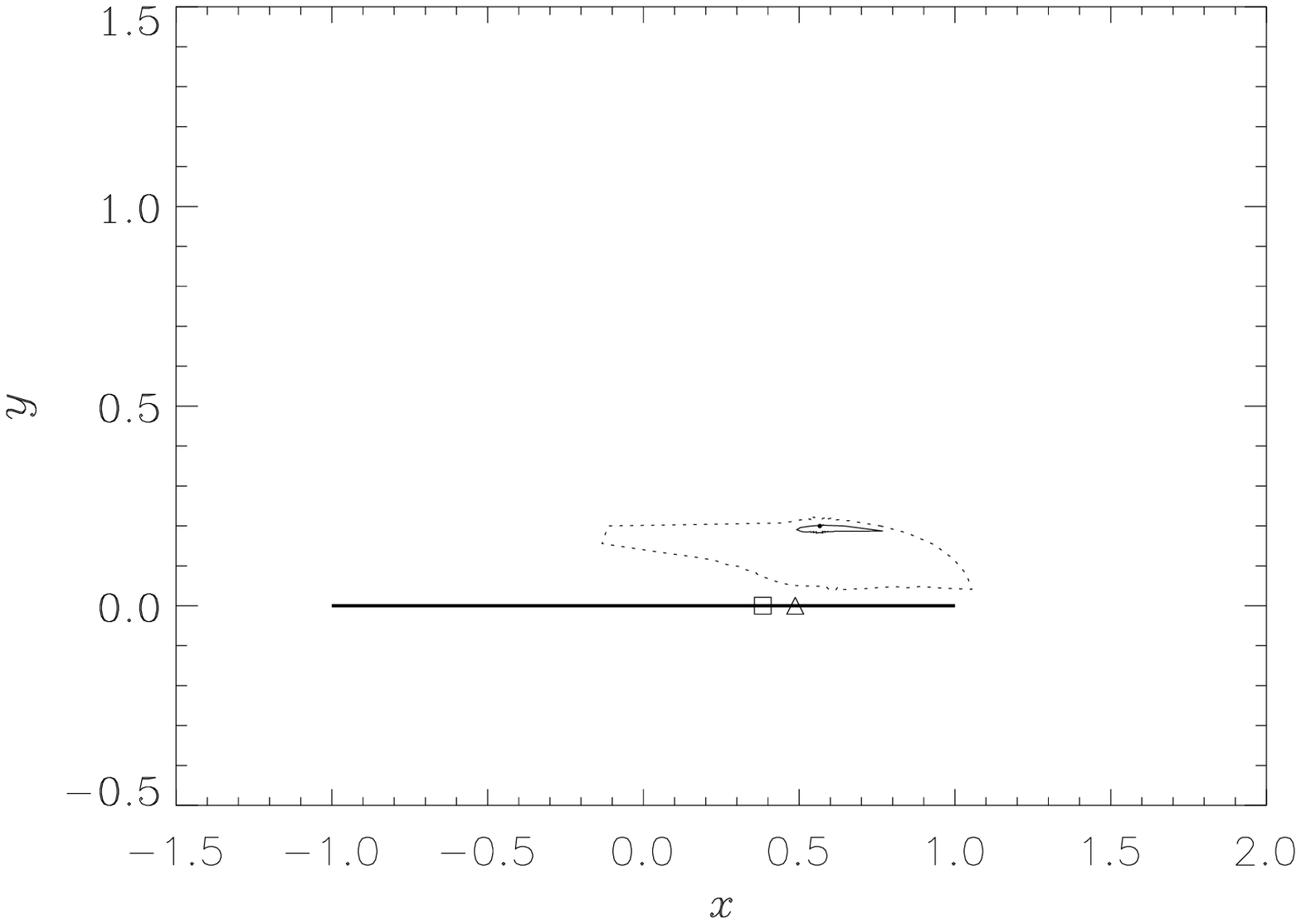}}\qquad
  \subfigure[case \#2]{\label{basin-m0-y06}\includegraphics[width=0.475\textwidth]{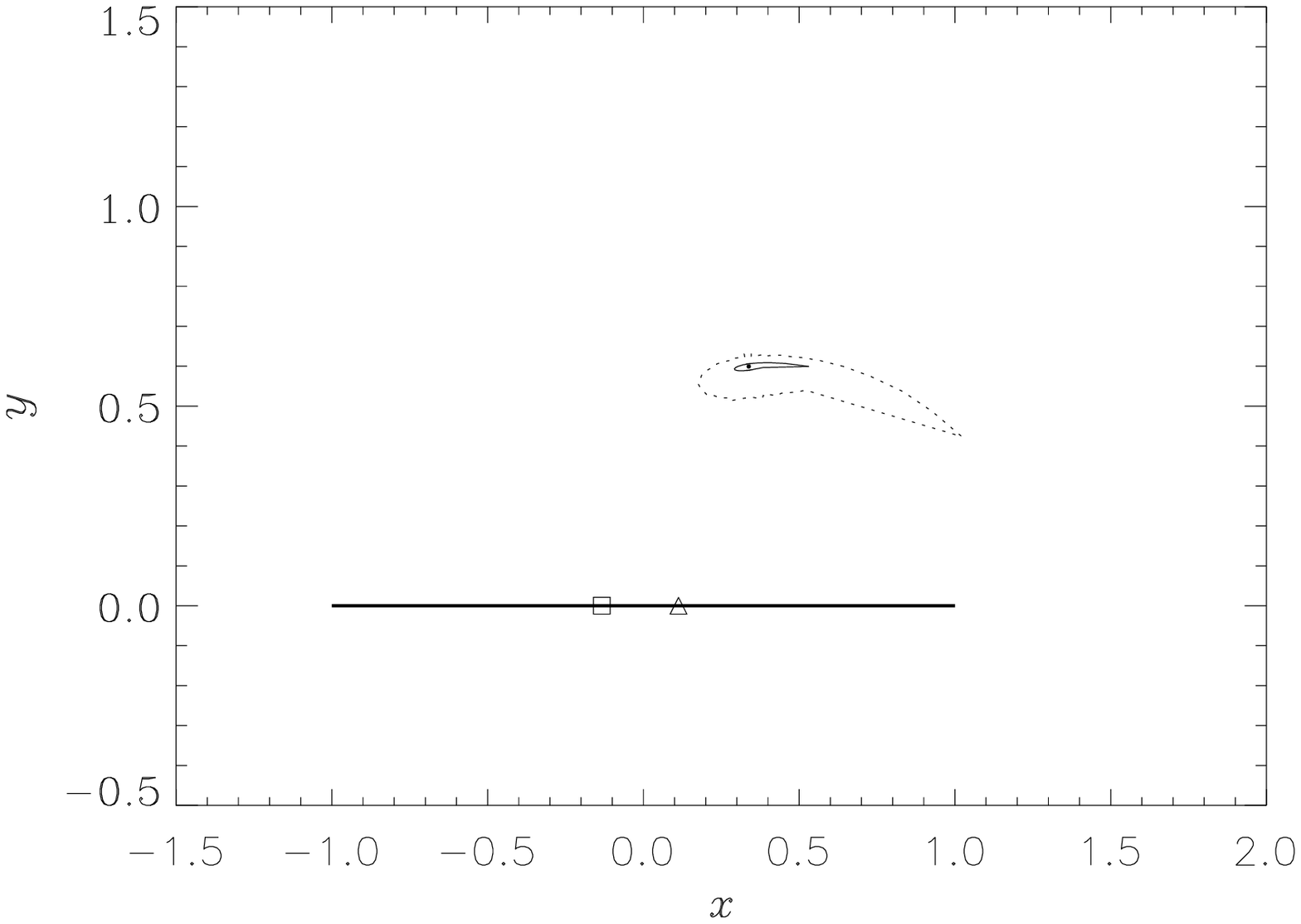}}
}
\mbox{
  \subfigure[case \#3]{\label{basin-kw1-y02}\includegraphics[width=0.475\textwidth]{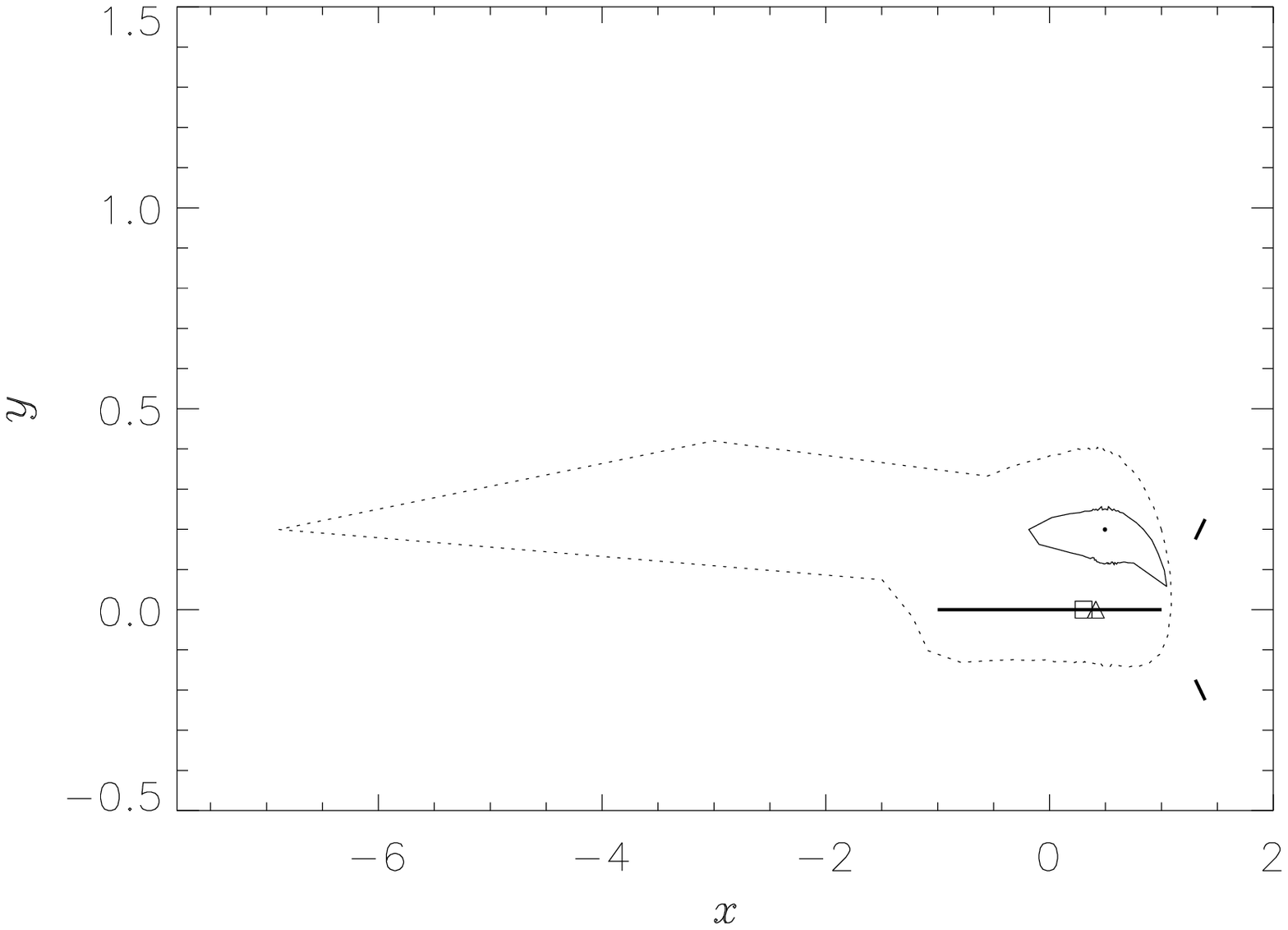}}\qquad
  \subfigure[case \#4]{\label{basin-kw1-y06}\includegraphics[width=0.475\textwidth]{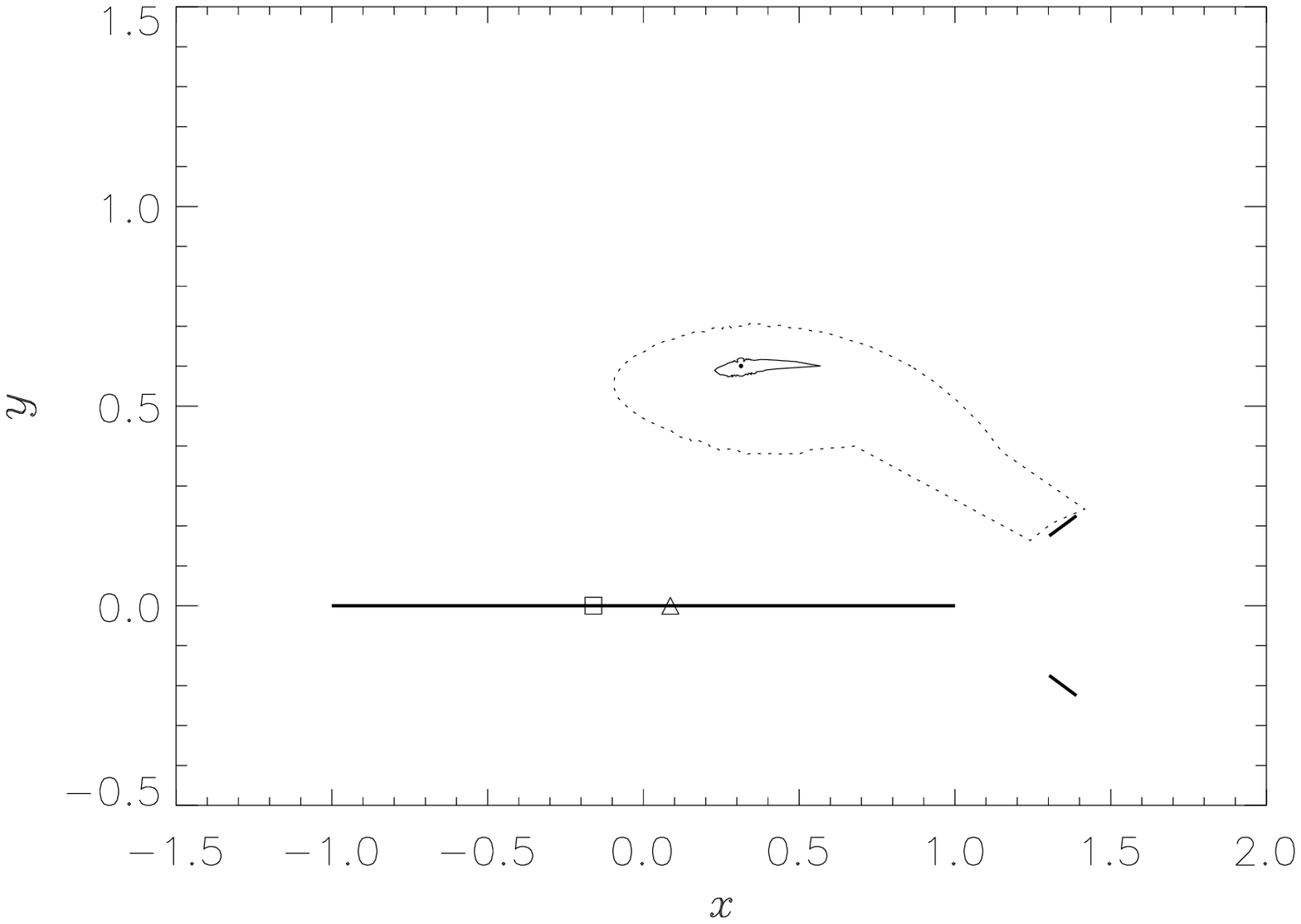}}
}
\mbox{
  \subfigure[case \#5]{\label{basin-kw2-y02}\includegraphics[width=0.475\textwidth]{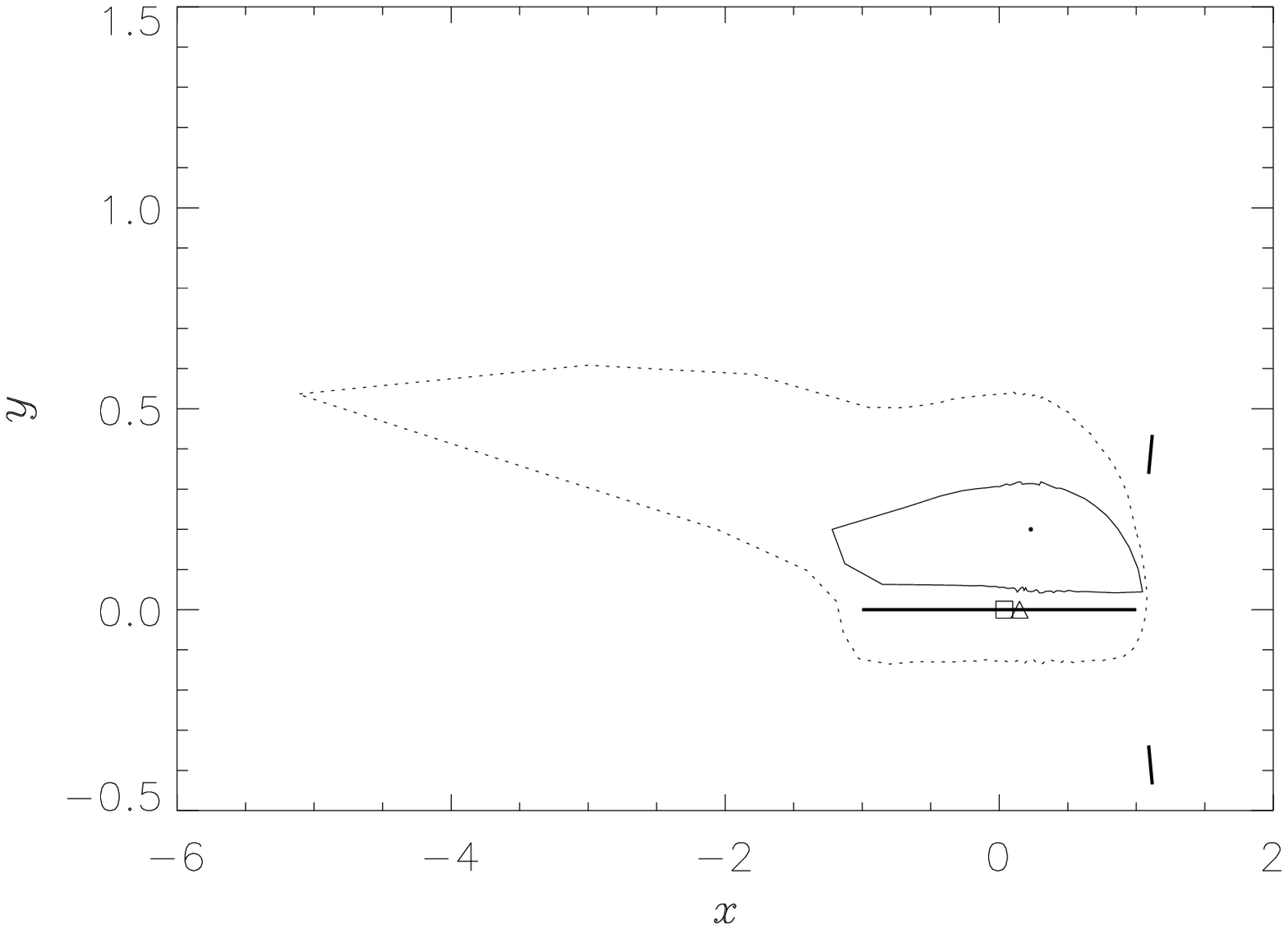}}\qquad
  \subfigure[case \#6]{\label{basin-kw2-y06}\includegraphics[width=0.475\textwidth]{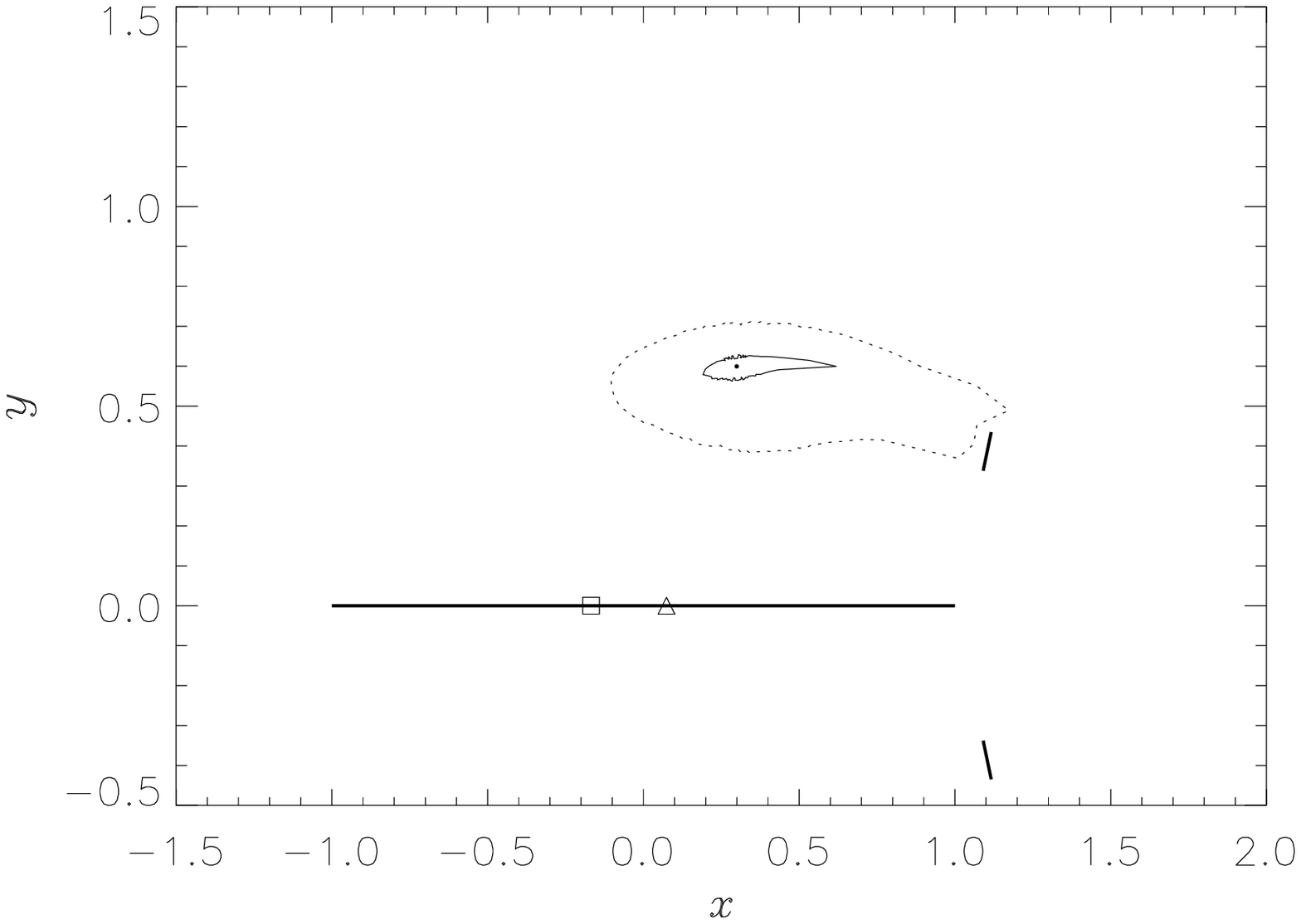}}
}
  \caption{(a)--(f) Basins of attraction for cases \#1--6
    respectively. Thick solid lines represent the plates, dotted
    curves the basins when $R=1$ and solid curves the basins when
    $R=100$ {(cf.~equation \eqref{eq:J})}. The actuator and sensor
    locations are indicated by the triangle and square symbols respectively.}
  \label{basins}
\end{figure}

\begin{figure}
 \centering
\mbox{
  \subfigure[]{\label{traj7}\includegraphics[width=0.475\textwidth]{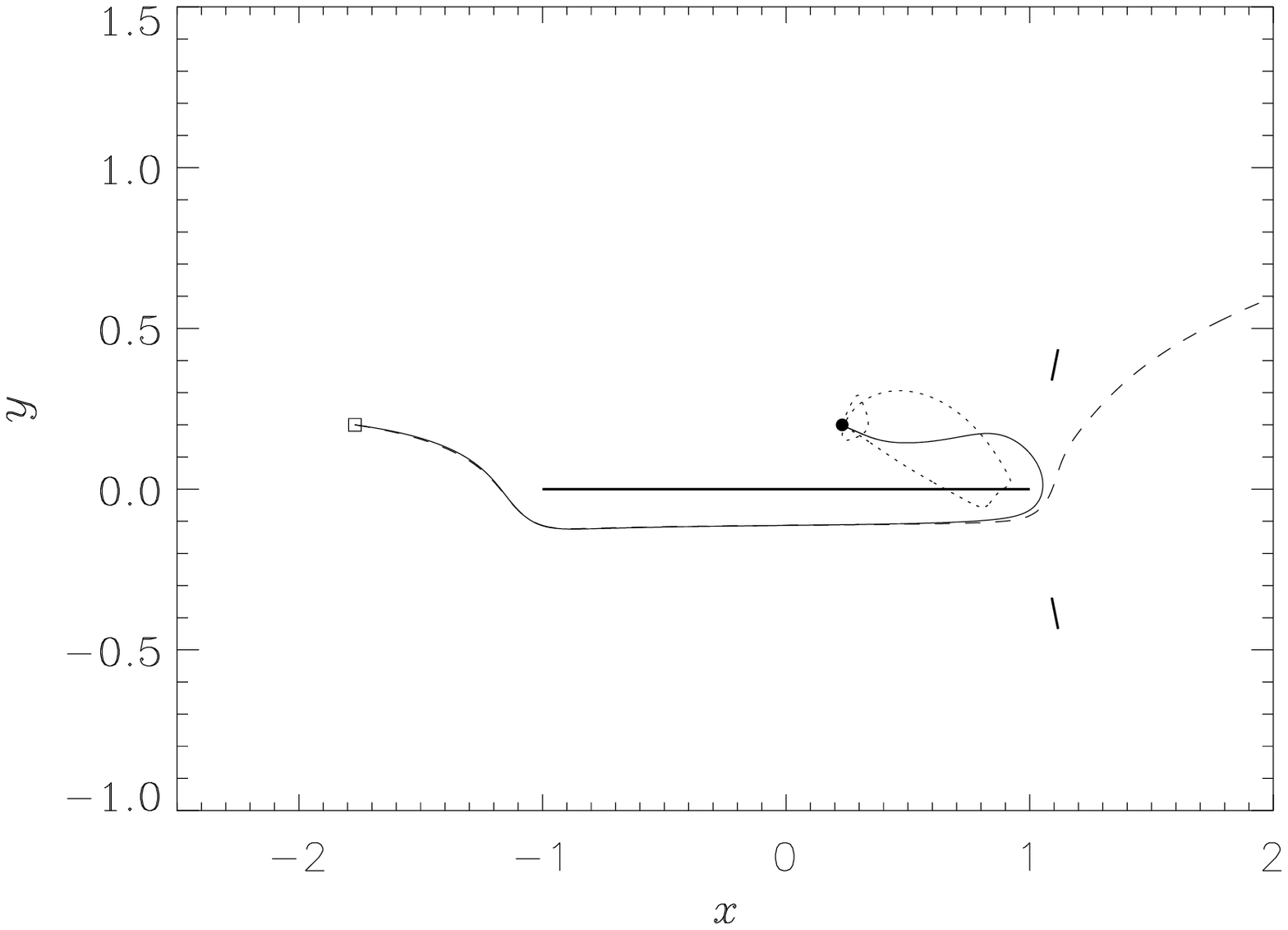}}\qquad
  \subfigure[]{\label{mhist7}\includegraphics[width=0.475\textwidth]{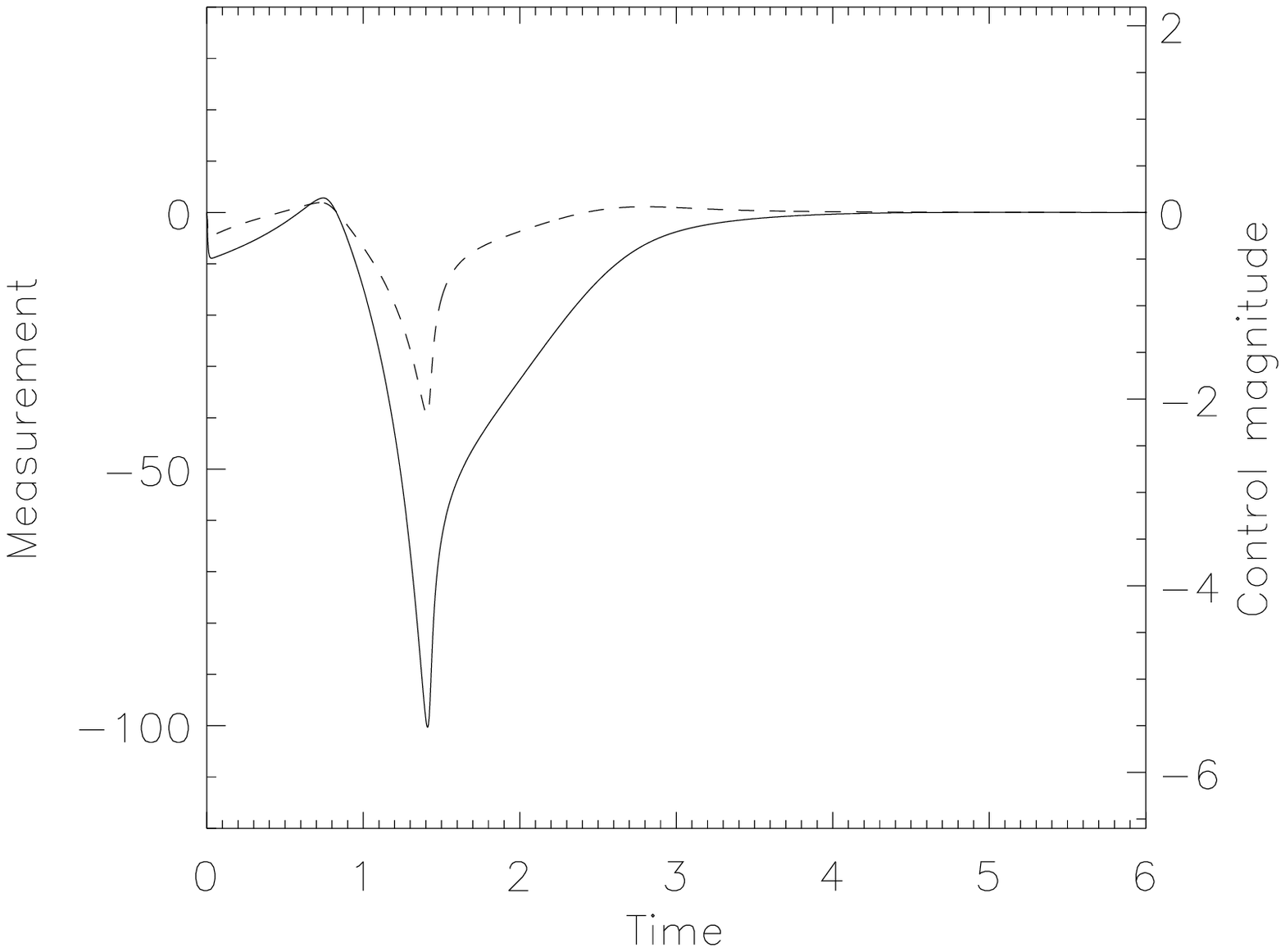}}
}
  \caption{Case \#5 with an initial perturbation of $\delta=-2.0$.
    (a) The vortex trajectory with LQG-based stabilization is
    represented by the solid curve, the estimator trajectory by the
    dotted curve and the uncontrolled trajectory by the dashed curve.
    The solid circular symbol represents the unperturbed equilibrium
    position and the square the initial perturbed position. (b) The
    corresponding measurement {$Y(t)$} (solid curve) and control
    {intensity $m(t)$} (dashed curve).}
  \label{kw2-y02-controlled2}
\end{figure}

To explore the {range of perturbations for} which LQG stabilization is
successful in each case, ``basins of attraction'' are computed.
{{They} represent the sets of initial perturbations which can be
  stabilized by the LQG compensator; trajectories corresponding to
  perturbations lying outside these basins escape to infinity}.  Each
basin is computed by discretizing perturbations $\delta \in\mathbb{C}$
about $z_{\alpha}$ such that
\begin{equation}
{\delta_{j}=r_{j}e^{2\pi{i}\frac{j}{J}}, \phantom{...} j=0,\dots,J-1,}
\end{equation}
where {{$r_{j}\in\mathbb{R}$} and $J=100$. The algorithm then
  calculates, for each $j$, the largest value {of} {$r_j$} {(to within
    an accuracy of 0.01)} such
  that{$|{z(t;\delta_j)}-z_{\alpha}|\rightarrow{0}$}} as
$t\rightarrow\infty$. Additionally, to demonstrate the effect of
changing the cost of control (see equation \eqref{eq:K}), basins are
computed for both $R=1$ and $R=100$. When $R=1$ control is ``cheap''
and {can be} used liberally. However, when the cost is increased to
$R=100$, control {is} ``expensive'' and {must be} used sparingly.
{Therefore, for {larger} values of $R$, the basins of attraction
  are expected to shrink. The basins computed} for cases \#1--6 are
shown in figure \ref{basins}. When comparing cases \#1, \#3 and \#5,
the basins of the two Kasper Wing configurations (for both values of
$R$) are significantly larger than those of the single-plate
configuration. Indeed, in cases \#3 and \#5 when $R=1$, due to the
trajectories the point vortex follows past the plates, the control is
capable of stabilizing some extreme perturbations. For the equilibria further from the plate,
basins of the Kasper Wing cases \#4 and \#6 are still noticeably
larger than those of the single-plate case \#2, but now to a lesser
extent.  It should also be pointed out that in cases \#3 and \#5 the
basin ``engulfs'' the main plate.  However, as the vortex approaches
the plate boundary, {the expression in} \eqref{fc1} becomes singular
and the {numerical computations} are no longer robust. Perturbations
placing the vortex very close to {one of the plates} cannot therefore
be claimed to be part of a basin of attraction.  Finally in this
section, figure \ref{kw2-y02-controlled2} presents a vortex trajectory
{together with the corresponding time-histories of the linearised
  measurements $Y(t)$ and control intensity $m(t)$} for a large
perturbation in case \#5.  This example shows {one of} the more
``exotic'' trajectories the LQG control is capable of stabilizing in
which the vortex {moves away from the equilibrium and} passes
under the plate prior to the control latching onto it and ``pulling''
it towards the equilibrium.

\section{Numerical results: {effects of disturbances}}
\label{stochastic}

In this section we examine how the LQG control performs {in the
  presence of} additional {disturbances affecting} the flow. Two forms
of {disturbances} are independently considered: the first of these
will be random {disturbances added} to the angle of attack ${\chi_0}$
{of the oncoming flow and then} a vortex shedding model will be
introduced {ensuring} that the Kutta conditions (see equations
\eqref{stabeq1a}--\eqref{stabeq1c}) are satisfied at discrete
{instances of} time throughout a simulation.  {Needless to say,
  the phenomenon of vortex shedding is not accounted for in the flow
  model (cf.~\S\ref{uc}) and hence may be interpreted as ``system
  uncertainty''. Therefore, even though it is deterministic in nature,
  it may be represented by the term proportional to $w$ in equation
  \eqref{eq:Xsa}.  Details concerning the two forms of disturbances}
are discussed in their respective subsections below.

\subsection{{Control in the presence of random disturbances of the
    angle of attack}}
\label{sec:random}

To check how the control performs in a fluctuating background flow,
the flow angle of attack {$\chi_0$} is periodically augmented
with a {{Gaussian random variable},  so that we obtain
\begin{equation}\label{eq:dchi}
\chi(t) = \chi_0 + \Delta \chi\left(\Big\lfloor \frac{t}{\Delta t}\Big\rfloor\right),
\end{equation}
where $\Delta t = 0.01$, $\lfloor \cdot \rfloor$ denotes the integer
part and $\Delta \chi(l)$ is the $l$-th sample of the random variable
with distribution $\mathcal{N}(\mu,\sigma^{2})$ {for some
  $\sigma, \mu \in \mathbb{R}$}. In other words, the stochastic
disturbance is frozen over the time window of length $\Delta t$ before
a new sample is drawn. We are interested here in a single realization
of the stochastic process, rather than in any statistic quantities, so
once the random variable has been sampled, the closed-loop system
\eqref{eq:F} with \eqref{eq:dchi} can be integrated as a deterministic
system. We do so with the approach introduced in \S\ref{nres}, i.e.,
Euler's explicit method with the time step {$\mbox{d}t = 0.001$.}}

\begin{figure}
 \centering
\mbox{
  \subfigure[]{\label{c1-skai0-traj}\includegraphics[width=0.475\textwidth]{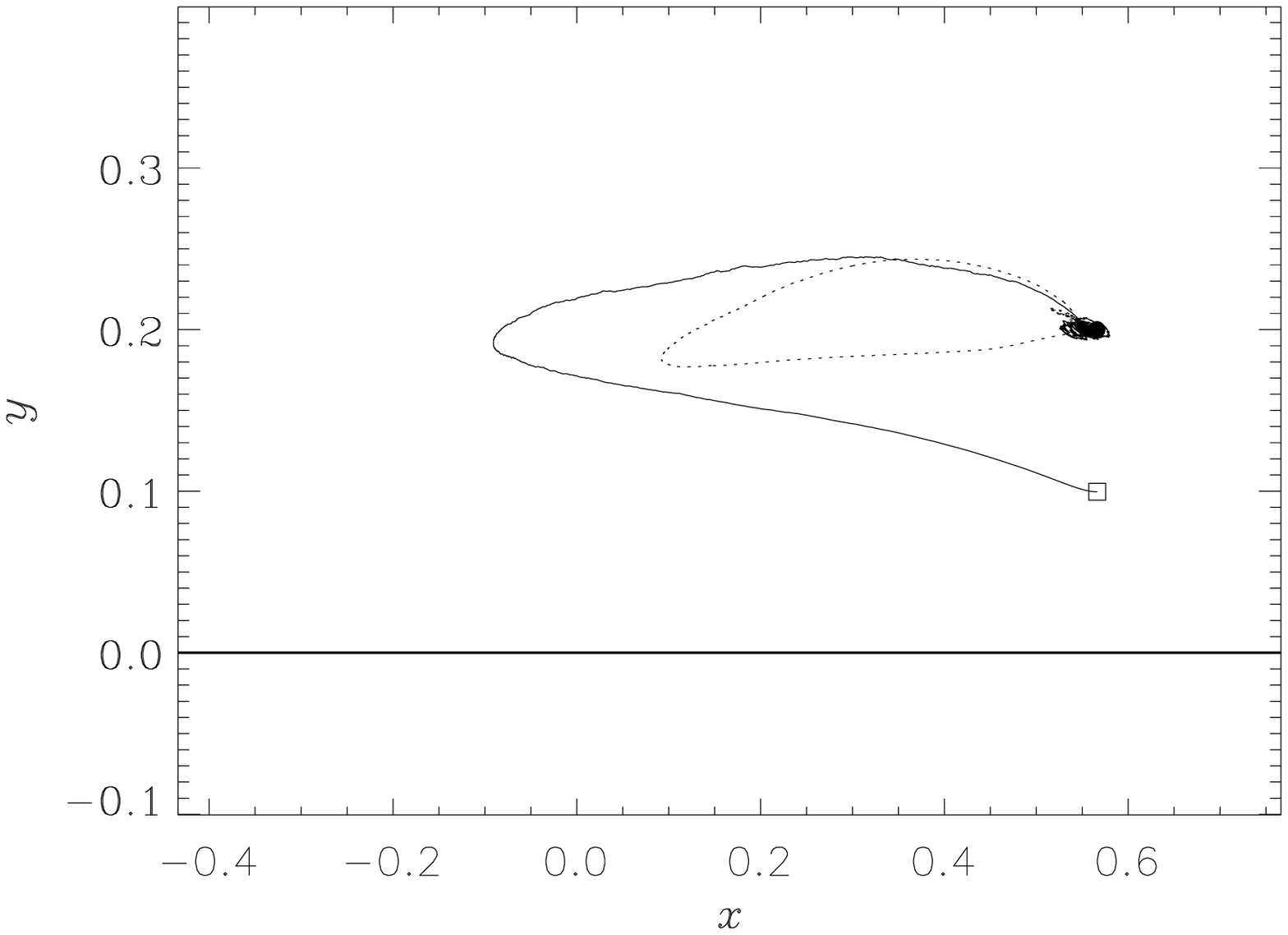}}\qquad
  \subfigure[]{\label{c1-skai0-mhist}\includegraphics[width=0.475\textwidth]{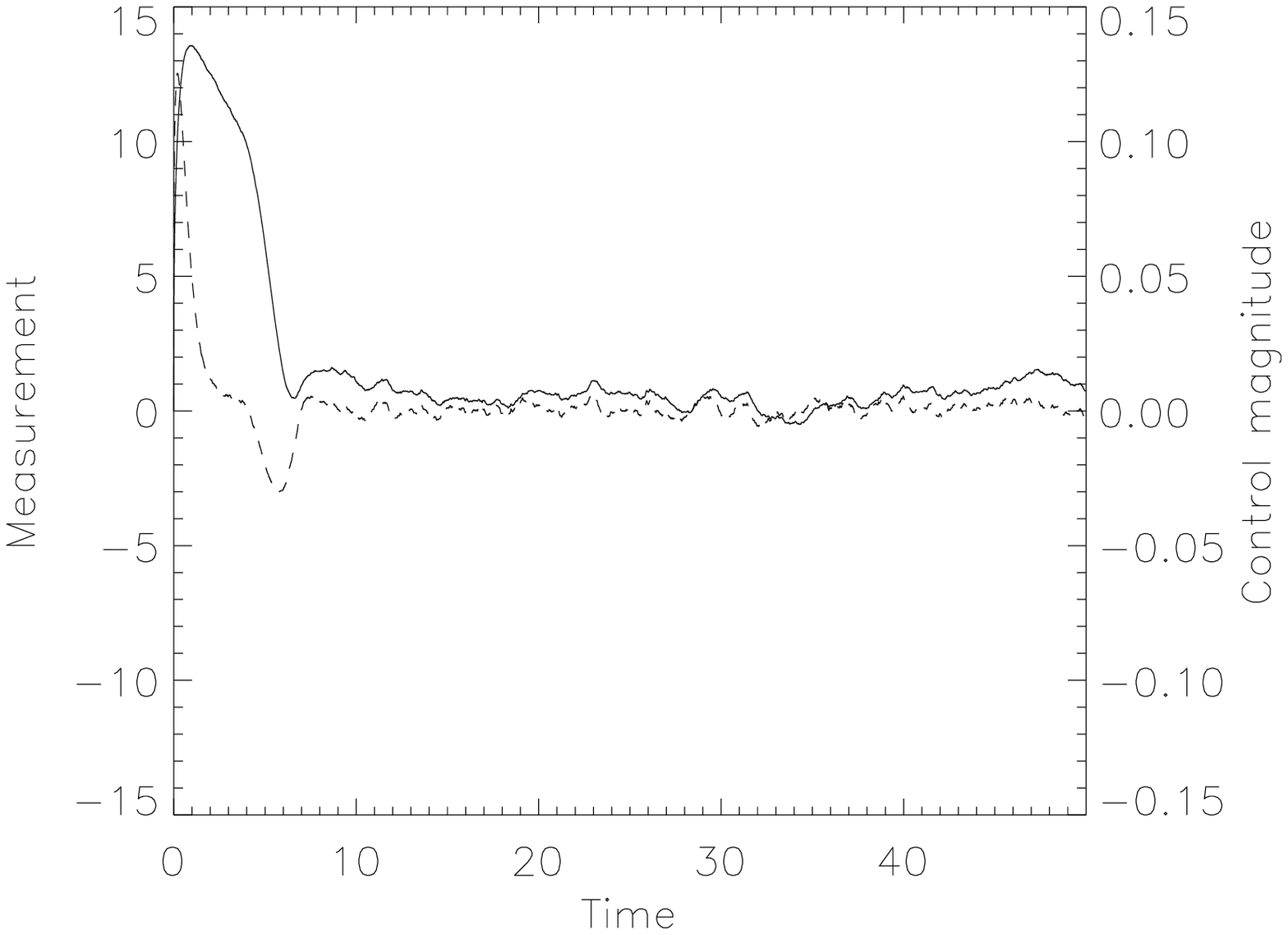}}
}
\mbox{
  \subfigure[]{\label{c4-skai0-traj}\includegraphics[width=0.475\textwidth]{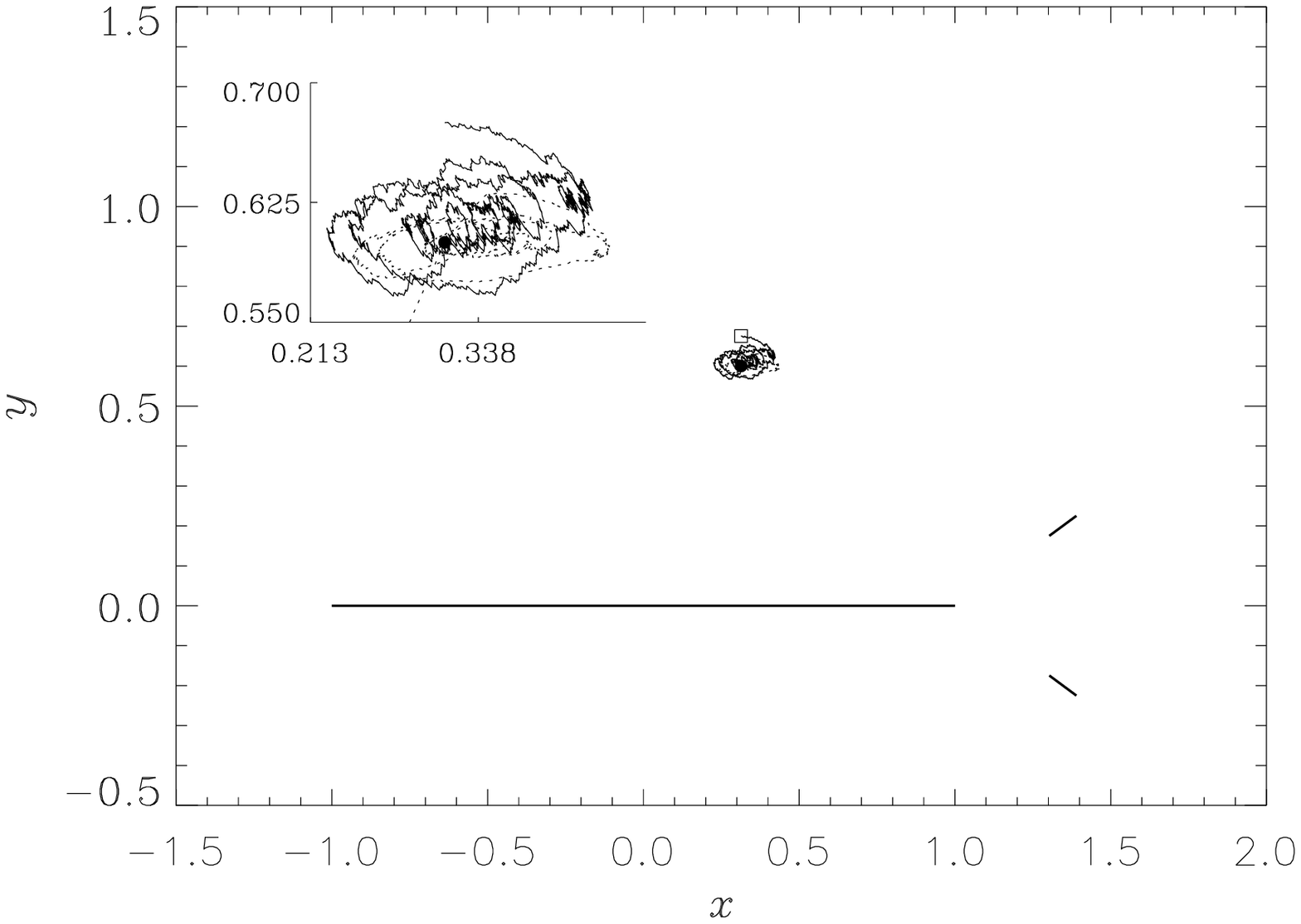}}\qquad
  \subfigure[]{\label{c4-skai0-mhist}\includegraphics[width=0.475\textwidth]{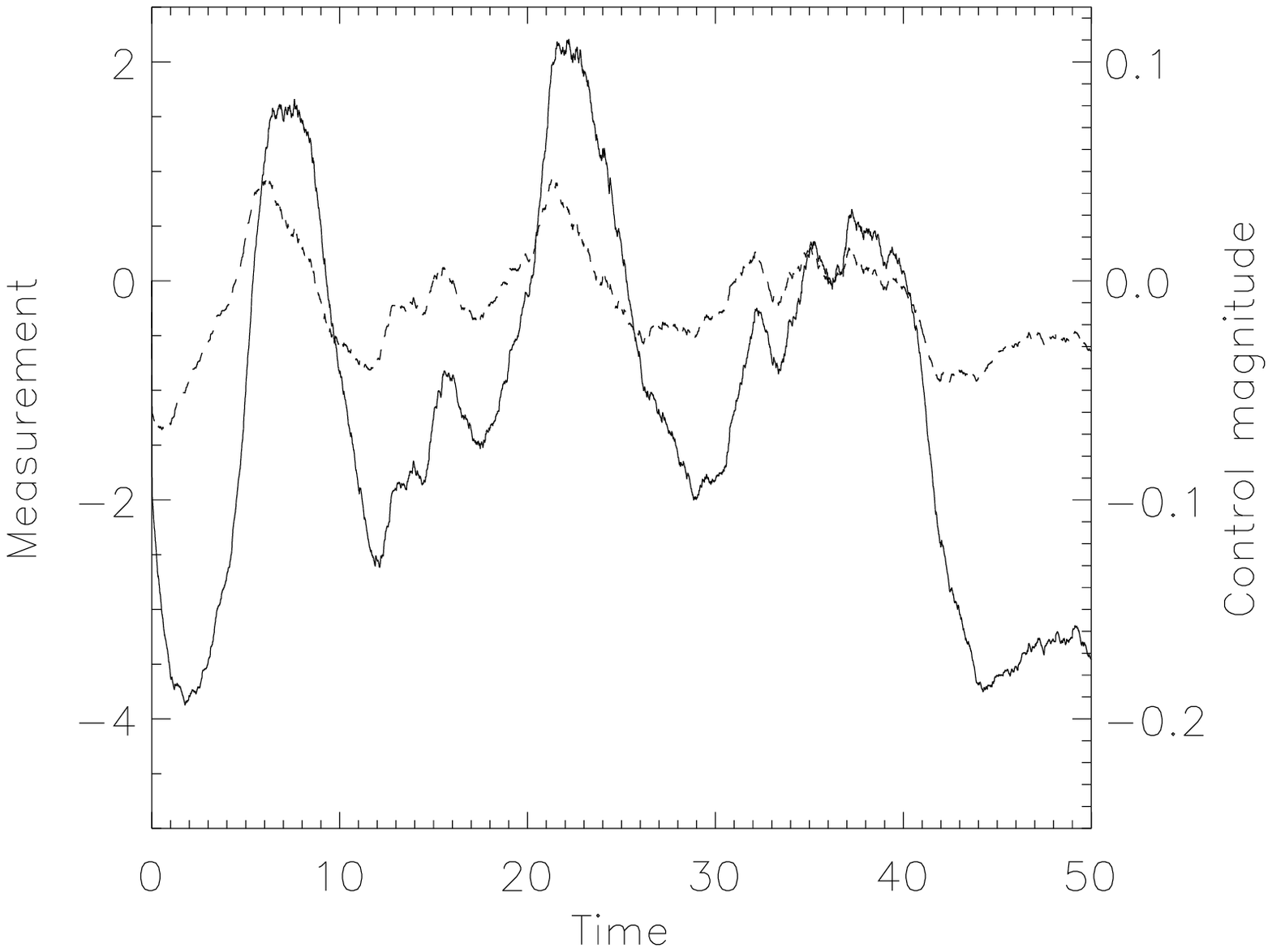}}
}
  \caption{(a) Vortex {trajectories} (solid curve) and estimator
    {trajectories} (dotted curve) for {initial} perturbation
    $\delta=-0.1i$ and {disturbances with $\Delta\chi \sim
      \mathcal{N}(0,0.2)$} in a case \#1. The square symbol indicates
    the initial position of the perturbed vortex and the solid circle
    the equilibrium position in the absence of any stochastic forcing.
    (b) The corresponding {time-histories of the} measurement (solid
    curve) and control magnitude (dashed curve).  (c) and (d): {same
      as} (a) and (b), but for a case \#4 configuration with
    $\delta=0.075i$ and $\Delta\chi \sim \mathcal{N}(0,0.4)$; {in
      panel (c) the inset represents a magnification of the
      neighbourhood of the equilibrium.}}
  \label{skai0}
\end{figure}

\begin{figure}
 \centering
\mbox{
  \subfigure[]{\label{c1-skai1-traj}\includegraphics[width=0.475\textwidth]{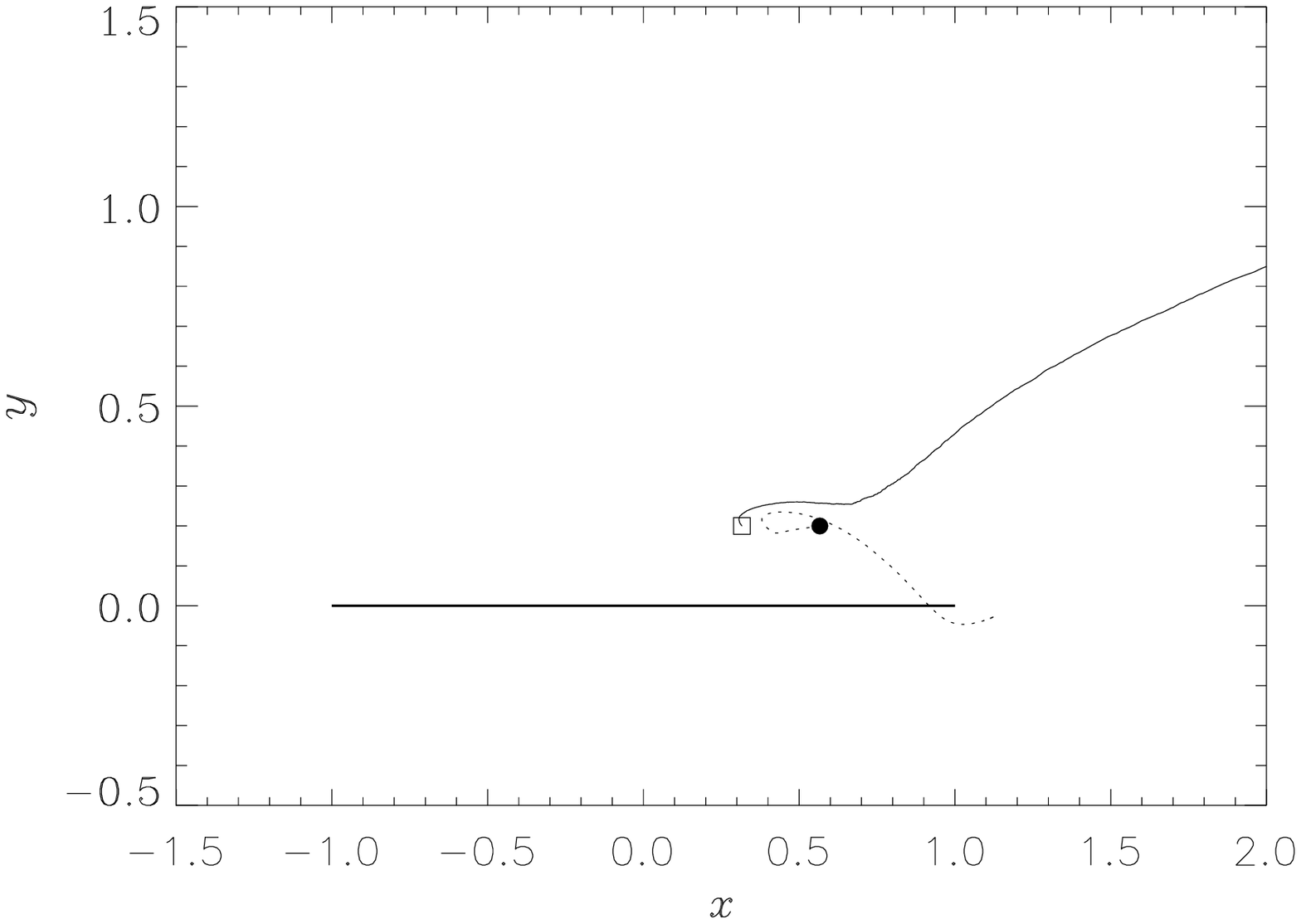}}\qquad
  \subfigure[]{\label{c1-skai1-mhist}\includegraphics[width=0.475\textwidth]{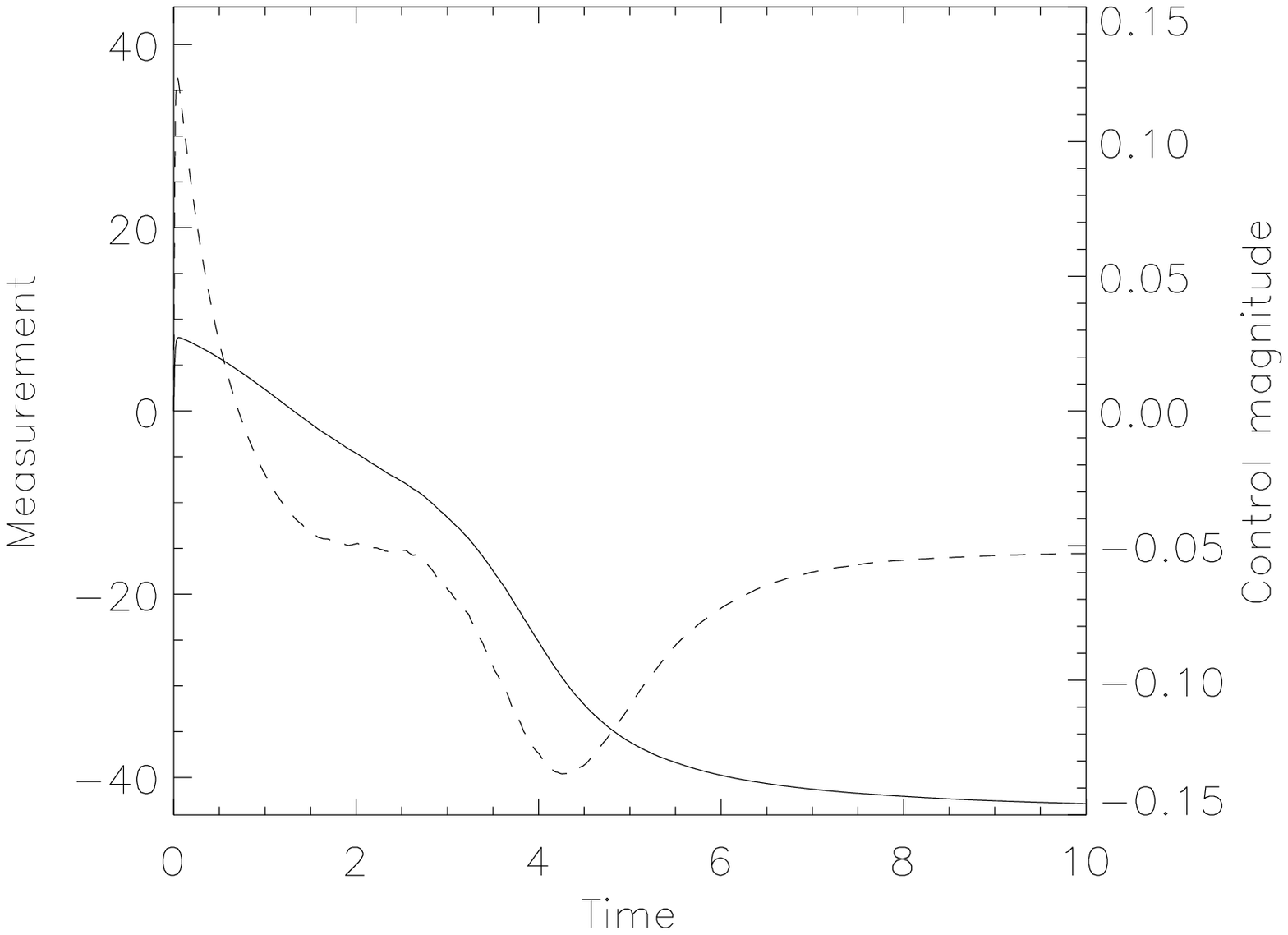}}
}
\mbox{
  \subfigure[]{\label{c5-skai1-traj}\includegraphics[width=0.475\textwidth]{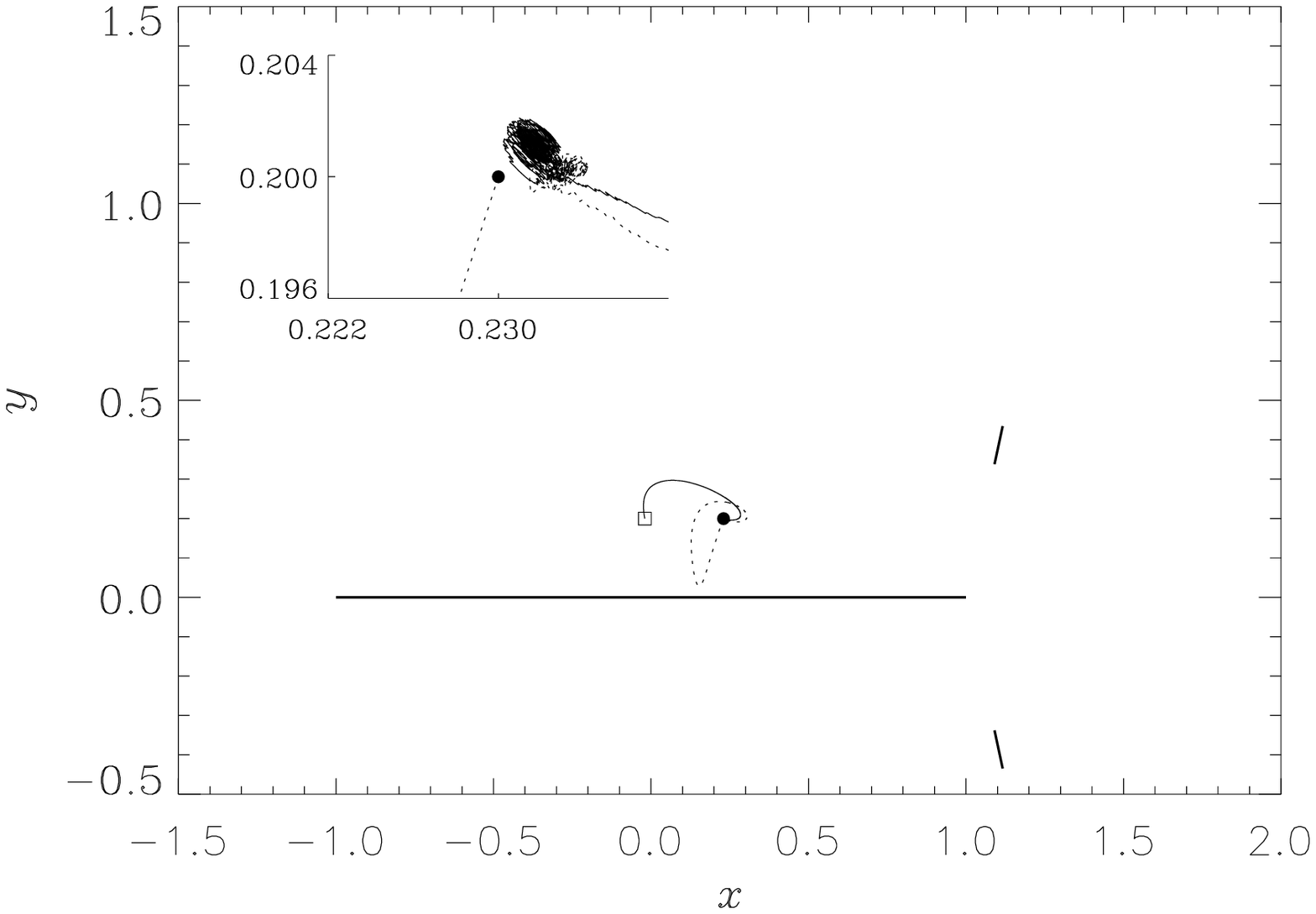}}\qquad
  \subfigure[]{\label{c5-skai1-mhist}\includegraphics[width=0.475\textwidth]{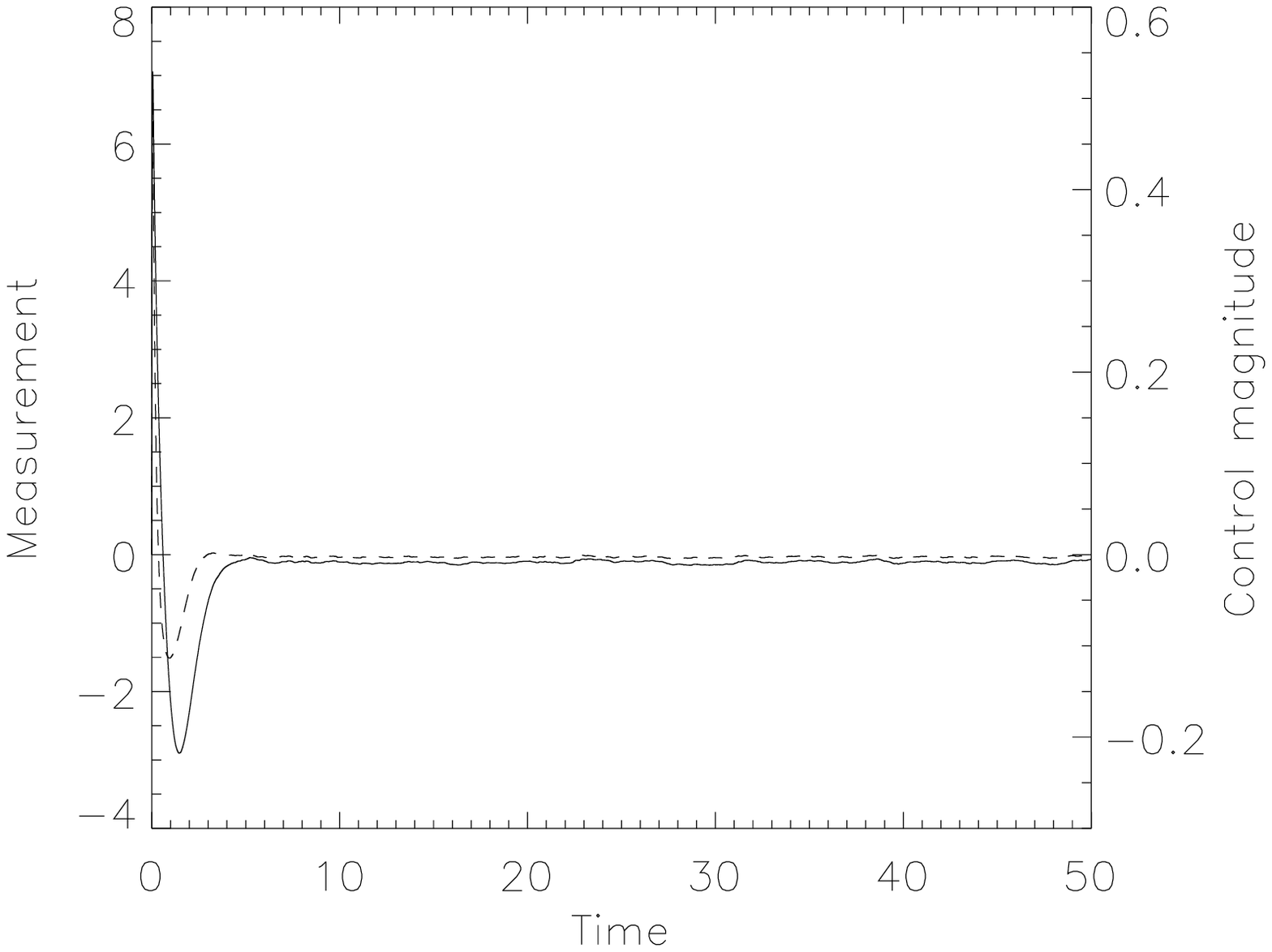}}
}
  \caption{(a) Vortex {trajectories} (solid curve) and estimator
    {trajectories} (dotted curve) for {initial} perturbation
    $\delta=-0.25$ and {disturbances with $\Delta\chi \sim
      \mathcal{N}(0.1,0.1)$} in a case \#1. The square symbol indicates
      the initial position of the perturbed vortex and the solid
      circle the equilibrium position in the absence of any stochastic
      forcing.  (b) The corresponding {time-histories of the}
      measurement (solid curve) and control magnitude (dashed curve).
      (c) and (d): {same as} (a) and (b), but for a case \#5. The
      inset in (c) shows a magnification of the region close the
      {uncontrolled} vortex equilibrium.}
  \label{skai1}
\end{figure}

In figure \ref{skai0} {we show the vortex trajectories together with
  the corresponding histories of the measurements and control
  intensities obtained for disturbances with $\mu = 0$ and two
  different values of $\sigma^2$ (0.2 and 0.4).} In both these
examples the {initial} perturbation has been chosen to lie close to
the edge of the basin of attraction {and in the absence of
  control the corresponding vortex trajectories are swept to
  infinity}. For these relatively large perturbations it is seen that
the control still performs well in the presence of a fluctuating
background flow, even for large values of $\sigma^{2}$.  Figure
\ref{skai1} then presents two examples {corresponding to cases \#1 and
  \#5 from {table} \ref{tab:cases} with disturbances for which
  $\mu>0$}. In both these {cases, $\Delta \chi \sim
  \mathcal{N}(0.1,0.1)$ in \eqref{eq:dchi}} and the same {initial}
perturbation with $\delta=-0.25$ was used, {chosen to lie} within each
of the respective basins of attraction. In the single-plate
configuration {(figures \ref{skai1}(a,b))} the control fails and the
vortex quickly escapes.  However, in the Kasper Wing configuration
{(figures \ref{skai1}(c,d))} the control is robust and the vortex
gradually moves to, and then undergoes a random motion about, a point
close to the {uncontrolled} equilibrium. {We note that in case
  \#5 in the absence of control the vortex trajectory also remained
  bounded, although in comparison with the controlled case, the
  departures from the equilibrium position were much larger (to avoid
  cluttering figures, these results are not shown here).}

Figures \ref{skai0} and \ref{skai1} {illustrate representative} types
of behaviour seen in the presence of {the stochastic forcing in the
  form \eqref{eq:dchi}}. The general behaviour can be summarized as
follows. When $\mu$ was set to zero, {the closed--loop control system
  was robust with respect to stochastic disturbances \eqref{eq:dchi}
  in all cases considered}, even for large values of $\sigma^{2}$.
{Perturbed trajectories would eventually perform a ``random walk''} in
a region near the {uncontrolled} equilibrium {whose {extent} depends}
on $\sigma^{2}$ {and in this regime the control magnitude
  oscillates around zero}. {On the other hand,} for values of $\mu>0$
{the closed-loop control was unable to stabilize the equilibrium in
  cases \#1 and \#2} even for small values of $\mu$.  However, in
cases \#3--6 it remained robust up to larger values of $\mu$.

\subsection{Control in the presence of vortex shedding}\label{sec:shedding}

We now {augment our model with} a vortex-shedding {mechanism} in which
point vortices are ``injected'' into the flow at locations close to
the rear tips of each plate at discrete {instances of} time. The
circulations of these injected vortices are chosen such that the Kutta
condition {\eqref{stabeq1c}} at {the rear tip of each plate} is
satisfied at the time of injection.  Integration is again carried out
using Euler's {explicit} method with a time-step of
$\mathrm{d}t=0.001$. Vortices are injected into the flow at $t=0$ and
then at intervals of $\Delta t=0.1$.  The injection location is chosen
to be at a distance of $0.1$ from each plate tip in the direction
tangent to the plate. That is, in cases \#1--6 a single vortex is
injected at the point
\begin{equation}\label{eq:ip1}
  z_{i1}=1+0.1=1.1.
\end{equation}
Additionally, in cases \#3--6 two further vortices are injected at
\begin{subequations}
\label{eq:ipe}
\begin{alignat}{1}
   &z_{i2}=1+(0.45+0.1)e^{i\phi}, \label{ip2} \\
   &z_{i3}=\overline{z_{i2}} \label{ip3}.
\end{alignat}
\end{subequations}
These injection points were chosen as they lie close to the rear tips
of each plate and can therefore be considered a fair approximation to
a vortex sheet {representing a separating boundary layer}. {They}
also provide numerical stability over long integration times.
Circulations of the injected vortices are calculated as follows. Prior
to injection, the total velocity field, including that owing to any
previously added vortices (see below), is calculated at each plate
tip. Vortices are then injected at the locations {given in
  \eqref{eq:ip1}--\eqref{eq:ipe}} with circulations {determined such
  that the total velocity field, including the contributions from both
  the vortices already present in the flow field and the newly created
  ones, vanishes} at each plate tip.  This corresponds to solving a
single linear equation for the circulation of the one added vortex in
single plate cases and to solving a set of three coupled linear
equations for the circulations of each of the three added vortices in
Kasper Wing cases.  Following this, the circulations of any newly
added vortices are {kept constant as the vortices} are allowed to move
freely, i.e., their positions evolve in time owing {to} the velocity
induced by other vortices plus that of the background flow.  In the
results that follow, after a vortex has been injected, it is then
dealt with explicitly for the remainder of the simulation, i.e., no
shed vortices are removed and no far-field averaging is employed.

\begin{figure}
 \centering
  \subfigure[]{\label{c1-shed-traj}\includegraphics[width=0.32\textwidth]{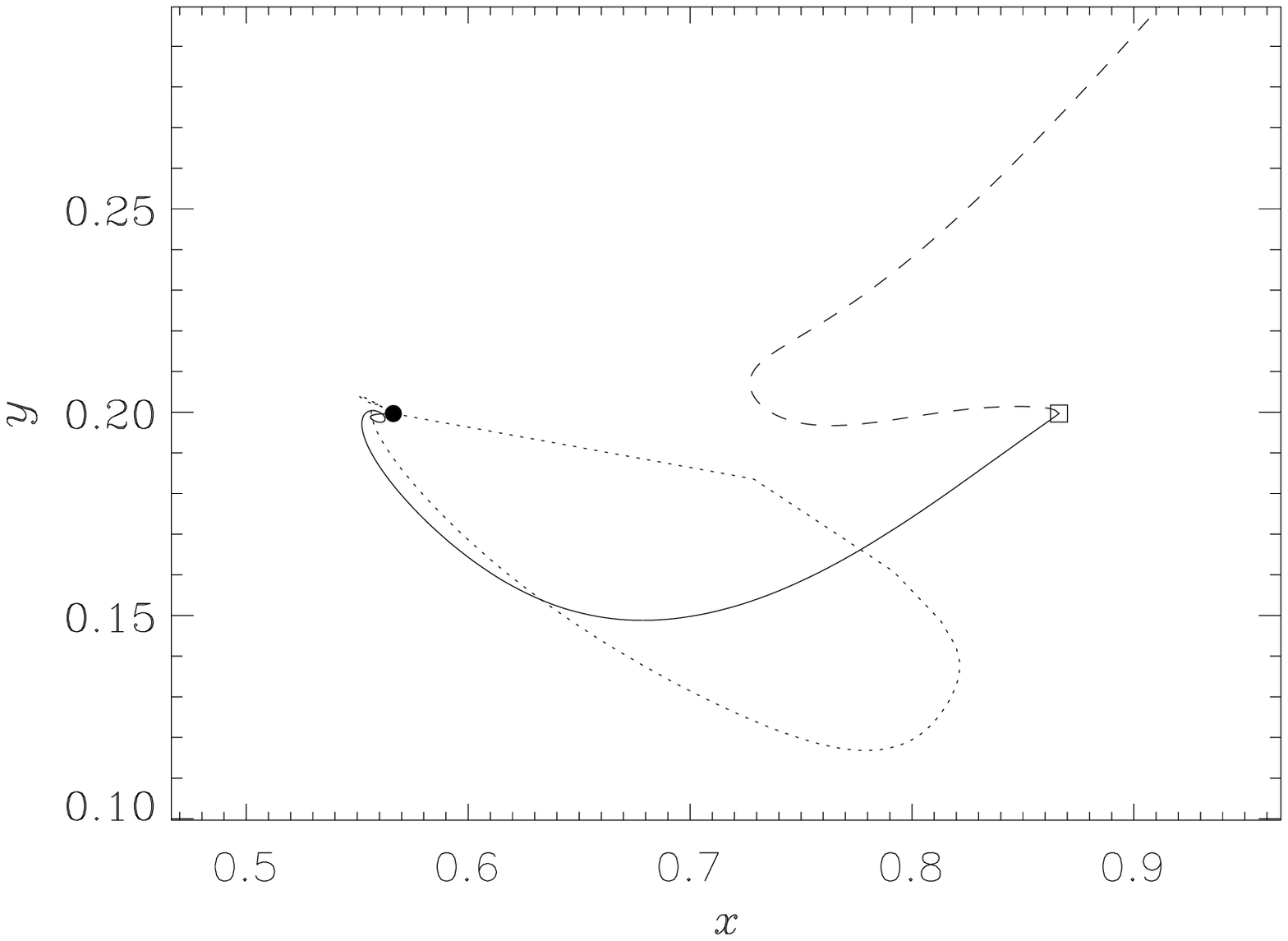}}
  \subfigure[]{\label{c1-shed-mhist}\includegraphics[width=0.33\textwidth]{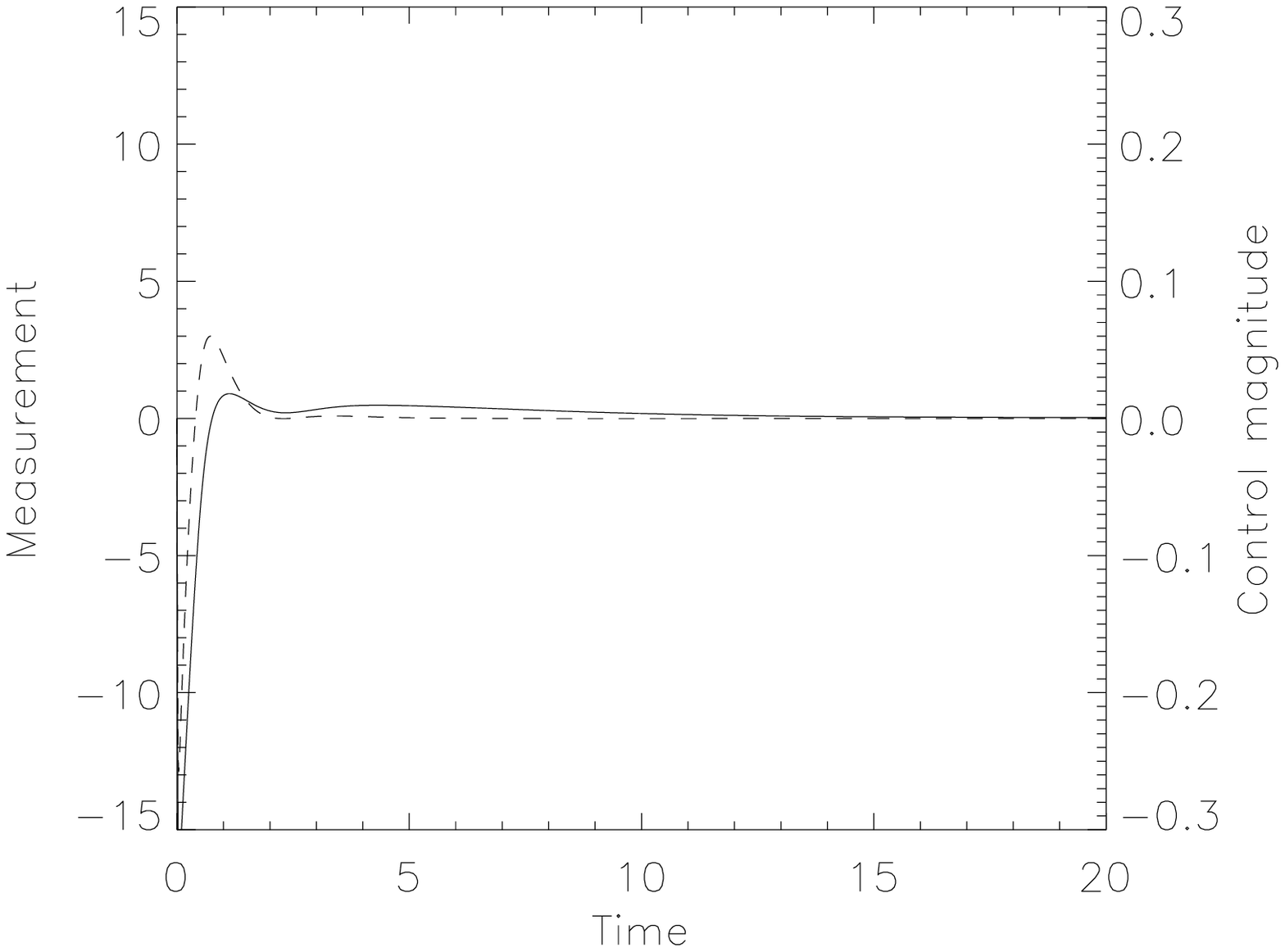}}
  \subfigure[]{\label{c1-shed-circ}\includegraphics[width=0.33\textwidth]{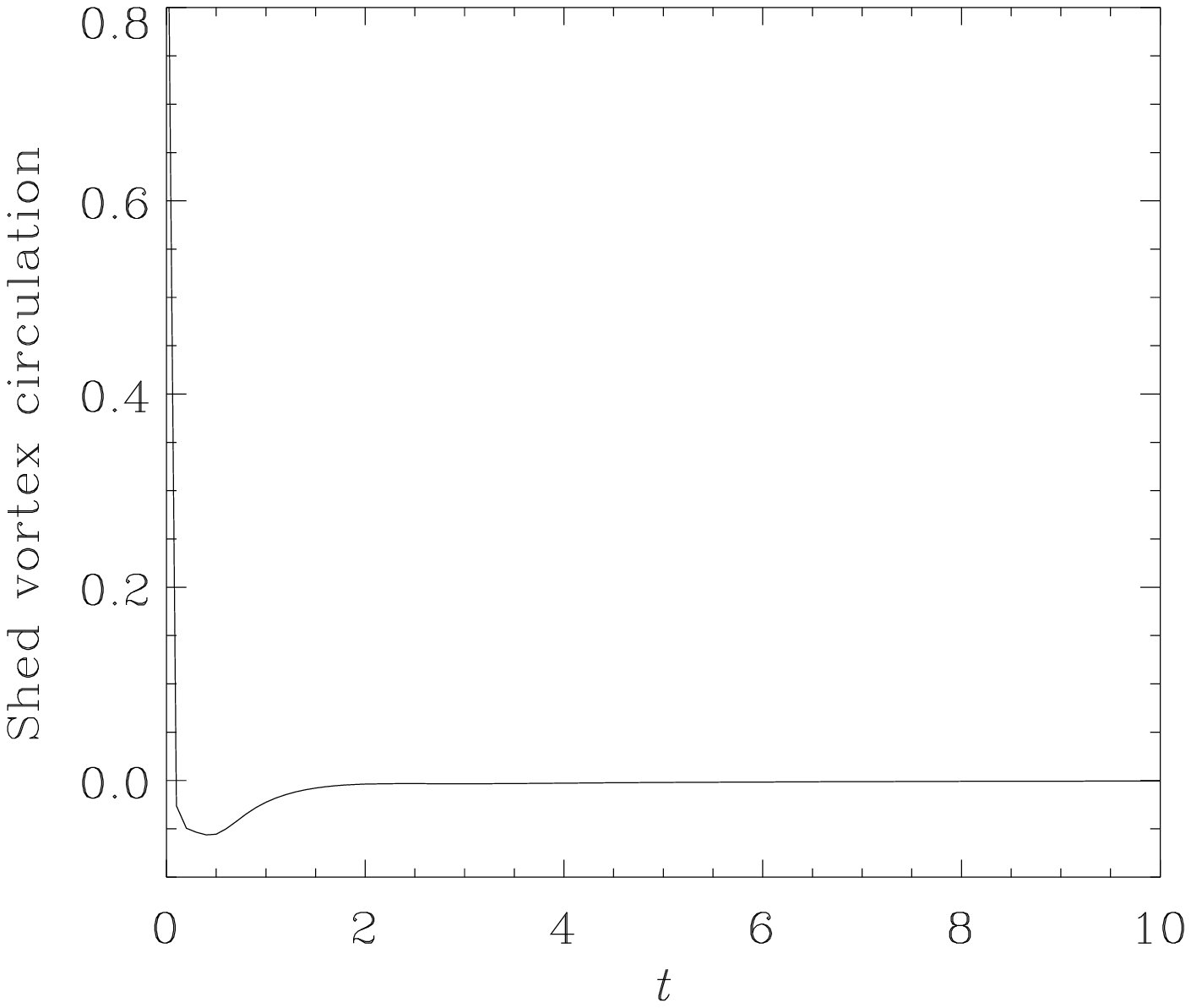}}
  \subfigure[]{\label{c5-shed-traj}\includegraphics[width=0.32\textwidth]{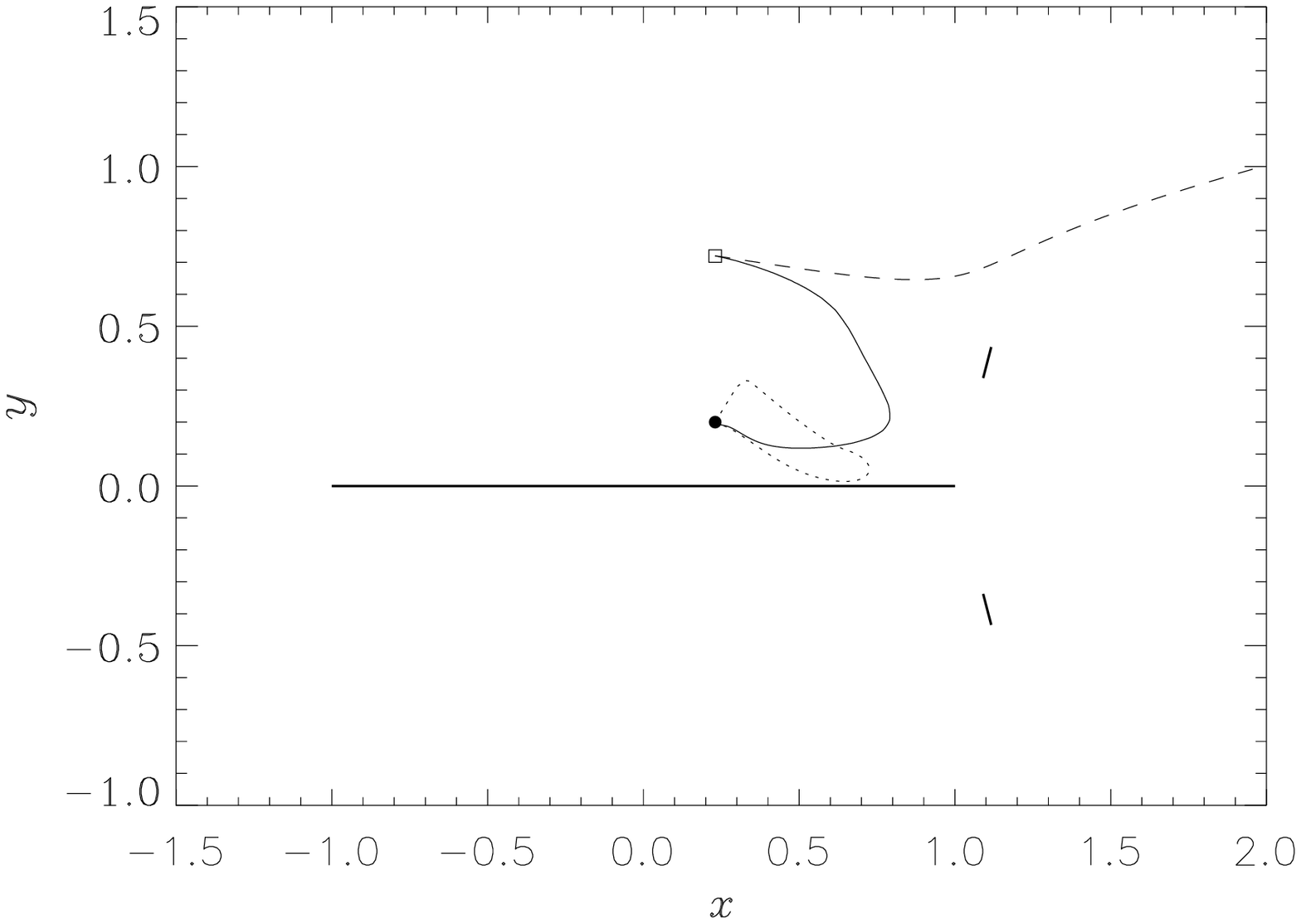}}
  \subfigure[]{\label{c5-shed-mhist}\includegraphics[width=0.33\textwidth]{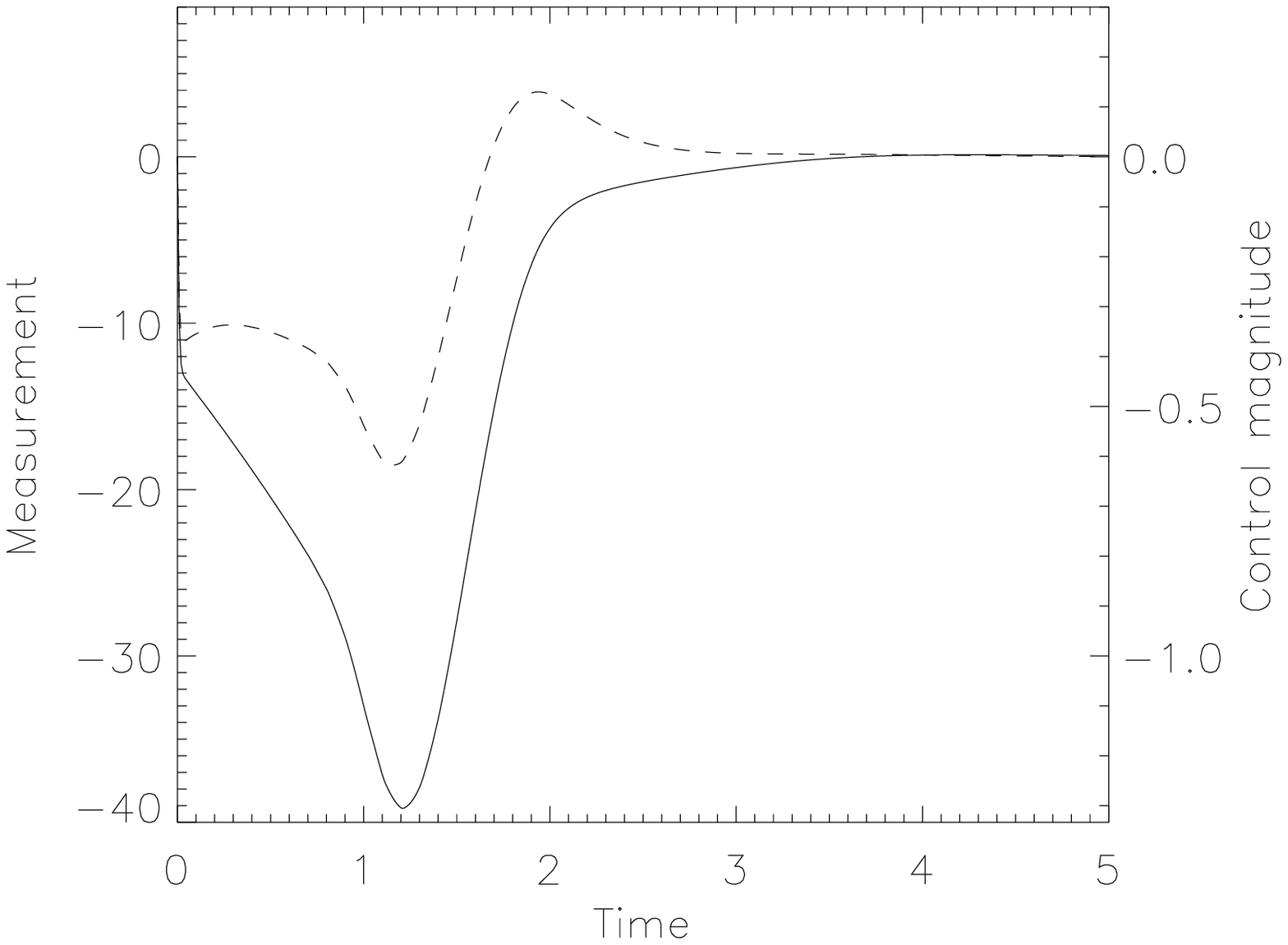}}
  \subfigure[]{\label{c5-shed-circ}\includegraphics[width=0.33\textwidth]{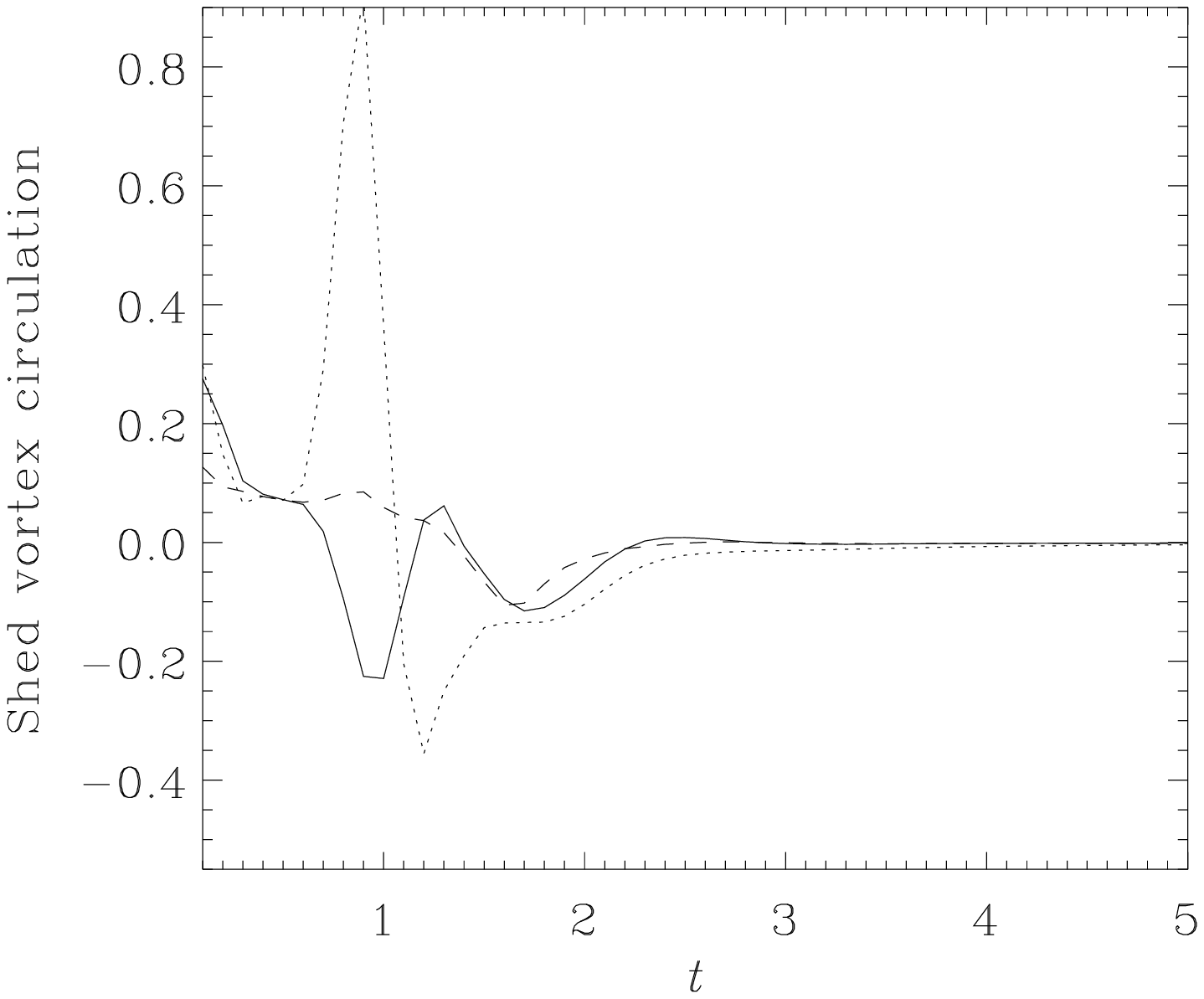}}
  \caption{(a) Case \#1 configuration with {initial} perturbation
    $\delta=0.3$.  The dashed curve represents the controlled vortex
    trajectory {starting from the outside of the basin of attraction} in the absence of shedding, the thin solid curve the
    controlled trajectory in the presence of shedding and the dotted
    curve the estimator trajectory in the presence of shedding. The
    solid circular symbol represents the unperturbed equilibrium
    position and the square the initial perturbed position.  (b) The
    corresponding {time-history of the} measurements
    {$Y(t)$} (solid curve) and control {intensity $m(t)$}
    (dashed curve) in the {presence} of shedding.  (c) The corresponding
    circulation of the most recently shed vortex at $z_{i1}$.
    (d)--(f): {same as} (a)--(c) but for a case \#5 configuration
    with $\delta=0.52i$.  Additionally, in (d) the thicker solid lines
    represent the plate boundaries and in (f) the solid, dotted and
    dashed curves represents the circulations of the most recently
    shed vortices at $z_{i1}$, $z_{i2}$ and $z_{i3}$ respectively.}
  \label{shed0}
\end{figure}

\begin{figure}
 \centering
  \subfigure[]{\label{c4-shed-traj}\includegraphics[width=0.32\textwidth]{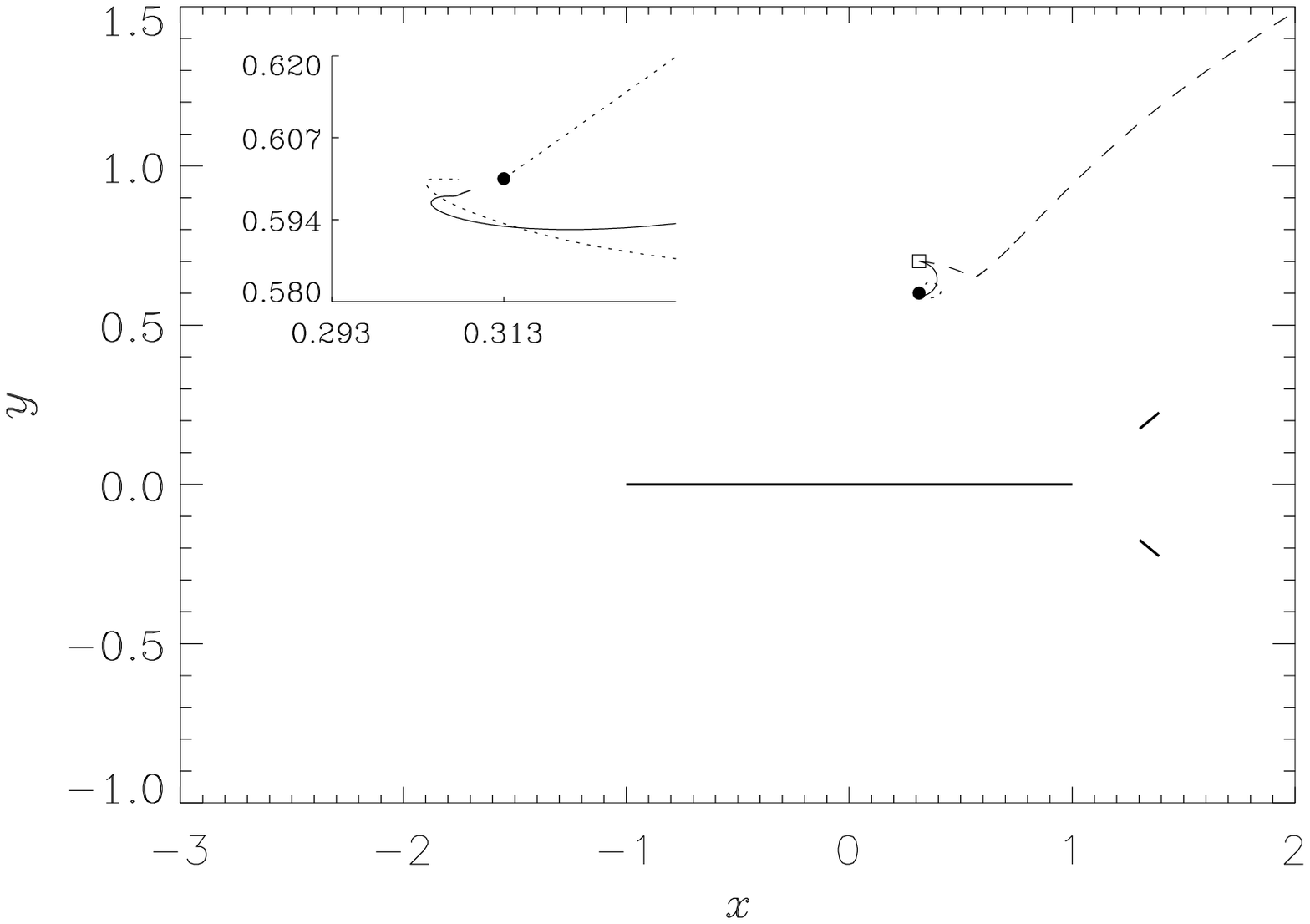}}
  \subfigure[]{\label{c4-shed-mhist}\includegraphics[width=0.33\textwidth]{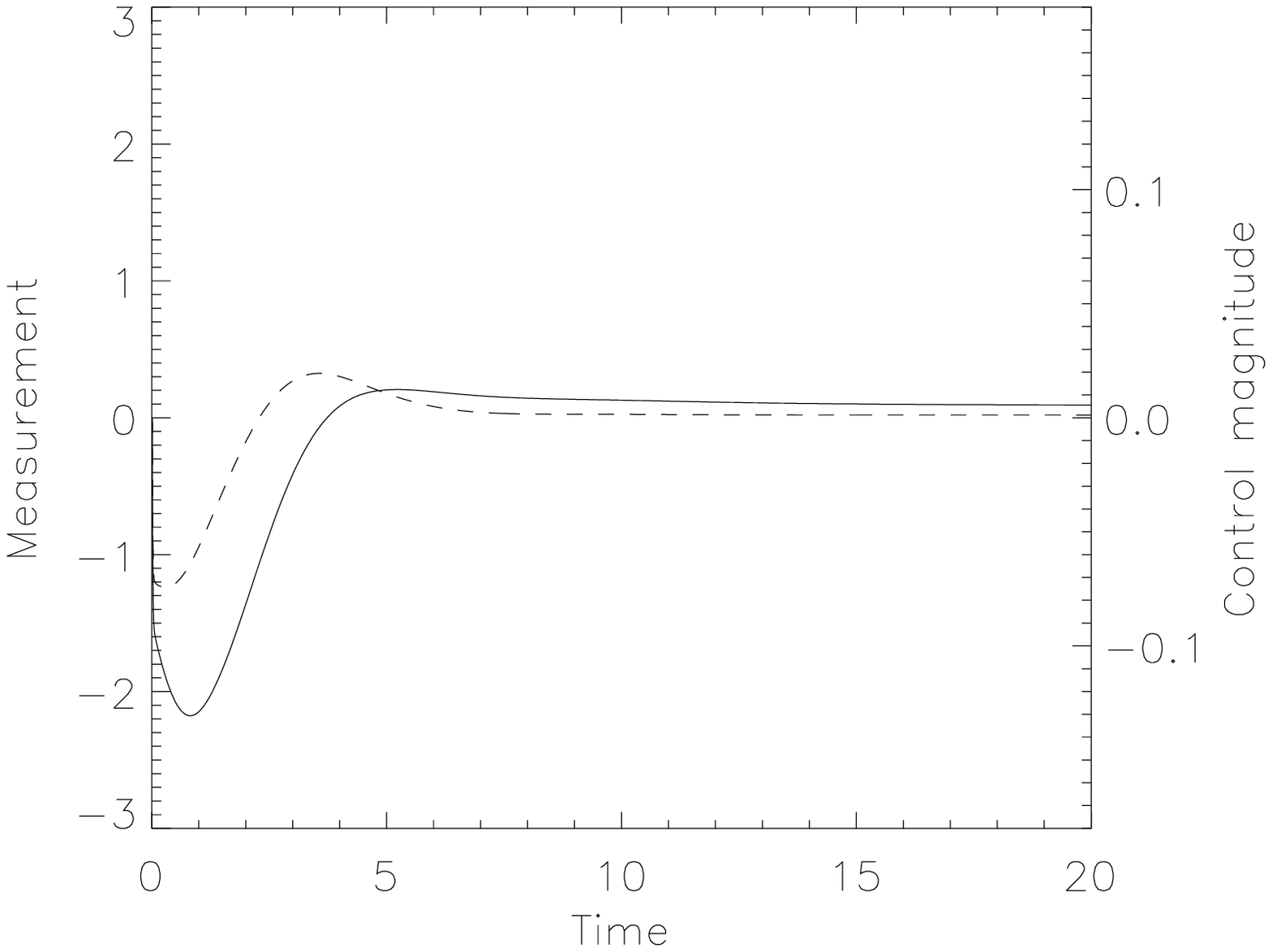}}
  \subfigure[]{\label{c4-shed-circ}\includegraphics[width=0.33\textwidth]{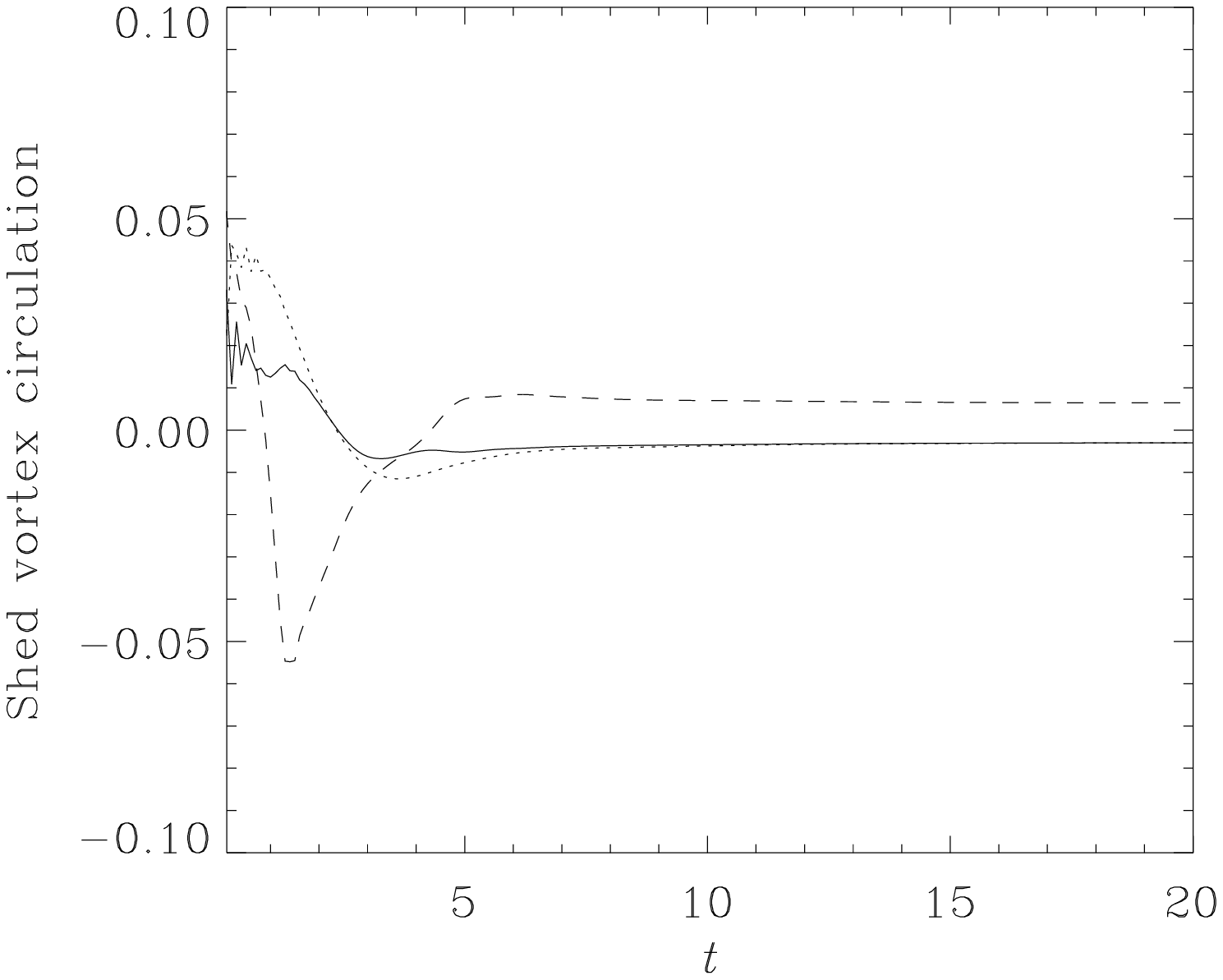}}
  \subfigure[]{\label{c1-chaotic-traj}\includegraphics[width=0.32\textwidth]{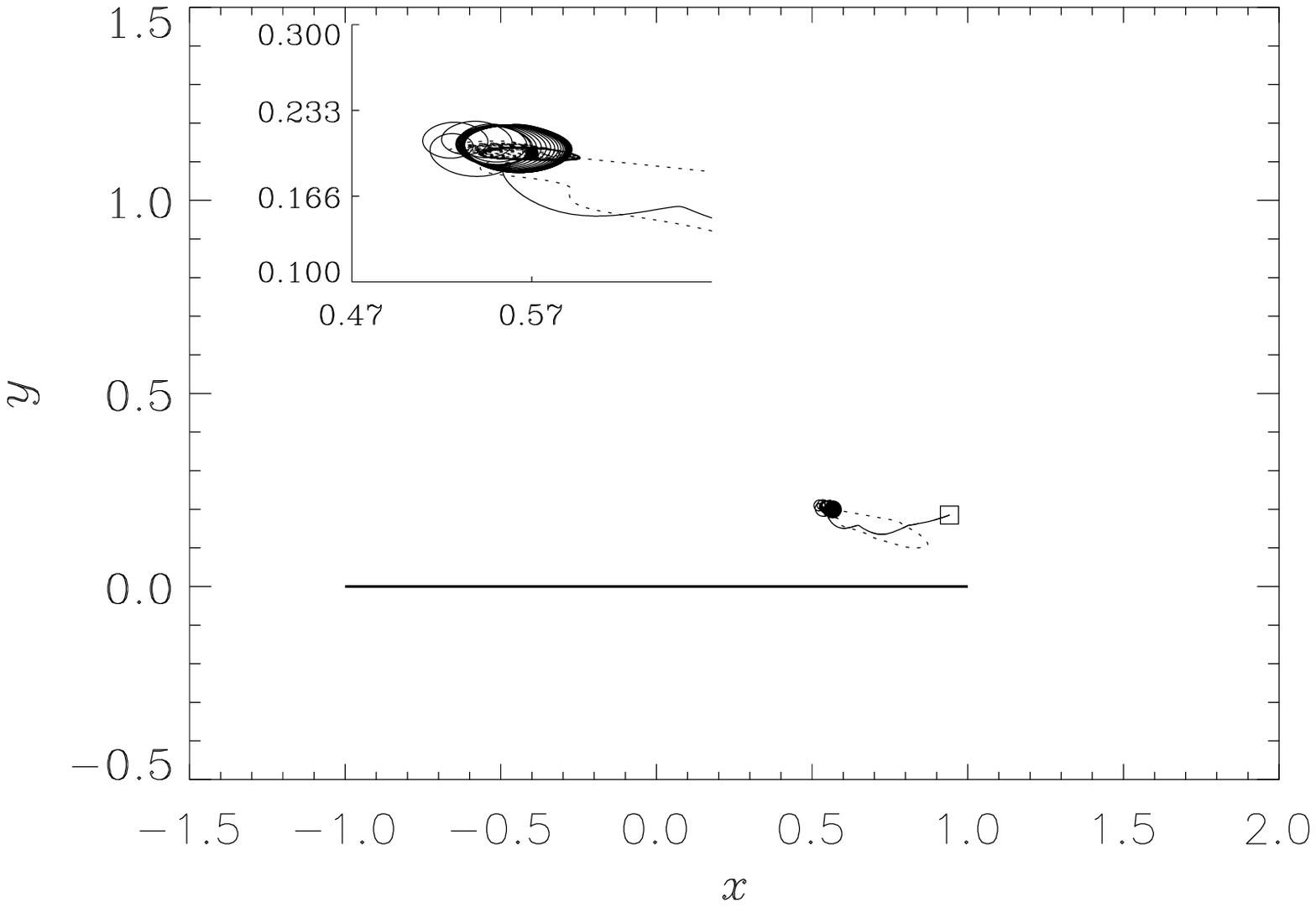}}
  \subfigure[]{\label{c1-chaotic-mhist}\includegraphics[width=0.33\textwidth]{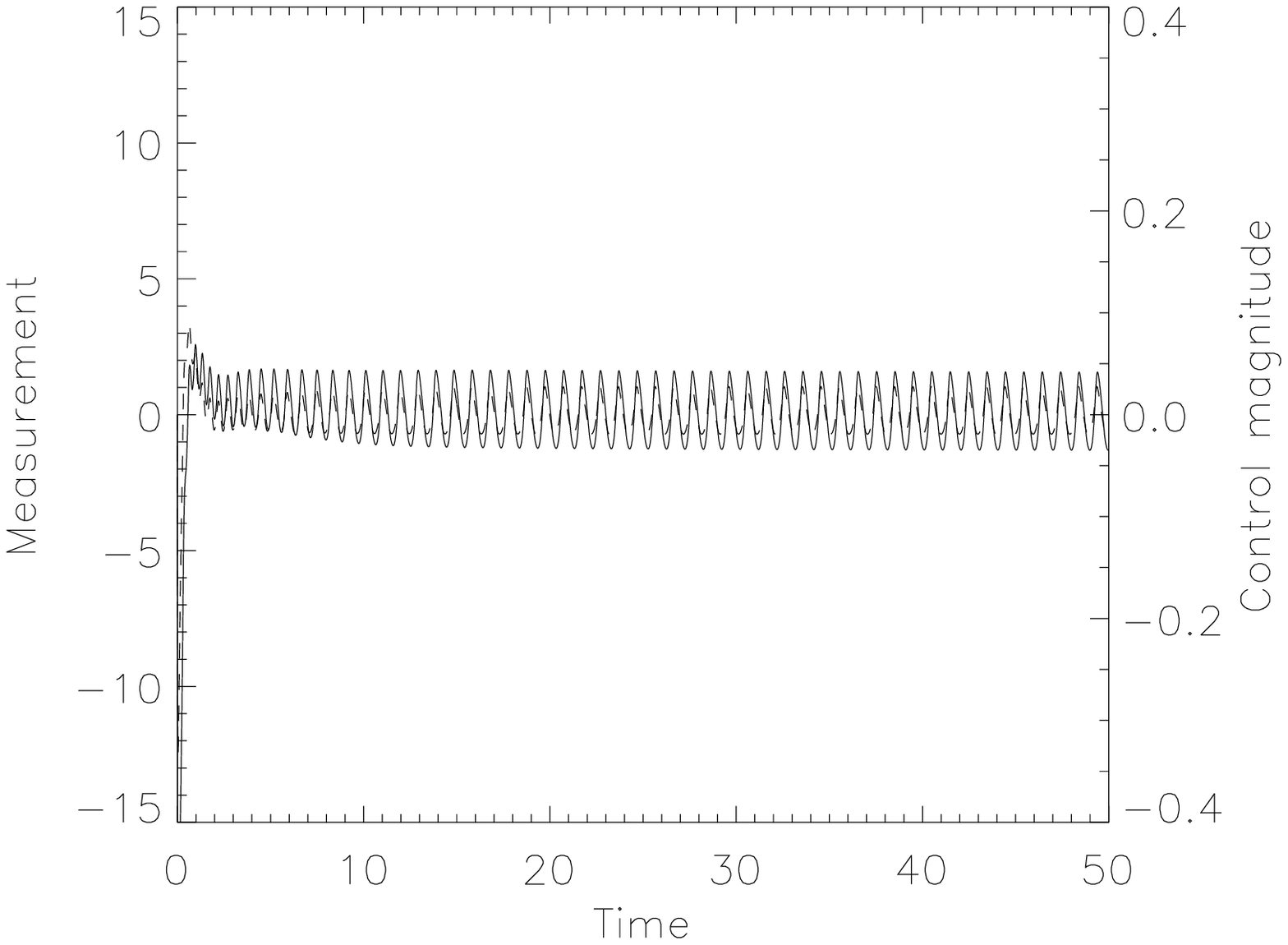}}
  \subfigure[]{\label{c1-chaotic-circ}\includegraphics[width=0.33\textwidth]{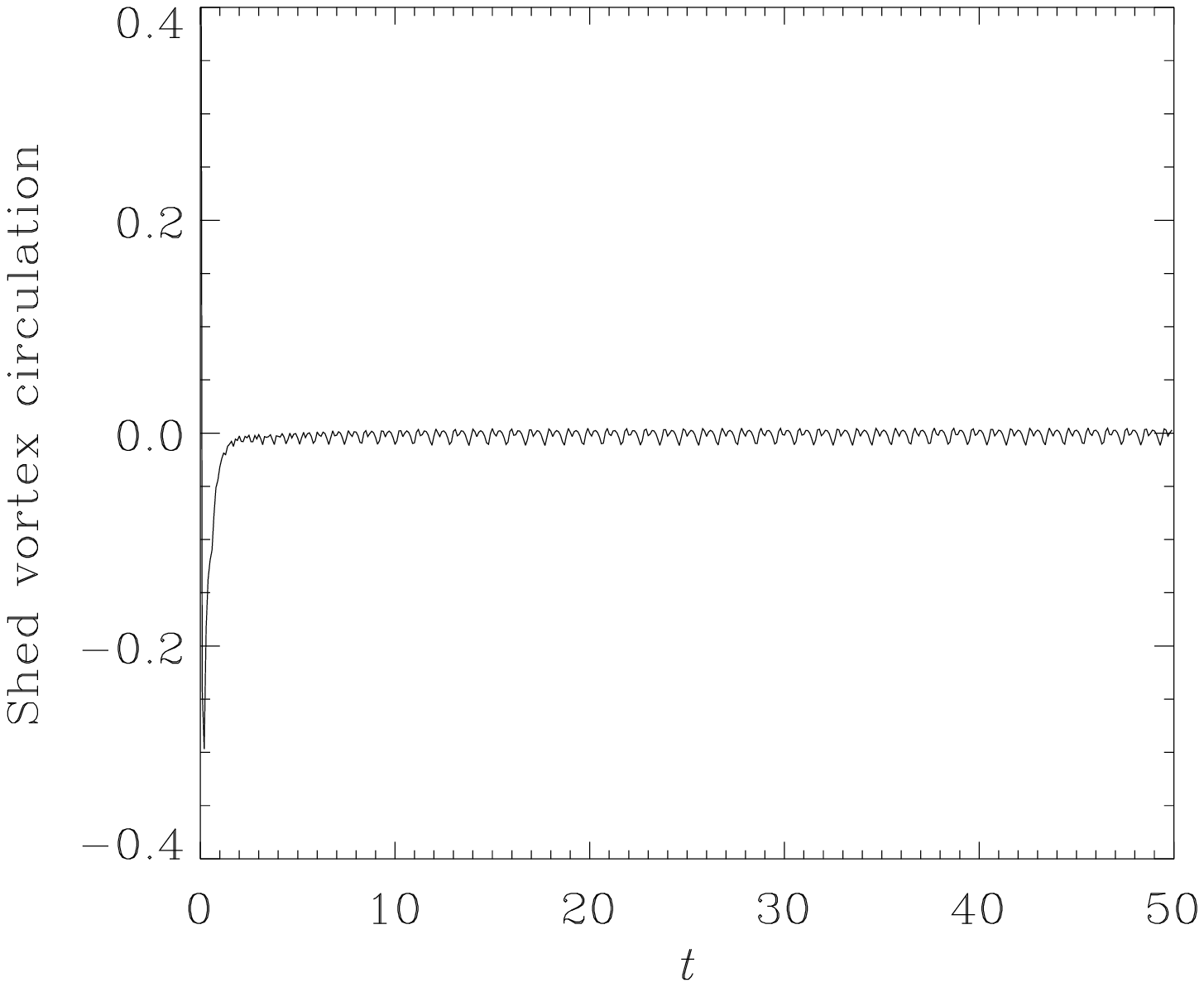}}
  \caption{(a) Case \#4 configuration with {initial} perturbation
    of $\delta=0.1i$. The dashed curve represents the controlled
    vortex trajectory in the absence of shedding, the thin solid curve
    the controlled trajectory in the presence of shedding and the
    dotted curve the estimator trajectory in the presence of shedding.
    The thicker solid lines represent the plate boundaries, the solid
    circular symbol the unperturbed equilibrium position and the
    square the initial perturbed position.  The inset shows a
    magnification of the region close the vortex equilibrium.  (b) The
    corresponding {time-history of the} measurements
    {$Y(t)$} (solid curve) and control {intensity $m(t)$}
    (dashed curve) in the {presence} of shedding.  (c) The corresponding
    circulations of the most recently shed vortices at $z_{i1}$ (solid
    curve), $z_{i2}$ (dotted curve) and $z_{i3}$ (dashed curve).
    (d)--(f): {same as} (a)--(c) but for a case \#1 configuration
    with $\delta=0.375\exp(-0.13i\pi)$. The inset shows the region close to the unperturbed equilibrium.}
  \label{shed1}
\end{figure}

Examples of the various behaviours observed with vortex shedding added
are presented in figures \ref{shed0} and \ref{shed1}. These behaviours
can be summarized as follows. In general, the addition of shed
vortices has a stabilizing effect. In almost all configurations
tested, perturbations larger than those in the absence of shedding
could be stabilized (exceptions to this rule will be mentioned below
and {are} discussed in a little further detail in the following
section). In figure \ref{shed0}, two examples are shown in which the
vortex is {initially} perturbed to outside the basins of
attraction {determined} in \S \ref{nres}. Despite this, the vortex
{trajectories} are rapidly stabilized and, after an initial increase,
the circulations of the most recently shed vortices decay to zero as
demonstrated in figures \ref{shed0}(c) and \ref{shed0}(f).

Figures \ref{shed1}(a)--(c) demonstrate a slightly different
behaviour. Now, owing to the larger circulation of the point vortex,
instead of returning to the unperturbed equilibrium position, the
vortex is held at a location close to the equilibrium and the
circulations of the shed vortices tend to constant values. That is, a
new equilibrium state {emerges}. Finally, figures
\ref{shed1}(d)--(f) show an example of a ``chaotic'' case. In
{situations} where the perturbation {$\delta$} is too large
for the vortex to be stabilized, but not large enough for the vortex
to escape, the vortex can undergo a complicated motion in the vicinity
of the equilibrium location over a long period of time.

Although the presence of vortex shedding has, {in general}, a
stabilizing effect in the majority of {cases}, two scenarios were
{identified} in which the effect was {in fact}
destabilizing. Firstly, in {the configurations corresponding to}
case \#2 equilibria would {not be stabilized for any initial
  perturbation $\delta$}. Additionally, in cases \#3--6 when {an
  initial} perturbation with a large positive real component is taken,
the extent to which {the corresponding evolutions} can be
controlled is reduced. Physical arguments for these scenarios are
discussed in the following section.

\section{Discussion}\label{discussion}

Before concluding, results presented in this paper are briefly
summarized and some additional points of discussion {are} raised.
Firstly, from a control design perspective, for the {given}
choice of actuation mechanism (sink-source singularity) and
measurement (pressure difference across the main plate), all
configurations demonstrated a similar responsiveness. That is, the
chosen configurations were completely controllable and observable for
all sensor and actuator locations and the corresponding residuals
displayed the same general features.

Placing the sink-source {actuator} at the location with maximal
controllability residual and the pressure sensor at the location {with
  maximal observability}, {the control} was seen to be effective for a
range of perturbations in all cases considered. Also, {as
  demonstrated in figure \ref{fig:pevol}(a),} for comparable setups
neutrally stable configurations could be stabilized in a shorter
amount of time than {linearly unstable} configurations. The range {of
  perturbations for} which the control was effective varied from cases
to case. Not unexpectedly, for a given configuration, a larger range
of perturbations could be stabilized when the vortex equilibrium was
located {closer} to the {main} plate.  Such equilibria are
comprised of vortices with a weaker circulation and they are therefore
more easily influenced by the sink-source actuator. However, whilst it
was expected that the presence the auxiliary ``flaps'' in Kasper Wing
configurations would enhance the robustness of the control, the
substantial increase in robustness was somewhat {surprising},
especially in cases \#3 and \#5.  Indeed, at least in the inviscid
setting, this demonstrates the large effect additional (small)
boundaries can have on such flows. {We add that a similar stabilizing
  effect achieved by a small obstacle placed in the wake of a larger
  body was observed experimentally by \cite{ss90}, and later explained
  in terms of intrinsic stability properties by \cite{gl07}.}

The effectiveness of the control was then examined in the presence of
additional {disturbances designed to mimic the uncertainty of the flow
  model and the flow configuration}. With a {randomly} varying angle
of attack ${\chi(t)}$, provided {its} mean value was {unchanged and
  equal to $\chi_0$, the system could be successfully controlled even
  for large oscillations of the angle of attack.} In such cases, the
vortex would undergo a random walk around its equilibrium location.
{On the other hand, as the expectation of the randomly perturbed angle
  of attack was allowed to deviate form $\chi_0$, for which the LQG
  compensator was designed, the control would quickly fail} (with
single-plate configurations breaking down quicker than their Kasper
Wing counterparts). When a vortex shedding {model} was introduced, the
robustness of the control was seen to {improve} in all cases for
almost all perturbations. An exception to this was when the vortex was
perturbed downstream of the equilibrium position in Kasper Wing
configurations (figures \ref{shed1}(a)--(c)). In such scenarios,
pushing the vortex close to the rear tip of {one of} the three plates
would result in three vortices with substantial circulations being
shed and the system {becoming unstable.  Furthermore,} with the
addition of vortex shedding, new ``exotic'' trajectories such as that
presented in figure \ref{shed1}(d) could be observed. In that
particular example, it appears that the vortex may enter a limit cycle
around a new equilibrium. Finally, although not discussed in \S
\ref{sec:shedding}, it is now noted that in the presence of shedding
case \#2 is seen to be uncontrollable for all perturbations.  This can
be understood by viewing the streamline pattern for this case shown in
figure \ref{sl-c1-7}(b) {in which} the recirculation region of the
vortex engulfs the plate. {As a result,} vortices {are shed} with far
weaker circulations than that of the main vortex {and} {are} trapped
in {the recirculation} region.  This eventually leads to a build up of
circulation around the plate, {a state which cannot be effectively
  controlled}. Indeed, it {may be} expected that this phenomenon will
occur for any equilibrium in which the vortex recirculation region
engulfs any boundary from which vortices are shed. {The relation
  between detachment of the leading-edge vortex and the global flow
  topology in real flows was investigated experimentally by
  \citet{Rival2014} and using a vortex-based model by
  \citet{rggoe14}.}

\section{Conclusions}\label{conclusions}

In this paper, an LQG compensator was designed to stabilize
point-vortex equilibria located above both an inclined flat plate and
Kasper Wing in the presence of an oncoming {uniform} flow.  A
sink-source singularity placed on the main plate acted as the
actuation mechanism and pressure difference across the main plate as
the system output. {Standard methods of Linear Control Theory were
  used to characterize this flow model and the} compensator was
applied to a range of systems, the results of which are summarized in
\S \ref{discussion}. Other forms of actuation could also be
considered, for example, applying an additional circulation around the
main plate and details regarding the derivation of such alternative
controls can be found in \citet{protas:lfs}.  However, the sink-source
{actuation} was chosen for this study as its effect on the vortex can
be considered analogous to that of a synthetic jet commonly used in
similar studies in which viscosity is included {(analogous models
  were also used to represent this form of flow actuation by
  \citet{control:c96,zannetti:suction})}.

{A key result of our} study is the demonstration of the large
effect the addition of small boundaries has on the {controlled}
flow. From a control design perspective, the effect of the Kasper
flaps is to ``push'' the vortex closer to the front of the plate and
also slightly reduce its circulation (for {vortex equilibria} of
corresponding elevation above the plate). Both these effects are
beneficial in terms of the effectiveness of the control. Further, for
nonlinearly unstable perturbations where in the absence of control the
vortex quickly travels downstream (cf.~figure \ref{trajuc}(a)), the
flaps in general have the effect of slightly slowing down the escape.
This is also beneficial in regard to the robustness of the control.

{Inviscid} flow models similar to that {studied} here have previously
been considered as reduced-order models of {real} flows governed by
the Navier-Stokes {system} \citep{p08a}.  The simplicity of such
reduced-order models, if they can be successfully utilized in the
design a control strategy for the full model, is of course very
appealing.  In comparison to the study of \citet{protas:lfs}, {where a
  controller designed based on an inviscid F\"oppl vortex model was
  used to stabilize a bluff-body wake flow at $Re=75$}, applying the
control strategy considered here to a {flow} governed by the full
Navier-Stokes {system} will require some additional careful
assumptions. This problem is currently being considered by the authors
of this paper (along with additional collaborators) with the aim of
{implementing control approaches capable of stabilizing vortex
  configurations which concomitantly increase the lift and decrease
  the drag experienced} by aerofoil configurations similar to those
considered here. {In particular, this investigation, which is
  based on the full Navier-Stokes model, will shed light on the
  significance of viscous effects always present in realistic flows,}
bringing {in this way} Kasper's vision closer to fruition.

\section*{Acknowledgements}

Rhodri Nelson and Takashi Sakajo would like to gratefully acknowledge
the support of JST-CREST who helped fund this study. {Bartosz
Protas was supported through an NSERC (Canada) Discovery Grant.}
{Takashi Sakajo was partially supported by Grants-in-Aid for Scientific Research
KAKENHI (B) No. 15TK0014 from JSPS.}

\end{document}